\documentclass{article}

\usepackage[nonatbib,dandb,preprint]{neurips_2025}

\usepackage[utf8]{inputenc} 
\usepackage[T1]{fontenc}    
\usepackage[backref=page]{hyperref}       
\usepackage{url}            
\usepackage{array}
\usepackage{booktabs}       
\usepackage{nicefrac}       
\usepackage{microtype}      
\usepackage[table]{xcolor}         

\usepackage{mathtools,amssymb,amsfonts}

\usepackage{amsmath}
\usepackage{lipsum}
\usepackage[numbers]{natbib}
\usepackage{enumitem}
\usepackage[capitalise]{cleveref}

\usepackage{makecell}
\newcolumntype{C}[1]{>{\centering\arraybackslash}m{#1}}
\newlength{\CellWidth}
\definecolor{metricblue}{HTML}{97CFFF}
\definecolor{myred}{HTML}{E93561}
\definecolor{high}{HTML}{97CFFF}  
\definecolor{low}{HTML}{FFFFFF}  
\newcommand*{\opacity}{90}

\newcommand{\ShadeMax}[3]{%
\begingroup
     \pgfmathparse{int(round(100*(#1-#2)/(#3-#2)))}
    \xdef\opa{\pgfmathresult}
    \cellcolor{metricblue!\opa}
  \endgroup
}

\newcommand{\ShadeMin}[3]{%
  \begingroup
     \pgfmathparse{int(round(100*(#3-#1)/(#3-#2)))+5}
    \xdef\opa{\pgfmathresult}
    \message{Opacity: \opa^^J}
    \cellcolor{metricblue!\opa}
  \endgroup
}

\usepackage[frozencache,cachedir=.]{minted}
\usepackage{tcolorbox}
\tcbuselibrary{minted,breakable,xparse,skins}

\usepackage[centerlast,small,sc]{caption}
\usepackage{newfloat}
\usepackage{listings}
\newenvironment{code}{\captionsetup{type=listing}}{}
\SetupFloatingEnvironment{listing}{name=Listing}

\definecolor{bg}{HTML}{FAFAFA}
\definecolor{orangeAPI}{HTML}{FCDE78}
\DeclareTCBListing{mintedbox}{O{}m!O{}}{%
  breakable=true,
  listing engine=minted,
  listing only,
  minted language=#2,
  minted style=default,
  minted options={%
    linenos,
    gobble=0,
    breaklines=true,
    breakafter=,,
    fontsize=\small,
    numbersep=8pt,
    #1},
  boxsep=0pt,
  left skip=0pt,
  right skip=0pt,
  left=25pt,
  right=0pt,
  top=3pt,
  bottom=3pt,
  arc=5pt,
  leftrule=0pt,
  rightrule=2pt,
  bottomrule=2pt,
  toprule=2pt,
  colback=bg!70,
  colframe=orangeAPI!70,
  enhanced,
  overlay={%
    \begin{tcbclipinterior}
    \fill[orangeAPI!20!white] (frame.south west) rectangle ([xshift=20pt]frame.north west);
    \end{tcbclipinterior}},
  #3}

\usepackage{wrapfig}
\usepackage{subcaption}   
\usepackage{pgffor}       

\providecommand{\DesignSpace}{\mathcal{X}}

\providecommand{\CondSpace}{\mathcal{C}}
\providecommand{\engibench}{\textsc{EngiBench}}
\providecommand{\engiopt}{\textsc{EngiOpt}}

\newcommand{\eg}{{\em e.g.}}

\newcommand{\etc}{{\em etc.}}

\newcommand{\revision}[1]{\textcolor{black}{#1}}

\title{EngiBench: A Framework for Data-Driven Engineering Design Research}

\author{%
  Florian Felten$^{1}$ \quad
    Gabriel Apaza$^{2,}$\thanks{Alphabetically ordered.} \quad
    Gerhard Bräunlich$^{1,*}$ \quad
    Cashen Diniz$^{1,*}$ \\
    \textbf{Xuliang Dong}$^{2,*}$ \quad
    \textbf{Arthur Drake}$^{2,*}$ \quad
    \textbf{Milad Habibi}$^{2,*}$ \quad
    \textbf{Nathaniel J. Hoffman}$^{2,*}$ \\
    \textbf{Matthew Keeler}$^{1,*}$ \quad
    \textbf{Soheyl Massoudi}$^{1,*}$ \quad
    \textbf{Francis G. VanGessel}$^{2,*}$ \quad
    \textbf{Mark Fuge}$^{1}$ \\
    $^{1}$ETH Zürich \quad $^{2}$University of Maryland, College Park\\
    \texttt{ffelten@mavt.ethz.ch} \quad \texttt{mafuge@ethz.ch}
    \vspace{-5mm}
}

\begin{document}

\maketitle

\begin{abstract}
Engineering design optimization seeks to automatically determine the shapes, topologies, or parameters of components that optimize performance under given conditions. This process often depends on physics-based simulations, which are difficult to install, computationally expensive, and require domain-specific expertise. To mitigate these challenges, we introduce EngiBench, the first open‐source library and datasets spanning diverse domains for data‐driven engineering design. EngiBench provides a unified API and a curated set of benchmarks\textemdash covering aeronautics, heat conduction, photonics, and more\textemdash that enable fair, reproducible comparisons of optimization and machine learning algorithms, such as generative or surrogate models. We also release EngiOpt, a companion library offering a collection of such algorithms compatible with the EngiBench interface. Both libraries are modular, letting users plug in novel algorithms or problems, automate end-to-end experiment workflows, and leverage built-in utilities for visualization, dataset generation, feasibility checks, and performance analysis. We demonstrate their versatility through experiments comparing state-of-the-art techniques across multiple engineering design problems, an undertaking that was previously prohibitively time-consuming to perform. Finally, we show that these problems pose significant challenges for standard machine learning methods due to highly sensitive and constrained design manifolds.\footnote{Documentation: \url{https://engibench.ethz.ch}.

~~~Benchmarks code: \url{https://github.com/IDEALLab/EngiBench}. 

~~~Datasets: \url{https://huggingface.co/IDEALLab}.

~~~Learning and optimization library code: \url{https://github.com/IDEALLab/EngiOpt}.}
\end{abstract}

\begin{figure}[H]
    \centering
    \includegraphics[width=0.9\linewidth]{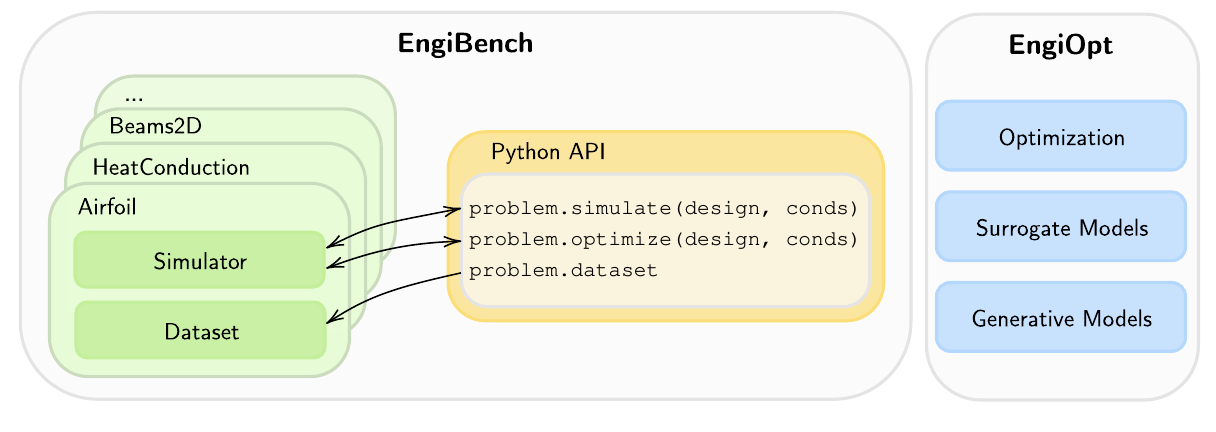}
    \caption{Overview of \engibench{} and \engiopt{} components.}
    \label{fig:engibench_overview}
\end{figure}

\section{Introduction}

Designing complex engineering systems has traditionally relied on iterative optimization processes and computationally expensive simulations. Recent advances in machine learning (ML)\textemdash particularly in inverse design (ID) and surrogate modeling (SM)\textemdash offer promising alternatives. Rather than repeatedly evaluating candidate designs through costly simulations, ID methods aim to generate designs directly from desired conditions and performance specifications, while surrogate models approximate simulation outputs. By significantly reducing the number of simulation calls, these approaches enable more efficient exploration of design spaces and lower computational costs. Engineering design researchers have explored ML methods for accelerating design exploration and optimization in many different directions: approximating optimal designs~\citep{chen2022inverse}, comparing neural operators or surrogate models~\citep{NEURIPS2024_013cf29a}, exploring design spaces/configurations~\citep{khan2023shiphullgan}, studying transfer learning~\citep{behzadi2022gantl}, among many others.

However, progress in data-driven engineering design has been significantly slowed by the lack of standardized simulation environments and publicly available, diverse datasets~\citep{regenwetter_deep_2022}. Most existing datasets are not shared, making it difficult to reproduce results or compare methods fairly. Even for datasets that do exist, different problem formulations often use different conventions or metrics, making it difficult or time consuming to compare algorithms across problems. Lastly, generating new datasets is not only computationally expensive, but also requires configuring and deploying complex simulation software\textemdash such as for computational fluid dynamics (CFD), finite element analysis (FEA), or circuit simulation\textemdash which can take weeks or even months and demands considerable domain expertise. This complexity presents a barrier to entry for ML researchers without an engineering background, narrowing research in these domains. As a result, researchers often focus on a single application domain, limiting broader generalization and hindering cross-domain benchmarking of algorithms.

To address this gap, we introduce \engibench{}\textemdash an open-source library supporting multiple domains in data-driven engineering design.\footnote{\engibench{} builds on the Maryland Inverse Design Benchmark (MIDBench) project (\url{https://ideal.umd.edu/midbench/}), which served as an initial prototype but did not reach the maturity needed for publication or widespread adoption.} It provides an intuitive and extensible Python API that unifies the modeling of diverse design problems and includes a suite of rigorously tested problem implementations. Each \engibench{} problem comprises a simulation engine and an associated dataset\textemdash encapsulating designs, objective values, attributes, and conditions\textemdash accessible via a standardized interface (see \cref{fig:engibench_overview}). To showcase the possibilities of using our framework, we also open-source a CleanRL-style~\cite{huang_cleanrl_2022} companion library called \engiopt{}, which contains a collection of ML and optimization algorithms integrated with \engibench{}. 

The key contributions of this paper are:

\begin{enumerate}[left=0mm]
    \item Novel challenging ML tasks that include hard constraints on what defines a feasible solution, and domain-specific metrics going further than statistical similarity, see \cref{subsec:preliminaries_background}. We show that the proposed problems are far from trivial, as these manifolds are highly sensitive to slight changes in the design space (\eg, keeping the continuity of an airfoil spline), making them difficult to learn with traditional methods, see \cref{sec:experiments}.
    \item A unified API which can be used in various optimization or ML paradigms, including inverse design, surrogate-based optimization, physics-informed neural networks (PINNs), etc. This API eases testing algorithms on a diverse set of problems for the engineering design community, see \cref{subsec:engibench_api}.
    \item Easy access to a diverse collection of real-world engineering design problems and associated physics simulators modeling different physical laws. The datasets collected on these problems are also made publicly available. These include problems across aerodynamics, heat transfer, power electronics design, photonics, etc., see \cref{subsec:engibench_problems}.
    \item A curated set of well-known generative and optimization algorithms implementations compatible with our API. This allows for benchmarking of multiple baseline algorithms across various design problems. In our proof-of-concept experiments, we observe that, contrary to popular belief, unconditional generative adversarial networks (GANs) may outperform their conditional counterparts and diffusion models on specific metrics relevant to engineering design; see \cref{sec:experiments}.
\end{enumerate}

\section{Preliminaries and Related Work}
\engibench{} aims at easing the work of researchers in ML for engineering design, but also proposes new challenges for the ML community. This section first defines the problem, its solutions, and how progress is generally measured in the field, it then discusses existing related works and how \engibench{} differs from those.


\subsection{Problem setting} 
\label{subsec:preliminaries_background}

The problem of finding optimal designs in engineering, often approached computationally, can be formulated as the following optimization problem. Let $ x \in \mathcal{X} \subseteq \mathbb{R}^d $ represent the design variables (\eg, spline parameters of an airfoil), $ a: \mathcal{X} \mapsto \mathcal{A} \subseteq \mathbb{R}^k $ be attributes computed for each design (\eg, mass), $ c \in \mathcal{C} \subseteq \mathbb{R}^l $ represent the environmental conditions (\eg, cruising velocity), $ f: \mathcal{X} \times \mathcal{C} \mapsto \mathcal{F} \subseteq \mathbb{R}^o $ be a (multi-)objective function (\eg, drag and lift), $ \sigma $ represent simulation assumptions and limitations (\eg, numerical approximations), and $ g $ and $ h $ represent inequality and equality constraints, respectively (\eg, a constraint ensuring the lift coefficient must exceed a given minimum). Then, without loss of generality,

\begin{equation*}
    \underset{x \in \mathcal{X}}{\text{minimize}} \, \tilde{f}(x, c, \sigma) \quad \text{subject to} \quad 
    \begin{aligned}
        & g_i(x, a(x), c, \sigma) \leq 0, \quad i = 1, \dots, q, \\
        & h_j(x, a(x), c, \sigma) = 0, \quad j = 1, \dots, r,
    \end{aligned}
\end{equation*}

where $ \tilde{f} $ only approximates the true objective $ f $ due to the simulation assumptions \& limitations $\sigma$.\footnote{For conciseness, we often omit $ \sigma $ and denote $ \tilde{f}(x, c, \sigma) $ as $ \tilde{f}(x, c)$.}

\textbf{The solutions} to this problem depend on the specific context and can take different forms: a single optimal design, a Pareto set of optimal designs when multiple objectives are considered, or a diverse set of near-optimal designs, allowing the human designer to select the most suitable one when exploring the design space.

\paragraph{Solving methods} 
In engineering design, optimization methods such as genetic algorithms (GAs) or adjoint solvers have traditionally been used to search for the optimal design variables $ x^* \in \mathcal{X} $. However, this process is computationally expensive, as it requires running simulation software to evaluate each design. ML approaches, such as inverse design and surrogate-based optimization, have emerged as popular alternatives. These methods leverage datasets $ \mathcal{D} = \{ (x_i, a(x_i), c_i, \tilde{f}(x_i, c_i)) \}_{i=1}^N $ of existing designs, their corresponding properties, and performance outcomes to reduce the computational burden of traditional optimization techniques.

\textbf{Inverse design} aims to learn a (generative) model $ m: \mathcal{A} \times \mathcal{C} \mapsto \mathcal{X} $ that can predict the optimal design variables $x^*$ given desired attributes $a$ and conditions $c$. The output of such a model is rarely optimal but is often used as an initial guess for the optimization process to speed up its convergence. The model can also be used as an operator in a classical search algorithm~\citep{regenwetter_deep_2022}.

\textbf{Surrogate-based optimization} approximates the objective function $ \tilde{f}(x, c) $ using a surrogate model $ m: \mathcal{X} \times \mathcal{C} \mapsto \mathcal{F} $. This model is then used to guide the optimization process (\eg, Bayesian optimization, GAs, or gradient-based methods when $ m $ is differentiable), replacing the computationally expensive simulations with faster approximations.


\paragraph{Performance metrics}  
Depending on the context and application, engineers prioritize different aspects of a model's output and often rely on performance metrics that go beyond mean squared error. In some cases, they may even prefer models that produce less detailed but easier-to-optimize outputs (\eg, blurry images) over sharper ones~\citep{habibi2025mean}. Below, we outline several metrics used in our experiments. For a more comprehensive overview, we refer the reader to~\citet{regenwetter_beyond_2023}.

\textbf{Similarity:} A model should generate results similar to the dataset. In inverse models, this means matching the dataset distribution, while surrogate models aim for accurate predictions of the objective values. A common metric is \emph{maximum mean discrepancy} (MMD,~\citep{gretton_kernel_2012}), which measures the distance between two distributions. Given a dataset $\mathcal{D} = \{z_i\}_{i=1}^{N}$ and generated samples $\mathcal{D}_g = \{\hat{z}_j\}_{j=1}^{M}$, MMD is defined as  
$\text{MMD}^2(k, \mathcal{D}, \mathcal{D}_g) = \mathbb{E}[k(z, z')] + \mathbb{E}[k(\hat{z}, \hat{z}')] - 2 \mathbb{E}[k(z, \hat{z})]$,  
where $k(\cdot, \cdot)$ is a kernel function (\eg, Gaussian).\footnote{Note that $z$ would denote designs ($x$) for inverse design models and objective values ($\tilde{f}(x,c)$) for surrogate models.} A lower MMD indicates better alignment between generated and real distributions.  

\textbf{Diversity:} In inverse problems, to allow for exploration of the design space, models should generate a diverse set of designs. \emph{Determinantal point processes} (DPP, \citep{kulesza_dpp_2012}) quantify diversity by favoring well-spread-out samples. Given a similarity kernel $K \in \mathbb{R}^{M \times M}$ with elements $K_{ij} = k(\hat{x}_i, \hat{x}_j)$, diversity is measured as $\text{DPP}(\mathcal{D}_g) = \det(K)$. A higher determinant value indicates greater diversity.  

\textbf{Optimality:} Engineers prefer models that generate high-performance designs. The \emph{cumulative optimality gap} (COG) measures how far a generated design lags behind the best known objective value \emph{throughout} the optimization process. Let $f^* = \max_{x \in \mathcal{X}} \tilde{f}(x, c)$ denote the optimal objective value under condition $c$, typically estimated using an adjoint solver. For a generated design $\hat{x} = x_0$ that is refined through $T$ optimization steps yielding a sequence $x_0, \ldots, x_T$, the COG is defined as $\text{COG}(x_0) = \sum_{t=0}^T ( \tilde{f}(x_t, c) - f^* )$. A lower COG implies the optimization path remained close to optimal throughout, and is thus preferred in practical settings.

\textbf{Feasibility:} Designs must be physically feasible and avoid constraint violations. The \emph{ratio of violated constraints} (RVC) measures how many generated samples fail to satisfy constraints on the designs. It is defined as  
$$\text{RVC}(\mathcal{D}_g) = \frac{1}{M} \sum_{k=1}^{M} \mathbb{I} \left( \exists i \in [1,q] \text{ s.t. } g_i(\hat{x}_k, a(\hat{x}_k), c) > 0 \text{ or } \exists j \in [1, r] \text{ s.t. } h_j(\hat{x}_k, a(\hat{x}_k), c) \neq 0 \right),$$  
where $\mathbb{I}(\cdot)$ is an indicator function. A lower RVC indicates that the model tends to generate designs that satisfy physical and problem-specific constraints more reliably. Another indicator used in this work is the \emph{ratio of failed simulations} (RF), which measures the number of generated designs that threw errors when trying to simulate.


\subsection{Related work}

\textbf{Datasets and Benchmarks:}  A wealth of well-known datasets have been used over the years to assess the performance of ML algorithms. Examples such as MNIST~\citep{lecun2010mnist} and CIFAR-10~\citep{Krizhevsky09learningmultiple} have streamlined the work of ML practitioners, enabling fair comparisons and reducing publication workloads. However, most of these tasks primarily focus on maximizing the similarity between model outputs and dataset labels, and are often unconstrained. In contrast, our problems require running heavyweight simulations (\eg, CFDs) to validate designs. Recently, a few general benchmarks have bridged new gaps in ML for complex physical systems. For example, PDEBench~\citep{takamoto2022pdebench} and LagrangeBench~\citep{toshev2023lagrangebench} propose a wide range of new datasets and metrics for ML applications in Partial Differential Equation (PDE) flow problems. While some valuable concepts are presented that may be extended to other domains, these works are limited to the single-physics domain of fluid flows. In contrast, \engibench{} provides a much wider variety of engineering datasets encompassing both single- and multi-physics domains. A few novel works that \textit{do} include significant multi-physics data are BubbleML~\citep{hassan2023bubbleml}, which benchmarks different neural operators and UNets for capturing phase change phenomena, and The Well~\citep{ohana2024well}, a large collection of numerical simulations for various spatiotemporal physical systems to train and evaluate surrogate ML models. Although the scale and variety of simulation data are impressive in both works, little emphasis is placed on generalized metrics or analysis across different problems, and no APIs or extensions to generative modeling are provided. In ML specifically for engineering design, most studies rely on custom datasets, often kept private, limiting reproducibility and benchmarking. Fortunately, this trend is changing, and recent efforts have made datasets publicly available for specific domains such as airfoils~\citep{chen_aerodynamic_GAN_2019}, beams~\citep{sosnovik_neural_2019}, bicycles~\citep{regenwetter_biked_2021}, car bodies~\cite{NEURIPS2024_013cf29a}, among many others. However, these datasets remain problem-specific, making cross-domain benchmarking difficult and slowing the study of the generalization of ML methods across engineering tasks.

\textbf{Simulation software in engineering design:}  
Unlike conventional ML tasks, engineering design requires evaluating solutions via physics-based simulations. Simulation software has been developed for decades to evaluate designs, with open-source examples such as MachAero for CFD in airplane design~\citep{Wu_pyoptsparse_2020,PyHyp,Mader_adflow_2020a}, Dolfin/FEniCS for FEAs~\citep{logg_DOLFIN_2010,logg_automated_2012}, and NgSpice~\citep{ngspice} in circuit simulation, along with many closed-source/commercial solver packages. However, these tools often present significant barriers to use: they can be difficult to install and parameterize, frequently involve long setup times, may produce inconsistent results due to differences in simulator versions or simulation configurations, or, in the case of commercial tools, cannot be easily adopted or scaled without significant cost. Moreover, their heterogeneous interfaces and problem representations further complicate integration into ML or optimization workflows, making it difficult to apply the same codebase across different design problems to conduct cross-domain studies. Collectively, these challenges create a high barrier to entry for ML practitioners, slow down research progress, and hinder reproducibility in the field.


\section{EngiBench}
\label{sec:engibench}

\begin{figure}[tb]
    \centering
    \includegraphics[width=0.95\linewidth]{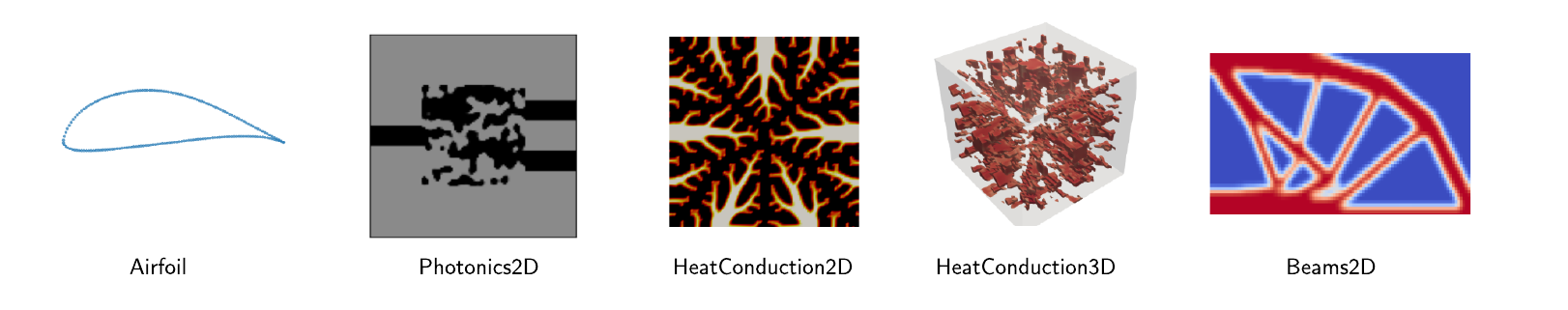}
    \caption{Visualization of some of the implemented problems.}
    \label{fig:problems_viz}
\end{figure}

To address these challenges, we propose \engibench{}, an open-source library designed for data-driven engineering design across multiple domains. It provides an intuitive and extensible Python API that unifies the modeling of diverse design problems. Conceptually, each \engibench{} problem consists of a simulation engine and an associated dataset containing designs, objectives, attributes, and conditions, all accessible through a standardized interface; see \cref{fig:engibench_overview}.

\engibench{} abstracts away simulation complexities behind its user-friendly API. This allows ML and engineering researchers to focus on modeling and optimization rather than spending weeks configuring simulation environments.
Additionally, \engibench{} provides a repository of ready-to-use problems and datasets generated from the supported simulators, see \cref{fig:problems_viz}. Researchers can directly leverage these datasets for training and validating models without needing to generate their own, significantly accelerating experimentation and enhancing reproducibility in ML for engineering design.

\subsection{Application Programming Interface}
\label{subsec:engibench_api}

\begin{code}
\begin{mintedbox}{python}
from engibench.problems.beams2d.v0 import Beams2D
problem = Beams2D() # Instantiate problem
problem.reset(seed=42)
# Inspect problem
problem.design_space  # Box(0.0, 1.0, (50, 100), float64)
problem.objectives  # (("compliance", "MINIMIZE"),)
problem.conditions  # (("volfrac", 0.35), ("forcedist", 0.0),...)
problem.dataset # A HuggingFace Dataset object
# inverse_model = train_inverse(problem.dataset)
desired_conds = {"volfrac": 0.7, "forcedist": 0.3}
# generated_design = inverse_model.predict(desired_conds)
random_design, _ = problem.random_design()
# check constraints on the design, config pair
violated_constraints = problem.check_constraints(random_design, desired_conds)
if not violated_constraints:
   # Only simulate to get objective values
   objs = problem.simulate(random_design, desired_conds)
   problem.reset(seed=41)
   # Or run a gradient-based optimizer to polish the design
   opt_design, history = problem.optimize(random_design, desired_conds)
   problem.render(opt_design)
\end{mintedbox}
\captionof{listing}{API example.\label{listing:api}}
\end{code}

\cref{listing:api} illustrates a high-level usage example of our API. Users can instantiate a problem (lines 1\textendash 2), extract relevant training data (lines 4\textendash 8), generate new designs (lines 9\textendash 12), check constraints for those designs (lines 14\textendash 15), validate or optimize these designs using a simulation engine (lines 16\textendash 20), and render a given design (line 21). A key feature of the API is its flexibility: switching to a different problem only requires modifying the first two lines of code, such as importing \textit{HeatConduction2D} instead of \textit{Beams2D}.

\subsection{Features}
\label{subsec:engibench_features}

\engibench{} introduces several features ensuring good reproducibility and ease of use.

\textbf{Versioning:} Similar to popular reinforcement learning libraries, such as Gymnasium and its variants~\cite{towers_gymnasium_2024,felten_toolkit_2023,felten_momaland_2024}, our problems' implementations and datasets are versioned (see line 1 in \cref{listing:api}), ensuring traceability when changes affect results.

\textbf{Problem metadata:} Problems expose attributes such as \textit{design\_space}, \textit{objectives}, and \textit{conditions}, containing information about the size, bounds, and shapes of the design components, the number of conditions and objectives. These allow automatic extraction of information like input and output dimensions for ML models, but also provide a generic way to extract relevant columns from the datasets (lines 4\textendash 8 in \cref{listing:api}).

\textbf{Constraint checking:} The library can verify constraint violations for a given design and conditions (line 14 in \cref{listing:api}). Constraints are categorized as theoretical (from the mathematical problem definition) or implementation-based (arising from simulation assumptions). Severity levels include errors (preventing simulation) and warnings (indicating possible inaccuracies). Although our implementations do not capture all possible kinds of errors, this allows users to pinpoint obvious issues with designs or configurations; see \cref{app:error_handling} for more information regarding error handling.

\textbf{Virtualized environment:} Some simulation software requires a native installation to run. To simplify the deployment and execution of our library in such cases, we support integration with Docker, Podman, and Apptainer. This drastically reduces the setup time for users as they can just \textit{pip install and run}.

\textbf{Adjoint optimization:} Most problems feature an adjoint optimizer (line 19 of \cref{listing:api}), enabling gradient-based optimization starting from generated designs. These state-of-the-art methods scale well to high-dimensional design spaces, though they may converge to local optima~\cite{mdobook}. This is notably useful to compute optimality-related metrics, such as COG.

\textbf{Integration with ML/optimization libraries:} \engibench{} seamlessly integrates with tools like HuggingFace Datasets and Diffusers~\citep{von_Platen_Diffusers_State-of-the-art_diffusion}, PyTorch~\citep{paszke_pytorch_2019}, BoTorch~\citep{balandat2020_botorch}, and Pymoo~\citep{blank_pymoo_2020}, streamlining ML and optimization workflows. We provide several examples of ML and optimization algorithms in 
\engiopt{}.

\textbf{Distributed computing:} \engibench{} supports distributed execution, enabling large-scale dataset generation through parallel optimizations or simulations on high-performance computing clusters using Slurm~\cite{yoo_slurm_2003}\textemdash \eg, we generated our Photonics2D dataset in under 30 minutes using such a tool.

\subsection{Implemented problems}

\label{subsec:engibench_problems}

\begin{table}
\caption{Problems currently implemented in EngiBench. Simulation times have been averaged across 10 runs on a MacBook Pro M3 Max (on the CPU cores).}
  \renewcommand{\arraystretch}{1.3}
    \centering
    \label{tab:problems}
\begin{tabular}{lllllcc}
\toprule
\textbf{Problem}            & \textbf{\begin{tabular}[c]{@{}c@{}}Design\\Repr.\end{tabular}} & \textbf{\begin{tabular}[c]{@{}c@{}}Design\\Space\end{tabular}} & \textbf{Opt. class}   & \textbf{Physics}     & \textbf{\begin{tabular}[c]{@{}c@{}}Simu.\\time\\(seconds)\end{tabular}} & \textbf{\begin{tabular}[c]{@{}c@{}}Num.\\Objs.\end{tabular}} \\ \midrule
\textit{Airfoil}                  & \begin{tabular}[c]{@{}l@{}}Vector +\\Scalar Tuple\end{tabular} & Cont.            & Shape            & Navier-Stokes                  &             9.85             & 1                                                             \\[4mm] 

\begin{tabular}[c]{@{}l@{}}\textit{Heat-}\\ \textit{Conduction2D}\end{tabular}          & Image                          & Disc.              & Topology        & Therm. diffusion              &       10.22                   & 1                                                             \\[4mm]

\begin{tabular}[c]{@{}l@{}}\textit{Heat-}\\ \textit{Conduction3D}\end{tabular}          & 3D Tensor                      & Disc.              & Topology        & Therm. diffusion              &     31.57                     & 1                                                             \\[4mm] 

\begin{tabular}[c]{@{}l@{}}\textit{Thermo-}\\ \textit{Elastic-}\\ \textit{Beams2D}\end{tabular} & Image                          & Disc.            & Topology        & \begin{tabular}[c]{@{}l@{}}Therm. diffusion\\ \& Lin. elasticity\end{tabular}  &     0.22                     & 3                                                             \\[7mm] 

\textit{Beams2D}             & Image                          & Disc.              & Topology         & Lin. elasticity           &          0.19                & 1                                                             \\[4mm]

\textit{Photonics2D}       & Image   & Disc.    & Topology      & Maxwell's eqs.  & 0.3 & 1\\[4mm]

\textit{PowerElectronics}           & Vector & Mixed                 & Tabular & Electr. dynamics          &           0.66               & 2                                                             \\                                                              \bottomrule
\end{tabular}
\end{table}

As mentioned earlier, \engibench{} includes a collection of implemented problems conforming to its API, listed in \cref{tab:problems} and partly illustrated in \cref{fig:problems_viz}. \revision{These were selected to reflect the diversity of real-world engineering design scenarios. They vary in design representation\textemdash ranging from vectors (\eg, electrical circuit parameters) to 2D images (\eg, pixel-based topology optimization for beams) and 3D tensors (\eg, voxel-based topology optimization for heat conduction)\textemdash as well as in physics domain, including aerodynamics, structural elasticity, thermal diffusion, and power electronics. The benchmark also covers both single- and multi-objective settings, and includes problems with a wide range of computational costs, from fast prototyping tasks to more demanding simulations. This range in simulation complexity is intentional: our goal is to provide a continuum of tasks that can serve both small labs with limited resources and large-scale teams who wish to address more complex, difficult, and realistic problems. Finally, several problems (and datasets) replicate setups from prior publications \citep{andreassen_88lines_2011,habibi2023actually,habibi2025mean,dinizdiffusion}, facilitating reproducibility and comparison with existing work.} Each problem is briefly described below and detailed explanations can be found in \cref{app:extended_description}.

\textbf{Airfoil:} The airfoil problem involves aerodynamic shape optimization based on the Reynolds-averaged Navier–Stokes (RANS) equations. The design is parameterized by 192 control points defining the 2D airfoil geometry, along with a scalar representing the angle of attack. The objective is to match a prescribed lift coefficient while minimizing drag. We use MachAero~\citep{Wu_pyoptsparse_2020,PyHyp,Mader_adflow_2020a} to perform the CFD simulations. A description of this problem and dataset is provided in~\citet{dinizdiffusion}.

\textbf{HeatConduction2D/HeatConduction3D:} These problems are a specific subset of topology optimization (TO) problems aimed at placing highly conductive material to minimize thermal compliance within a unit square (2D) or unit cube (3D), subject to: a constraint on the volume of material used, and given conditions, particularly the location of the adiabatic region. Our code implementation is a modified version of the open-source Dolfin-adjoint example to solve the 2D and 3D heat conduction topology optimization problems~\citep{mitusch2019dolfin,logg_DOLFIN_2010,logg_automated_2012}.

\textbf{ThermoElasticBeams2D:} The ThermoElasticBeams2D problem specifies a multi-physics TO problem governed by linear elasticity and steady-state heat conduction. The goal is to find an optimal placement of material such that both structural compliance and thermal compliance are minimized while respecting a volume fraction constraint. Our implementation is based on the well-known 88-line MATLAB code for compliance minimization~\citep{andreassen_88lines_2011}, adapted to Python and extended to the thermoelastic multi-physics case.

\textbf{Beams2D:} Beams2D is a structural TO problem that optimizes a beam under bending. The beam has a force applied at the top, which can be parameterized through conditions. The goal is to optimize the distribution of solid material within a rectangular grid to minimize compliance while satisfying constraints on material usage and minimum feature size. 
Our implementation is also based on the 88-line MATLAB code for compliance minimization~\citep{andreassen_88lines_2011}, adapted to Python.

\textbf{Photonics2D:} Photonics2D is an electromagnetic waveguide optimization problem where the goal is to demultiplex an incoming port such that light under two different wavelengths exits the device at different locations~\cite{hughes2018adjoint, piggott2015inverse}. The goal is to maximize the electromagnetic field present at the desired exit location for the target wavelength. Our implementation is based on the examples from the \texttt{ceviche} finite-difference frequency-domain (FDFD) library~\cite{hughes2019forward}.

\textbf{PowerElectronics:} This problem models a DC-DC power converter circuit with a fixed circuit topology. The circuit comprises 5 switches, 4 diodes, 3 inductors, and 6 capacitors. We vary circuit parameters, such as capacitance, then use the NgSpice simulator \citep{ngspice} to approximate the circuit and compute performance metrics including $\textit{DcGain}$ and $\textit{Voltage Ripple}$. Despite the fixed topology, optimizing the circuit parameters to minimize the objectives remains a challenging task for surrogate models, as we show in \cref{sec:experiments}.

\subsection{Implemented algorithms in EngiOpt}

\label{subsec:engiopt_algorithms}

\revision{We provide baselines that are well understood by ML practitioners\textemdash Generative Adversarial Networks (GANs)~\cite{goodfellow_generative_2014,mirza_conditional_2014,radford_2016_dcgan} and diffusion models~\cite{ho_denoising_2020,von_Platen_Diffusers_State-of-the-art_diffusion} for inverse design, and MLP+NSGA‑II~\cite{deb_fast_2002} for surrogate‑assisted optimization. The implementations are single‑file (CleanRL-style~\cite{huang_cleanrl_2022}), easy to debug, and avoid heavy simulator‑specific dependencies or complex physics priors to maximize reproducibility and pedagogical value. They also provide a baseline framework that can easily extend to more complex algorithms in future versions of \engiopt{}.
}

\revision{For harder cases (Sec.~\ref{sec:zoomed_exp}), we also include more specialized algorithms when warranted: (i) B\'ezierGAN uses an airfoil‑specific B\'ezier parameterization that enforces smoothness and reduces meshing failures~\cite{chen_airfoil_2020}; (ii) our surrogate stack combines robust preprocessing, Bayesian hyperparameter search, implicit deep ensembles, and NSGA‑II for multi‑objective search. Notably, these models have matched simulator Pareto fronts and were experimentally validated in turbomachinery~\cite{massoudi_dimensionless_2025,massoudi_integrated_2024,massoudi_robust_2025}. Even with these advanced models, several problems remain challenging (\eg, PowerElectronics), underscoring the intrinsic difficulty of many engineering design tasks.}

\section{Proof-Of-Concept Experiments}
\label{sec:experiments}

In this section, we use \engibench{} and \engiopt{} to conduct a series of proof-of-concept empirical experiments and discuss the corresponding results. Our aim is to illustrate the types of experiments, comparisons, and analyses\textemdash across different algorithms, problem domains, and performance metrics\textemdash that our framework supports. Given the potentially high computational cost of training, optimization, and evaluation, we limit each algorithm-problem pair to 10 independent runs, reporting the mean and standard deviation for each metric. While more extensive experiments with reduced variance could be conducted, our focus here is not on providing a comprehensive benchmark of existing algorithms, but rather on demonstrating the practical capabilities and flexibility of our framework. A complete description of the conducted experiments, additional results, hyperparameters, discussions, and illustrations is available in \cref{app:additional_info_expes}.

\subsection{Cross-domain study}
\label{sec:3by3_exp}

\begin{table}[ht]
\centering
\caption{Averaged metrics (mean $\pm$ std) per problem and model over 10 seeds. Gradient colors indicate best by linear interpolation between the min. and max. values across algorithms for a given metric on a given problem.}

\caption*{
\small
Each cell displays
\centering
\begin{tabular}{cc}
COG ($\downarrow$) & RVC ($\downarrow$) \\
\hline
MMD ($\downarrow$) & DPP ($\uparrow$) \\
\end{tabular}. B2D=Beams2D, HC2D=HeatConduction2D, P2D=Photonics2D.
}

\label{tab:3by3}
\addtolength{\tabcolsep}{-0.3em}
\begin{tabular}{cccc}
\toprule
 & GAN2D & CGAN2D & CDiffusion2D \\
\specialrule{\heavyrulewidth}{0pt}{0pt}
\raisebox{-1.5\height}{B2D}   & 
  {\renewcommand{\arraystretch}{1.9}\begin{tabular}[t]{@{}cl@{}}
     \makecell{
         \ShadeMin{16}{15}{25} 1.56e+08 $\pm$\\
         \ShadeMin{16}{15}{25} 6.92e+07
     } 
      & \makecell{
        \ShadeMin{94}{68}{100} 9.44e-01 $\pm$\\
        \ShadeMin{94}{68}{100} 3.50e-02
    } \\ \hline
    \makecell{
        \ShadeMin{11}{11}{19} 1.07e-01 $\pm$\\
        \ShadeMin{11}{11}{19} 1.95e-02
    } 
    & \makecell{
        \ShadeMax{100}{0}{100} 8.08e-07 $\pm$\\
        \ShadeMax{100}{0}{100} 2.56e-06
    } 
   \end{tabular}} &
  {\renewcommand{\arraystretch}{1.9}\begin{tabular}[t]{@{}cl@{}}
    \makecell{
        \ShadeMin{15}{15}{25} 1.51e+08 $\pm$\\
        \ShadeMin{15}{15}{25} 1.74e+08
    }
    & \makecell{
        \ShadeMin{68}{68}{100} 6.76e-01$\pm$\\
        \ShadeMin{66}{68}{100} 1.50e-01
    } \\ \hline
    \makecell{
        \ShadeMin{13}{11}{19} 1.25e-01 $\pm$\\
        \ShadeMin{13}{11}{19} 1.02e-01
    }
    & \makecell{
        \ShadeMax{0}{0}{100} 3.62e-19 $\pm$\\
        \ShadeMax{0}{0}{100} 1.14e-18
    }
   \end{tabular}} &
  {\renewcommand{\arraystretch}{1.9}\begin{tabular}[t]{@{}cl@{}}
    \makecell{
         \ShadeMin{25}{15}{25} 2.45e+08 $\pm$\\
         \ShadeMin{25}{15}{25} 1.09e+08 
    }
    & \makecell{
        \ShadeMin{100}{68}{100} 9.98e-01 $\pm$\\ 
        \ShadeMin{100}{68}{100} 6.32e-03
    } \\ \hline
    \makecell{
        \ShadeMin{19}{11}{19} 1.90e-01 $\pm$\\
        \ShadeMin{19}{11}{19} 4.17e-02
    }
    & \makecell{
        \ShadeMax{1}{0}{100} 3.50e-19 $\pm$\\
        \ShadeMax{1}{0}{100} 7.87e-19 
    }
   \end{tabular}} \\
\specialrule{\heavyrulewidth}{0pt}{0pt}
\raisebox{-1.5\height}{HC2D} & 
  {\renewcommand{\arraystretch}{1.9}\begin{tabular}[t]{@{}cl@{}}
    \makecell{
        \ShadeMin{26}{13}{64} 2.56e-03 $\pm$\\
        \ShadeMin{26}{13}{64} 2.77e-04 
    }
    & \makecell{
        \ShadeMin{96}{75}{100} 9.60e-01 $\pm$\\
        \ShadeMin{96}{75}{100} 1.89e-02
    } \\ \hline
    \makecell{
        \ShadeMin{8}{8}{37} 8.43e-02 $\pm$\\ 
        \ShadeMin{8}{8}{37} 1.06e-02 
    }
    & \makecell{
        \ShadeMax{100}{0}{100} 7.73e-01 $\pm$\\
        \ShadeMax{100}{0}{100} 1.30e-01 
    } 
   \end{tabular}} &
  {\renewcommand{\arraystretch}{1.9}\begin{tabular}[t]{@{}cl@{}}
    \makecell{
        \ShadeMin{13}{13}{64} 1.38e-03 $\pm$\\
        \ShadeMin{13}{13}{64} 3.15e-04
    }
    & \makecell{
        \ShadeMin{75}{75}{100} 7.52e-01 $\pm$\\
        \ShadeMin{75}{75}{100} 1.68e-01
    } \\ \hline
    \makecell{
        \ShadeMin{37}{8}{37} 3.68e-01 $\pm$\\
        \ShadeMin{37}{8}{37} 4.23e-02
    }
    & \makecell{
        \ShadeMax{0}{0}{100} 2.65e-23 $\pm$\\
        \ShadeMax{0}{0}{100} 7.57e-23 
    }
   \end{tabular}} &
  {\renewcommand{\arraystretch}{1.9}\begin{tabular}[t]{@{}cl@{}}    
    \makecell{
        \ShadeMin{64}{13}{64} 6.44e-03 $\pm$\\ 
        \ShadeMin{64}{13}{64} 3.03e-03 
    }
    & \makecell{
        \ShadeMin{100}{75}{100} 1.00e+00 $\pm$\\
        \ShadeMin{100}{75}{100} 0.00e+00
    } \\ \hline
    \makecell{
        \ShadeMin{18}{8}{37} 1.83e-01 $\pm$\\
        \ShadeMin{18}{8}{37} 3.52e-02 
    }
    & \makecell{
        \ShadeMax{1}{0}{100} 3.70e-03 $\pm$\\
        \ShadeMax{1}{0}{100} 6.58e-03 
    }
   \end{tabular}} \\
\specialrule{\heavyrulewidth}{0pt}{0pt}
\raisebox{-1.5\height}{P2D} & 
  {\renewcommand{\arraystretch}{1.9}\begin{tabular}[t]{@{}cl@{}}
    \makecell{
        \ShadeMin{44}{19}{74} 9.44e+02 $\pm$\\
        \ShadeMin{44}{19}{74} 1.82e+01 
    }
    & \makecell{
        N/A 
    } \\ \hline
    \makecell{
        \ShadeMin{31}{5}{90} 3.11e-01 $\pm$\\
        \ShadeMin{31}{5}{90} 1.22e-01
    }
    & \makecell{
        \ShadeMax{0}{0}{100} 5.11e-07 $\pm$\\
        \ShadeMax{0}{0}{100} 1.52e-06 
    }
   \end{tabular}} &
  {\renewcommand{\arraystretch}{1.9}\begin{tabular}[t]{@{}cl@{}}
    \makecell{
        \ShadeMin{74}{19}{74} 9.74e+02 $\pm$\\
        \ShadeMin{74}{19}{74} 7.72e+00 
    }
    & \makecell{
        N/A
    }  \\ \hline
    \makecell{
        \ShadeMin{90}{5}{90} 9.00e-01 $\pm$\\
        \ShadeMin{90}{5}{90} 4.61e-02 
    }
    & \makecell{
        \ShadeMax{0}{0}{100} 6.24e-89 $\pm$\\
        \ShadeMax{0}{0}{100} 1.97e-88
    }
   \end{tabular}} &
  {\renewcommand{\arraystretch}{1.9}\begin{tabular}[t]{@{}cl@{}}
    \makecell{
        \ShadeMin{19}{19}{74} 9.19e+02 $\pm$\\
        \ShadeMin{19}{19}{74} 2.38e+01 
    }
    & \makecell{
        N/A 
    } \\ \hline
    \makecell{
        \ShadeMin{5}{5}{90} 5.30e-02 $\pm$\\
        \ShadeMin{5}{5}{90} 6.58e-03
    }
    & \makecell{ 
        \ShadeMax{100}{0}{100} 8.23e-01 $\pm$\\
        \ShadeMax{100}{0}{100} 1.46e-01 
    }
   \end{tabular}} \\
\bottomrule
\end{tabular}
\end{table}

We showcase a new type of experiment that was previously difficult to conduct: a comparative study of different algorithms across multiple engineering design problems using a variety of performance metrics. With our tools, this evaluation became straightforward\textemdash different problems could be explored simply by changing the \texttt{problem\_id} argument from the command line for each algorithm. We trained several generative models for inverse design over 100 epochs, including a deep convolutional generative adversarial network (GAN2D), a conditional deep convolutional GAN (CGAN2D), and a conditional diffusion model (CDiffusion2D).
 
Results are shown in \cref{tab:3by3}. Interestingly, the unconditioned GAN model performs well overall in terms of COG and MMD, despite not producing well-defined shapes (see \cref{fig:beams,fig:heatconduction,fig:photonics} in \cref{app:additional_info_expes}). In terms of RVC\textemdash \revision{specifically, maintaining the volume fraction below the specified threshold for B2D and HC2D (P2D has no such constraint)\textemdash all algorithms perform rather poorly.} The CGAN model achieves the best performance, whereas the CDiffusion model performs worst, which is somewhat unexpected. Regarding design diversity, the GAN model performs best on the first two problems, likely because it does not restrict itself to condition-specific regions. In contrast, the diffusion model excels on the photonics task, suggesting that the ``best'' generative model may depend on the nature of the design problem at hand. \revision{Finally, raw model outputs are often not directly usable in practice (\eg, disconnected beam members, \cref{fig:beams}) and typically require post-processing or the injection of domain-specific priors, as illustrated in the next section\textemdash highlighting a common limitation of current generative approaches.}

\subsection{Hard cases: Airfoils and PowerElectronics}

\label{sec:zoomed_exp}

\begin{wrapfigure}[14]{R}{0.45\textwidth}
\vspace{-2.5em}
\centering
    \includegraphics[width=0.45\textwidth]{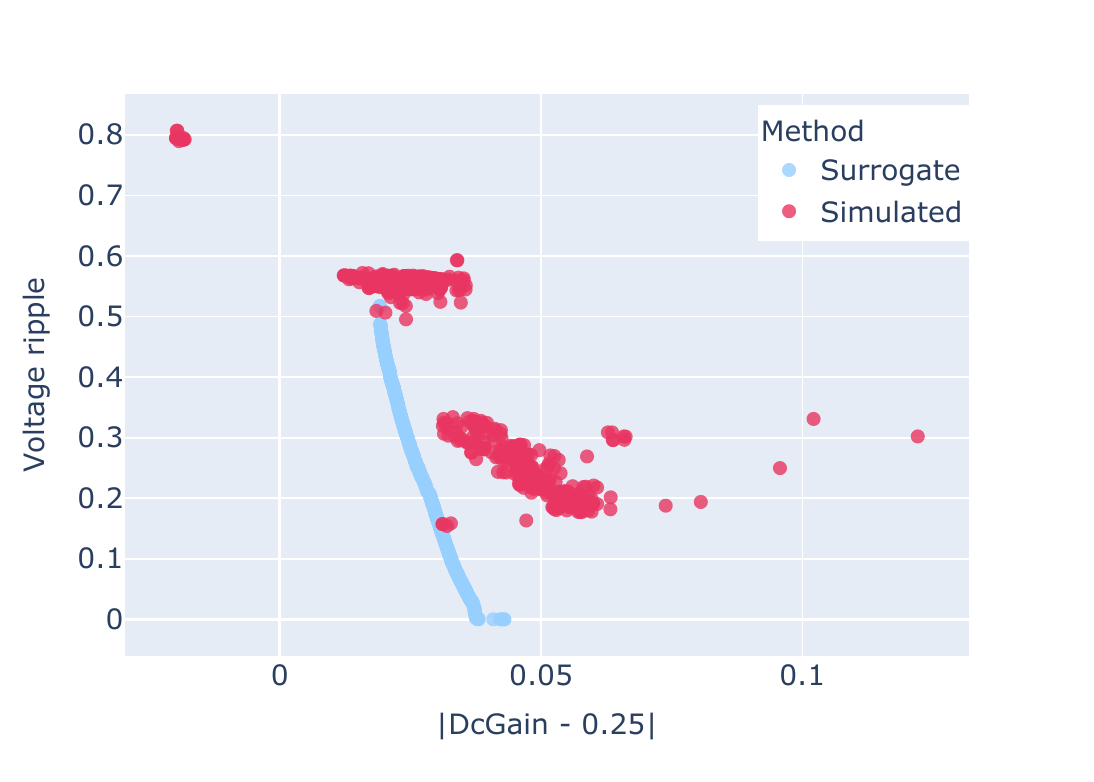}
\caption{Difference between the Pareto fronts based on the surrogate models and the simulator.}
\label{fig:pf_diff_power_elec}
\end{wrapfigure}

\textbf{Surrogate models for PowerElectronics:} We trained two robust-scaled MLP surrogates\textemdash one for \textit{DcGain} and one for \textit{Voltage Ripple}\textemdash using Bayesian hyperparameter search and implicit deep ensembles~\cite{xue_deep_2021,ganaie_ensemble_2022}. We then ran NSGA-II using our surrogate models to find Pareto optimal designs. Despite careful tuning and the variance reduction afforded by the ensembles, all the surrogate-estimated Pareto fronts diverge sharply from NgSpice when re-evaluated (\eg, see \cref{fig:pf_diff_power_elec}); across our 10 runs, all squared MMD tests rejected distributional equality.

We attribute the failure largely to the stiff, outlier-prone \textit{Voltage Ripple} response (see \cref{app:additional_info_expes} for more details), which remains difficult to approximate even after log transformation. In contrast, \textit{DcGain} exhibits smoother behavior. These results underscore the difficulty of substituting physical simulators with black-box surrogates in systems with mixed time scales and sparse, high-variance regimes. Richer experimental design and physics guided feature transformations may be essential for reliable optimization.

\begin{figure*}
  \centering
  \def\figwidth{0.3\textwidth}

  \begin{subfigure}{\figwidth}
    \centering
    \includegraphics[width=\linewidth]{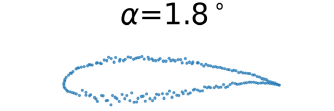}
    \caption*{\footnotesize GAN}
  \end{subfigure}\hfill
  \begin{subfigure}{\figwidth}
    \centering
    \includegraphics[width=\linewidth]{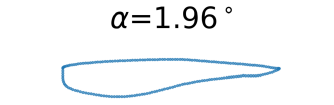}
    \caption*{\footnotesize Diffusion}
  \end{subfigure}\hfill
  \begin{subfigure}{\figwidth}
    \centering
    \includegraphics[width=\linewidth]{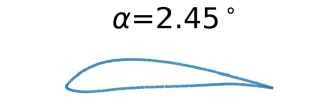}
    \caption*{\footnotesize B\'ezierGAN}
  \end{subfigure}

  \caption{Example outputs of our generative models on the Airfoil problem.}
  \label{fig:airfoil_main_paper_examples}
\end{figure*}

\textbf{Generative models for Airfoil:} We trained 3 generative models to perform inverse design for airfoils: a standard GAN, a diffusion model, and a B\'ezierGAN, which is specifically tailored to airfoil parameterization. Each model was trained for 2500 epochs. For every algorithm and random seed, we generated 50 candidate designs and attempted to simulate them, reporting the resulting ratio of failed simulations (RF). The GAN yielded an RF of $0.304 \pm 0.167$, the Diffusion model $0.034 \pm 0.038$, and the B\'ezierGAN $0.014 \pm 0.027$. Notably, the Diffusion models exhibited mode collapse, generating highly similar designs across samples, which impacts RF (see \cref{app:additional_info_expes}). These results support the intuition that domain-informed generative models improve performance.
Simulation failures were typically caused by issues during meshing. Such failures were often due to violations of geometric continuity in the point-based spline representation (see \cref{fig:airfoil_main_paper_examples}).


\section{Use Cases, Limitations, and Future Work of \engibench{}}

\label{sec:limitations}

While our experiments highlight only a few possibilities, \engibench{} supports a broader range of research workflows. For the ML community, these contributions can serve as new testbeds for assessing ML model performance on problems that differ from traditional image- or text-based datasets since our problems come from engineering applications, are backed by real-world physics, and are constrained. For the engineering design community, it enables cross-domain algorithm evaluation via a simple \texttt{problem\_id} switch. Conversely, engineers can contribute new problems conforming to the API and immediately benefit from existing algorithm implementations in \engiopt{}. Several datasets include not only input–output pairs but also full field data (\eg, flow fields), making them suitable for PINNs and neural operators. The framework also supports multi-resolution or multi-fidelity studies (\eg, we provide multiple Beams2D datasets with different resolutions), enabling training on cheap data and generalization to higher-fidelity scenarios. Additionally, it facilitates transfer learning across problems and the development of a ``foundation design model.'' Finally, the API can support latent-space optimization via autoencoders and integration with reinforcement learning to guide optimization or data generation.

Despite its breadth, \engibench{} does not yet support unstructured meshes or grammar-based representations. Additionally, our benchmarks focus on static simulations, whereas real-world components are often dynamic\textemdash morphing or changing position during operation. \revision{Moreover, it may be beneficial to make problem configurations more flexible, such as varying the number of heat sources or sinks in the heat conduction problem. We have begun addressing this in the Beams2D problem by providing datasets at multiple image resolutions.} Currently, \engibench{} does not include multi-part or assembly-level design tasks, such as the joint optimization of interdependent components within a mechanical system. \revision{Finally, simulator assumptions can introduce biases that propagate to learned models, and quantifying these biases without physical experiments remains challenging\textemdash a common limitation in simulation-based ML. However, \engibench{}'s diversity of problems and simulators helps reveal such issues: models exploiting simulator-specific artifacts tend to regress to the mean in cross-task evaluations, exposing overfitting. Creating multiple versions of problems with different simulators and fidelities could further enable systematic bias assessment.} Addressing these gaps will guide future developments. We do not anticipate any negative societal impacts of this work.


\section{Conclusion}

We introduced \engibench{} and \engiopt{}, two modular, open-source libraries for reproducible research in engineering design. \engibench{} provides diverse physics-driven benchmarks and datasets under a unified interface, while \engiopt{} offers compatible implementations of ML algorithms including GANs, diffusion models, and surrogate-based optimization. Through experiments, we demonstrate how these tools enable rigorous comparisons of algorithms across domains for different performance metrics and introduce new challenges for ML models when applied to constrained, highly-sensitive, real-world design problems.

\section*{Acknowledgments and disclosure of funding}

We would like to thank the open-source community, notably the developers of NumPy~\citep{harris_array_2020}, PyTorch~\citep{paszke_pytorch_2019}, HuggingFace Datasets and Diffusers~\cite{Lhoest_Datasets_A_Community_2021,von_Platen_Diffusers_State-of-the-art_diffusion}, Weights and Biases~\citep{biewald_wandb_2020}, BoTorch~\citep{balandat2020_botorch}, Pymoo~\citep{blank_pymoo_2020}, Gymnasium~\citep{towers_gymnasium_2024}, MachAero~\citep{Wu_pyoptsparse_2020,PyHyp,Mader_adflow_2020a}, Dolfin/FEniCS ~\citep{mitusch2019dolfin,logg_DOLFIN_2010,logg_automated_2012}, Slurm~\citep{yoo_slurm_2003}, Docker~\citep{merkel_docker_2014}, Singularity/Apptainer~\citep{kurtzer_hpcngsingularity_2021}. We also acknowledge useful discussions with both Dr. Jun Wang and Dr. Quiyi Chen who helped refine core concepts that led to \href{https://ideal.umd.edu/midbench/}{MIDBench}, an earlier precursor of \engibench{}. The individual problem implementations in \cref{tab:problems} as well as baseline models in \engiopt{} were developed over 5-6 years at UMD and were funded in part by the following grants: DARPA-16-63-YFA-FP-059, NSF CAREER 1943699, ARPA-E DE-AR0001216, ARPA-E DE-AR0001200, \& ARL CA W911NF2320040.

{
\small
\bibliographystyle{plainnat}
\bibliography{ref}

\begin{thebibliography}{76}
\providecommand{\natexlab}[1]{#1}
\providecommand{\url}[1]{\texttt{#1}}
\expandafter\ifx\csname urlstyle\endcsname\relax
  \providecommand{\doi}[1]{doi: #1}\else
  \providecommand{\doi}{doi: \begingroup \urlstyle{rm}\Url}\fi

\bibitem[ngs(2025)]{ngspice}
Ngspice: A mixed-level/mixed-signal circuit simulator.
\newblock \url{http://ngspice.sourceforge.net}, 2025.
\newblock Accessed: 2025-03-31.

\bibitem[Andreassen et~al.(2011)Andreassen, Clausen, Schevenels, Lazarov, and Sigmund]{andreassen_88lines_2011}
Erik Andreassen, Anders Clausen, Mattias Schevenels, Boyan~S. Lazarov, and Ole Sigmund.
\newblock Efficient topology optimization in {MATLAB} using 88 lines of code.
\newblock \emph{Structural and Multidisciplinary Optimization}, 43\penalty0 (1):\penalty0 1--16, January 2011.
\newblock ISSN 1615-1488.
\newblock \doi{10.1007/s00158-010-0594-7}.
\newblock URL \url{https://doi.org/10.1007/s00158-010-0594-7}.

\bibitem[Balandat et~al.(2020)Balandat, Karrer, Jiang, Daulton, Letham, Wilson, and Bakshy]{balandat2020_botorch}
Maximilian Balandat, Brian Karrer, Daniel Jiang, Samuel Daulton, Ben Letham, Andrew~G Wilson, and Eytan Bakshy.
\newblock Botorch: A framework for efficient monte-carlo bayesian optimization.
\newblock In H.~Larochelle, M.~Ranzato, R.~Hadsell, M.F. Balcan, and H.~Lin, editors, \emph{Advances in Neural Information Processing Systems}, volume~33, 2020.

\bibitem[Behzadi and Ilie{\c{s}}(2022)]{behzadi2022gantl}
Mohammad~Mahdi Behzadi and Horea~T Ilie{\c{s}}.
\newblock Gantl: Toward practical and real-time topology optimization with conditional generative adversarial networks and transfer learning.
\newblock \emph{Journal of Mechanical Design}, 144\penalty0 (2):\penalty0 021711, 2022.

\bibitem[Bends{\o}e(1989)]{bendsoe1989optimal}
Martin~P Bends{\o}e.
\newblock Optimal shape design as a material distribution problem.
\newblock \emph{Structural optimization}, 1\penalty0 (4):\penalty0 193--202, 1989.

\bibitem[Bendsoe and Sigmund(2013)]{bendsoe2013topology}
Martin~Philip Bendsoe and Ole Sigmund.
\newblock \emph{Topology optimization: theory, methods, and applications}.
\newblock Springer Science \& Business Media, 2013.

\bibitem[Biewald(2020)]{biewald_wandb_2020}
Lukas Biewald.
\newblock Experiment tracking with weights and biases, 2020.
\newblock URL \url{https://www.wandb.com/}.
\newblock Software available from wandb.com.

\bibitem[Blank and Deb(2020)]{blank_pymoo_2020}
J.~Blank and K.~Deb.
\newblock pymoo: {Multi}-{Objective} {Optimization} in {Python}.
\newblock \emph{IEEE Access}, 8:\penalty0 89497--89509, 2020.

\bibitem[Chen et~al.(2022)Chen, Wang, Pope, Chen, and Fuge]{chen2022inverse}
Qiuyi Chen, Jun Wang, Phillip Pope, Wei Chen, and Mark Fuge.
\newblock Inverse design of two-dimensional airfoils using conditional generative models and surrogate log-likelihoods.
\newblock \emph{Journal of Mechanical Design}, 144\penalty0 (2):\penalty0 021712, 2022.

\bibitem[Chen et~al.(2019)Chen, Chiu, and Fuge]{chen_aerodynamic_GAN_2019}
Wei Chen, Kevin Chiu, and Mark Fuge.
\newblock \emph{Aerodynamic Design Optimization and Shape Exploration using Generative Adversarial Networks}.
\newblock 2019.
\newblock \doi{10.2514/6.2019-2351}.
\newblock URL \url{https://arc.aiaa.org/doi/abs/10.2514/6.2019-2351}.

\bibitem[Chen et~al.(2020)Chen, Chiu, and Fuge]{chen_airfoil_2020}
Wei Chen, Kevin Chiu, and Mark~D. Fuge.
\newblock Airfoil {Design} {Parameterization} and {Optimization} {Using} {Bézier} {Generative} {Adversarial} {Networks}.
\newblock \emph{AIAA Journal}, 58\penalty0 (11):\penalty0 4723--4735, 2020.
\newblock ISSN 0001-1452.
\newblock \doi{10.2514/1.J059317}.
\newblock URL \url{https://doi.org/10.2514/1.J059317}.
\newblock Publisher: American Institute of Aeronautics and Astronautics \_eprint: https://doi.org/10.2514/1.J059317.

\bibitem[Deb et~al.(2002)Deb, Pratap, Agarwal, and Meyarivan]{deb_fast_2002}
K.~Deb, A.~Pratap, S.~Agarwal, and T.~Meyarivan.
\newblock A fast and elitist multiobjective genetic algorithm: {NSGA}-{II}.
\newblock \emph{IEEE Transactions on Evolutionary Computation}, 6\penalty0 (2):\penalty0 182--197, April 2002.
\newblock ISSN 1941-0026.
\newblock \doi{10.1109/4235.996017}.

\bibitem[Diniz and Fuge(2024)]{dinizdiffusion}
Cashen Diniz and Mark Fuge.
\newblock Optimizing diffusion to diffuse optimal designs.
\newblock In \emph{AIAA SCITECH 2024 Forum}. American Institute of Aeronautics and Astronautics, 2024.
\newblock \doi{10.2514/6.2024-2013}.
\newblock URL \url{https://arc.aiaa.org/doi/abs/10.2514/6.2024-2013}.

\bibitem[Eimer et~al.(2023)Eimer, Lindauer, and Raileanu]{eimer_hyperparameters_2023}
Theresa Eimer, Marius Lindauer, and Roberta Raileanu.
\newblock Hyperparameters in {Reinforcement} {Learning} and {How} {To} {Tune} {Them}.
\newblock In \emph{Proceedings of the 40th {International} {Conference} on {Machine} {Learning} ({ICML} 2023)}, June 2023.
\newblock URL \url{https://openreview.net/forum?id=0Vm8Ghcxmp}.

\bibitem[Elrefaie et~al.(2024)Elrefaie, Morar, Dai, and Ahmed]{NEURIPS2024_013cf29a}
Mohamed Elrefaie, Florin Morar, Angela Dai, and Faez Ahmed.
\newblock Drivaernet++: A large-scale multimodal car dataset with computational fluid dynamics simulations and deep learning benchmarks.
\newblock In \emph{Advances in Neural Information Processing Systems}, volume~37, pages 499--536, 2024.

\bibitem[Fawaz et~al.(2022)Fawaz, Hua, Le~Corre, Fan, and Luo]{fawaz2022topology}
Ahmad Fawaz, Yuchao Hua, Steven Le~Corre, Yilin Fan, and Lingai Luo.
\newblock Topology optimization of heat exchangers: A review.
\newblock \emph{Energy}, 252:\penalty0 124053, 2022.

\bibitem[Felten et~al.(2023{\natexlab{a}})Felten, Alegre, Nowe, Bazzan, Talbi, Danoy, and Silva]{felten_toolkit_2023}
Florian Felten, Lucas~Nunes Alegre, Ann Nowe, Ana L.~C. Bazzan, El~Ghazali Talbi, Grégoire Danoy, and Bruno Castro~da Silva.
\newblock A {Toolkit} for {Reliable} {Benchmarking} and {Research} in {Multi}-{Objective} {Reinforcement} {Learning}.
\newblock In \emph{Proceedings of the 37th {Conference} on {Neural} {Information} {Processing} {Systems} ({NeurIPS})}, 2023{\natexlab{a}}.

\bibitem[Felten et~al.(2023{\natexlab{b}})Felten, Gareev, Talbi, and Danoy]{felten_hyperparameter_2023}
Florian Felten, Daniel Gareev, El-Ghazali Talbi, and Grégoire Danoy.
\newblock Hyperparameter {Optimization} for {Multi}-{Objective} {Reinforcement} {Learning}, October 2023{\natexlab{b}}.
\newblock URL \url{http://arxiv.org/abs/2310.16487}.
\newblock arXiv:2310.16487 [cs].

\bibitem[Felten et~al.(2024)Felten, Ucak, Azmani, Peng, Röpke, Baier, Mannion, Roijers, Terry, Talbi, Danoy, Nowé, and Rădulescu]{felten_momaland_2024}
Florian Felten, Umut Ucak, Hicham Azmani, Gao Peng, Willem Röpke, Hendrik Baier, Patrick Mannion, Diederik~M. Roijers, Jordan~K. Terry, El-Ghazali Talbi, Grégoire Danoy, Ann Nowé, and Roxana Rădulescu.
\newblock {MOMAland}: {A} {Set} of {Benchmarks} for {Multi}-{Objective} {Multi}-{Agent} {Reinforcement} {Learning}, July 2024.
\newblock URL \url{http://arxiv.org/abs/2407.16312}.
\newblock arXiv:2407.16312 [cs].

\bibitem[Ganaie et~al.(2022)Ganaie, Hu, Malik, Tanveer, and Suganthan]{ganaie_ensemble_2022}
M.~A. Ganaie, Minghui Hu, A.~K. Malik, M.~Tanveer, and P.~N. Suganthan.
\newblock Ensemble deep learning: {A} review.
\newblock \emph{Engineering Applications of Artificial Intelligence}, 115:\penalty0 105151, October 2022.
\newblock ISSN 0952-1976.
\newblock \doi{10.1016/j.engappai.2022.105151}.

\bibitem[Giraldo-Londo{\~n}o et~al.(2020)Giraldo-Londo{\~n}o, Mirabella, Dalloro, and Paulino]{giraldo2020multi}
Oliver Giraldo-Londo{\~n}o, Lucia Mirabella, Livio Dalloro, and Glaucio~H Paulino.
\newblock Multi-material thermomechanical topology optimization with applications to additive manufacturing: Design of main composite part and its support structure.
\newblock \emph{Computer Methods in Applied Mechanics and Engineering}, 363:\penalty0 112812, 2020.

\bibitem[Goodfellow et~al.(2014)Goodfellow, Pouget-Abadie, Mirza, Xu, Warde-Farley, Ozair, Courville, and Bengio]{goodfellow_generative_2014}
Ian~J. Goodfellow, Jean Pouget-Abadie, Mehdi Mirza, Bing Xu, David Warde-Farley, Sherjil Ozair, Aaron Courville, and Yoshua Bengio.
\newblock Generative adversarial nets.
\newblock \emph{Advances in neural information processing systems}, 27, 2014.
\newblock URL \url{https://proceedings.neurips.cc/paper_files/paper/2014/hash/f033ed80deb0234979a61f95710dbe25-Abstract.html}.

\bibitem[Gretton et~al.(2012)Gretton, Borgwardt, Rasch, Sch{\"o}lkopf, and Smola]{gretton_kernel_2012}
Arthur Gretton, Karsten~M Borgwardt, Malte~J Rasch, Bernhard Sch{\"o}lkopf, and Alexander Smola.
\newblock A kernel two-sample test.
\newblock \emph{The Journal of Machine Learning Research}, 13\penalty0 (1):\penalty0 723--773, 2012.

\bibitem[Habibi et~al.(2023)Habibi, Wang, and Fuge]{habibi2023actually}
Milad Habibi, Jun Wang, and Mark Fuge.
\newblock When is it actually worth learning inverse design?
\newblock In \emph{International Design Engineering Technical Conferences and Computers and Information in Engineering Conference}, volume 87301, page V03AT03A025. American Society of Mechanical Engineers, 2023.

\bibitem[Habibi et~al.(2025)Habibi, Bernard, Wang, and Fuge]{habibi2025mean}
Milad Habibi, Shai Bernard, Jun Wang, and Mark Fuge.
\newblock Mean squared error may lead you astray when optimizing your inverse design methods.
\newblock \emph{Journal of Mechanical Design}, 147\penalty0 (2):\penalty0 021701, 2025.

\bibitem[Hajdik et~al.(2023)Hajdik, Yildirim, Wu, Brelje, Seraj, Mangano, Anibal, Jonsson, Adler, Mader, Kenway, and Martins]{PyGeo}
Hannah~M. Hajdik, Anil Yildirim, Neil Wu, Benjamin~J. Brelje, Sabet Seraj, Marco Mangano, Joshua~L. Anibal, Eirikur Jonsson, Eytan~J. Adler, Charles~A. Mader, Gaetan K.~W. Kenway, and Joaquim R. R.~A. Martins.
\newblock {pyGeo}: A geometry package for multidisciplinary design optimization.
\newblock \emph{Journal of Open Source Software}, 8\penalty0 (87):\penalty0 5319, 2023.
\newblock \doi{10.21105/joss.05319}.

\bibitem[Harris et~al.(2020)Harris, Millman, Walt, Gommers, Virtanen, Cournapeau, Wieser, Taylor, Berg, Smith, Kern, Picus, Hoyer, Kerkwijk, Brett, Haldane, Río, Wiebe, Peterson, Gérard-Marchant, Sheppard, Reddy, Weckesser, Abbasi, Gohlke, and Oliphant]{harris_array_2020}
Charles~R. Harris, K.~Jarrod Millman, Stéfan J. van~der Walt, Ralf Gommers, Pauli Virtanen, David Cournapeau, Eric Wieser, Julian Taylor, Sebastian Berg, Nathaniel~J. Smith, Robert Kern, Matti Picus, Stephan Hoyer, Marten H.~van Kerkwijk, Matthew Brett, Allan Haldane, Jaime Fernández~del Río, Mark Wiebe, Pearu Peterson, Pierre Gérard-Marchant, Kevin Sheppard, Tyler Reddy, Warren Weckesser, Hameer Abbasi, Christoph Gohlke, and Travis~E. Oliphant.
\newblock Array programming with {NumPy}.
\newblock \emph{Nature}, 585\penalty0 (7825):\penalty0 357--362, September 2020.
\newblock \doi{10.1038/s41586-020-2649-2}.
\newblock URL \url{https://doi.org/10.1038/s41586-020-2649-2}.
\newblock Publisher: Springer Science and Business Media LLC.

\bibitem[Hassan et~al.(2023)Hassan, Feeney, Dhruv, Kim, Suh, Ryu, Won, and Chandramowlishwaran]{hassan2023bubbleml}
Sheikh Md~Shakeel Hassan, Arthur Feeney, Akash Dhruv, Jihoon Kim, Youngjoon Suh, Jaiyoung Ryu, Yoonjin Won, and Aparna Chandramowlishwaran.
\newblock Bubbleml: A multi-physics dataset and benchmarks for machine learning.
\newblock \emph{arXiv preprint arXiv:2307.14623}, 2023.

\bibitem[He et~al.(2019)He, Li, Mader, Yildirim, and Martins]{He2019c}
Xiaolong He, Jichao Li, Charles~A. Mader, Anil Yildirim, and Joaquim R. R.~A. Martins.
\newblock Robust aerodynamic shape optimization---from a circle to an airfoil.
\newblock \emph{Aerospace Science and Technology}, 87:\penalty0 48--61, April 2019.
\newblock \doi{10.1016/j.ast.2019.01.051}.

\bibitem[Ho et~al.(2020)Ho, Jain, and Abbeel]{ho_denoising_2020}
Jonathan Ho, Ajay Jain, and Pieter Abbeel.
\newblock Denoising {Diffusion} {Probabilistic} {Models}.
\newblock In \emph{Advances in {Neural} {Information} {Processing} {Systems}}, volume~33, pages 6840--6851. Curran Associates, Inc., 2020.
\newblock URL \url{https://proceedings.neurips.cc/paper/2020/hash/4c5bcfec8584af0d967f1ab10179ca4b-Abstract.html}.

\bibitem[Huang et~al.(2022)Huang, Dossa, Ye, Braga, Chakraborty, Mehta, and Araújo]{huang_cleanrl_2022}
Shengyi Huang, Rousslan Fernand~Julien Dossa, Chang Ye, Jeff Braga, Dipam Chakraborty, Kinal Mehta, and João G.~M. Araújo.
\newblock {CleanRL}: {High}-quality {Single}-file {Implementations} of {Deep} {Reinforcement} {Learning} {Algorithms}.
\newblock \emph{Journal of Machine Learning Research}, 23\penalty0 (274):\penalty0 1--18, 2022.
\newblock ISSN 1533-7928.
\newblock URL \url{http://jmlr.org/papers/v23/21-1342.html}.

\bibitem[Hughes et~al.(2018)Hughes, Minkov, Williamson, and Fan]{hughes2018adjoint}
Tyler~W Hughes, Momchil Minkov, Ian~AD Williamson, and Shanhui Fan.
\newblock Adjoint method and inverse design for nonlinear nanophotonic devices.
\newblock \emph{ACS Photonics}, 5\penalty0 (12):\penalty0 4781--4787, 2018.

\bibitem[Hughes et~al.(2019)Hughes, Williamson, Minkov, and Fan]{hughes2019forward}
Tyler~W Hughes, Ian~AD Williamson, Momchil Minkov, and Shanhui Fan.
\newblock Forward-mode differentiation of maxwell’s equations.
\newblock \emph{ACS Photonics}, 6\penalty0 (11):\penalty0 3010--3016, 2019.

\bibitem[Kenway et~al.(2019)Kenway, Mader, He, and Martins]{ADFLOWADJOINT}
Gaetan K.~W. Kenway, Charles~A. Mader, Ping He, and Joaquim R. R.~A. Martins.
\newblock Effective adjoint approaches for computational fluid dynamics.
\newblock \emph{Progress in Aerospace Sciences}, 110:\penalty0 100542, October 2019.
\newblock \doi{10.1016/j.paerosci.2019.05.002}.

\bibitem[Khan et~al.(2023)Khan, Goucher-Lambert, Kostas, and Kaklis]{khan2023shiphullgan}
Shahroz Khan, Kosa Goucher-Lambert, Konstantinos Kostas, and Panagiotis Kaklis.
\newblock Shiphullgan: A generic parametric modeller for ship hull design using deep convolutional generative model.
\newblock \emph{Computer Methods in Applied Mechanics and Engineering}, 411:\penalty0 116051, 2023.

\bibitem[Krizhevsky(2009)]{Krizhevsky09learningmultiple}
Alex Krizhevsky.
\newblock Learning multiple layers of features from tiny images.
\newblock Technical report, 2009.

\bibitem[Kulesza and Taskar(2012)]{kulesza_dpp_2012}
Alex Kulesza and Ben Taskar.
\newblock Determinantal point processes for machine learning.
\newblock \emph{Foundations and Trends® in Machine Learning}, 5\penalty0 (2–3):\penalty0 123--286, 2012.
\newblock ISSN 1935-8237.
\newblock \doi{10.1561/2200000044}.
\newblock URL \url{http://dx.doi.org/10.1561/2200000044}.

\bibitem[Kurtzer et~al.(2021)Kurtzer, {cclerget}, Bauer, Kaneshiro, Trudgian, and Godlove]{kurtzer_hpcngsingularity_2021}
Gregory~M. Kurtzer, {cclerget}, Michael Bauer, Ian Kaneshiro, David Trudgian, and David Godlove.
\newblock hpcng/singularity: {Singularity} 3.7.3, April 2021.
\newblock URL \url{https://doi.org/10.5281/zenodo.4667718}.

\bibitem[LeCun et~al.(2010)LeCun, Cortes, and Burges]{lecun2010mnist}
Yann LeCun, Corinna Cortes, and CJ~Burges.
\newblock Mnist handwritten digit database.
\newblock \emph{ATT Labs [Online]. Available: http://yann.lecun.com/exdb/mnist}, 2, 2010.

\bibitem[Lhoest et~al.(2021)Lhoest, Villanova~del Moral, von Platen, Wolf, Šaško, Jernite, Thakur, Tunstall, Patil, Drame, Chaumond, Plu, Davison, Brandeis, Sanh, Le~Scao, Canwen~Xu, Patry, Liu, McMillan-Major, Schmid, Gugger, Raw, Lesage, Lozhkov, Carrigan, Matussière, von Werra, Debut, Bekman, and Delangue]{Lhoest_Datasets_A_Community_2021}
Quentin Lhoest, Albert Villanova~del Moral, Patrick von Platen, Thomas Wolf, Mario Šaško, Yacine Jernite, Abhishek Thakur, Lewis Tunstall, Suraj Patil, Mariama Drame, Julien Chaumond, Julien Plu, Joe Davison, Simon Brandeis, Victor Sanh, Teven Le~Scao, Kevin Canwen~Xu, Nicolas Patry, Steven Liu, Angelina McMillan-Major, Philipp Schmid, Sylvain Gugger, Nathan Raw, Sylvain Lesage, Anton Lozhkov, Matthew Carrigan, Théo Matussière, Leandro von Werra, Lysandre Debut, Stas Bekman, and Clément Delangue.
\newblock {Datasets: A Community Library for Natural Language Processing}.
\newblock In \emph{Proceedings of the 2021 Conference on Empirical Methods in Natural Language Processing: System Demonstrations}, pages 175--184. Association for Computational Linguistics, November 2021.
\newblock URL \url{https://aclanthology.org/2021.emnlp-demo.21}.

\bibitem[Logg and Wells(2010)]{logg_DOLFIN_2010}
Anders Logg and Garth~N. Wells.
\newblock Dolfin: Automated finite element computing.
\newblock \emph{ACM Trans. Math. Softw.}, 37\penalty0 (2), April 2010.
\newblock ISSN 0098-3500.
\newblock \doi{10.1145/1731022.1731030}.
\newblock URL \url{https://doi.org/10.1145/1731022.1731030}.

\bibitem[Logg et~al.(2012)Logg, Mardal, and Wells]{logg_automated_2012}
Anders Logg, Kent-Andre Mardal, and Garth Wells, editors.
\newblock \emph{Automated {Solution} of {Differential} {Equations} by the {Finite} {Element} {Method}: {The} {FEniCS} {Book}}, volume~84 of \emph{Lecture {Notes} in {Computational} {Science} and {Engineering}}.
\newblock Springer, Berlin, Heidelberg, 2012.
\newblock ISBN 978-3-642-23098-1 978-3-642-23099-8.
\newblock \doi{10.1007/978-3-642-23099-8}.
\newblock URL \url{https://link.springer.com/10.1007/978-3-642-23099-8}.

\bibitem[Mader et~al.(2020)Mader, Kenway, Yildirim, and Martins]{Mader_adflow_2020a}
Charles~A. Mader, Gaetan K.~W. Kenway, Anil Yildirim, and Joaquim R. R.~A. Martins.
\newblock {ADflow}---an open-source computational fluid dynamics solver for aerodynamic and multidisciplinary optimization.
\newblock \emph{Journal of Aerospace Information Systems}, 2020.
\newblock \doi{10.2514/1.I010796}.

\bibitem[Martins(2022)]{martinsreview}
Joaquim Martins.
\newblock Aerodynamic design optimization: Challenges and perspectives.
\newblock \emph{Computers \& Fluids}, 239:\penalty0 105391, 03 2022.
\newblock \doi{10.1016/j.compfluid.2022.105391}.

\bibitem[Martins and Ning(2022)]{mdobook}
Joaquim R. R.~A. Martins and Andrew Ning.
\newblock \emph{Engineering Design Optimization}.
\newblock Cambridge University Press, Cambridge, UK, January 2022.
\newblock ISBN 9781108833417.
\newblock \doi{10.1017/9781108980647}.
\newblock URL \url{https://mdobook.github.io}.

\bibitem[Massoudi and Schiffmann(2025)]{massoudi_dimensionless_2025}
Soheyl Massoudi and Jürg Schiffmann.
\newblock Dimensionless group-driven ensemble neural networks for robust design optimization in engineering.
\newblock \emph{Journal of Computational Design and Engineering}, 12\penalty0 (7):\penalty0 61--95, July 2025.
\newblock ISSN 2288-5048.
\newblock \doi{10.1093/jcde/qwaf056}.

\bibitem[Massoudi et~al.(2024)Massoudi, Picard, and Schiffmann]{massoudi_integrated_2024}
Soheyl Massoudi, Cyril Picard, and Jürg Schiffmann.
\newblock An {Integrated} {Approach} to {Designing} {Robust} {Gas}-{Bearing} {Supported} {Turbocompressors} {Through} {Surrogate} {Modeling} and {Constrained} {All}-{At}-{Once} {Multi}-{Objective} {Optimization}.
\newblock \emph{Journal of Mechanical Design}, 146\penalty0 (121706), July 2024.
\newblock ISSN 1050-0472.
\newblock \doi{10.1115/1.4065823}.

\bibitem[Massoudi et~al.(2025)Massoudi, Bush, and Schiffmann]{massoudi_robust_2025}
Soheyl Massoudi, Cameron Bush, and Jürg Schiffmann.
\newblock Robust design optimization of gas-lubricated herringbone grooved journal bearings: {Surrogate} modeling and experimental validation.
\newblock \emph{Tribology International}, 204:\penalty0 110429, April 2025.
\newblock ISSN 0301-679X.
\newblock \doi{10.1016/j.triboint.2024.110429}.

\bibitem[Meng et~al.(2020)Meng, Zhang, Quan, Shi, Tang, Hou, Breitkopf, Zhu, and Gao]{meng2020topology}
Liang Meng, Weihong Zhang, Dongliang Quan, Guanghui Shi, Lei Tang, Yuliang Hou, Piotr Breitkopf, Jihong Zhu, and Tong Gao.
\newblock From topology optimization design to additive manufacturing: Today’s success and tomorrow’s roadmap.
\newblock \emph{Archives of Computational Methods in Engineering}, 27\penalty0 (3):\penalty0 805--830, 2020.

\bibitem[Merkel(2014)]{merkel_docker_2014}
Dirk Merkel.
\newblock Docker: lightweight linux containers for consistent development and deployment.
\newblock \emph{Linux J.}, 2014\penalty0 (239), March 2014.
\newblock ISSN 1075-3583.

\bibitem[Mirza and Osindero(2014)]{mirza_conditional_2014}
Mehdi Mirza and Simon Osindero.
\newblock Conditional {Generative} {Adversarial} {Nets}, November 2014.
\newblock URL \url{http://arxiv.org/abs/1411.1784}.
\newblock arXiv:1411.1784 [cs].

\bibitem[Mitusch et~al.(2019)Mitusch, Funke, and Dokken]{mitusch2019dolfin}
Sebastian Mitusch, Simon Funke, and J{\o}rgen Dokken.
\newblock dolfin-adjoint 2018.1: automated adjoints for fenics and firedrake.
\newblock \emph{Journal of Open Source Software}, 4\penalty0 (38):\penalty0 1292, 2019.

\bibitem[Mo et~al.(2021)Mo, Zhi, Xiao, Hua, and He]{mo2021topology}
Xiaobao Mo, Hui Zhi, Yizhi Xiao, Haiyu Hua, and Liang He.
\newblock Topology optimization of cooling plates for battery thermal management.
\newblock \emph{International Journal of Heat and Mass Transfer}, 178:\penalty0 121612, 2021.

\bibitem[Ohana et~al.(2024)Ohana, McCabe, Meyer, Morel, Agocs, Beneitez, Berger, Burkhart, Dalziel, Fielding, et~al.]{ohana2024well}
Ruben Ohana, Michael McCabe, Lucas Meyer, Rudy Morel, Fruzsina Agocs, Miguel Beneitez, Marsha Berger, Blakesly Burkhart, Stuart Dalziel, Drummond Fielding, et~al.
\newblock The well: a large-scale collection of diverse physics simulations for machine learning.
\newblock \emph{Advances in Neural Information Processing Systems}, 37:\penalty0 44989--45037, 2024.

\bibitem[Parker-Holder et~al.(2022)Parker-Holder, Rajan, Song, Biedenkapp, Miao, Eimer, Zhang, Nguyen, Calandra, Faust, Hutter, and Lindauer]{parker-holder_automated_2022}
Jack Parker-Holder, Raghu Rajan, Xingyou Song, André Biedenkapp, Yingjie Miao, Theresa Eimer, Baohe Zhang, Vu~Nguyen, Roberto Calandra, Aleksandra Faust, Frank Hutter, and Marius Lindauer.
\newblock Automated {Reinforcement} {Learning} ({AutoRL}): {A} {Survey} and {Open} {Problems}.
\newblock \emph{Journal of Artificial Intelligence Research}, 74, September 2022.
\newblock ISSN 1076-9757.
\newblock \doi{10.1613/jair.1.13596}.
\newblock URL \url{https://dl.acm.org/doi/10.1613/jair.1.13596}.

\bibitem[Paszke et~al.(2019)Paszke, Gross, Massa, Lerer, Bradbury, Chanan, Killeen, Lin, Gimelshein, Antiga, Desmaison, Kopf, Yang, DeVito, Raison, Tejani, Chilamkurthy, Steiner, Fang, Bai, and Chintala]{paszke_pytorch_2019}
Adam Paszke, Sam Gross, Francisco Massa, Adam Lerer, James Bradbury, Gregory Chanan, Trevor Killeen, Zeming Lin, Natalia Gimelshein, Luca Antiga, Alban Desmaison, Andreas Kopf, Edward Yang, Zachary DeVito, Martin Raison, Alykhan Tejani, Sasank Chilamkurthy, Benoit Steiner, Lu~Fang, Junjie Bai, and Soumith Chintala.
\newblock {PyTorch}: {An} {Imperative} {Style}, {High}-{Performance} {Deep} {Learning} {Library}.
\newblock In \emph{Advances in {Neural} {Information} {Processing} {Systems} 32}, pages 8024--8035. Curran Associates, Inc., 2019.
\newblock URL \url{http://papers.neurips.cc/paper/9015-pytorch-an-imperative-style-high-performance-deep-learning-library.pdf}.

\bibitem[Piggott et~al.(2015)Piggott, Lu, Lagoudakis, Petykiewicz, Babinec, and Vu{\v{c}}kovi{\'c}]{piggott2015inverse}
Alexander~Y Piggott, Jesse Lu, Konstantinos~G Lagoudakis, Jan Petykiewicz, Thomas~M Babinec, and Jelena Vu{\v{c}}kovi{\'c}.
\newblock Inverse design and demonstration of a compact and broadband on-chip wavelength demultiplexer.
\newblock \emph{Nature photonics}, 9\penalty0 (6):\penalty0 374--377, 2015.

\bibitem[Radford et~al.(2016)Radford, Metz, and Chintala]{radford_2016_dcgan}
Alec Radford, Luke Metz, and Soumith Chintala.
\newblock Unsupervised representation learning with deep convolutional generative adversarial networks.
\newblock In Yoshua Bengio and Yann LeCun, editors, \emph{4th International Conference on Learning Representations, {ICLR} 2016, San Juan, Puerto Rico, May 2-4, 2016, Conference Track Proceedings}, 2016.
\newblock URL \url{http://arxiv.org/abs/1511.06434}.

\bibitem[Regenwetter et~al.(2021)Regenwetter, Curry, and Ahmed]{regenwetter_biked_2021}
Lyle Regenwetter, Brent Curry, and Faez Ahmed.
\newblock {BIKED}: {A} {Dataset} for {Computational} {Bicycle} {Design} {With} {Machine} {Learning} {Benchmarks}.
\newblock \emph{Journal of Mechanical Design}, 144\penalty0 (031706), October 2021.
\newblock ISSN 1050-0472.
\newblock \doi{10.1115/1.4052585}.
\newblock URL \url{https://doi.org/10.1115/1.4052585}.

\bibitem[Regenwetter et~al.(2022)Regenwetter, Nobari, and Ahmed]{regenwetter_deep_2022}
Lyle Regenwetter, Amin~Heyrani Nobari, and Faez Ahmed.
\newblock Deep generative models in engineering design: {A} review.
\newblock \emph{Journal of Mechanical Design}, 144\penalty0 (7), 2022.

\bibitem[Regenwetter et~al.(2023)Regenwetter, Srivastava, Gutfreund, and Ahmed]{regenwetter_beyond_2023}
Lyle Regenwetter, Akash Srivastava, Dan Gutfreund, and Faez Ahmed.
\newblock Beyond {Statistical} {Similarity}: {Rethinking} {Metrics} for {Deep} {Generative} {Models} in {Engineering} {Design}.
\newblock \emph{Computer-Aided Design}, 165:\penalty0 103609, December 2023.
\newblock ISSN 0010-4485.
\newblock \doi{10.1016/j.cad.2023.103609}.
\newblock URL \url{https://www.sciencedirect.com/science/article/pii/S0010448523001410}.

\bibitem[Secco et~al.(2021)Secco, Kenway, He, Mader, and Martins]{PyHyp}
Ney Secco, Gaetan K.~W. Kenway, Ping He, Charles~A. Mader, and Joaquim R. R.~A. Martins.
\newblock Efficient mesh generation and deformation for aerodynamic shape optimization.
\newblock \emph{AIAA Journal}, 2021.
\newblock \doi{10.2514/1.J059491}.

\bibitem[Sigmund(2007)]{sigmund2007morphology}
Ole Sigmund.
\newblock Morphology-based black and white filters for topology optimization.
\newblock \emph{Structural and Multidisciplinary Optimization}, 33\penalty0 (4):\penalty0 401--424, 2007.

\bibitem[Snoek et~al.(2012)Snoek, Larochelle, and Adams]{snoek_practical_2012}
Jasper Snoek, Hugo Larochelle, and Ryan~P Adams.
\newblock Practical {Bayesian} {Optimization} of {Machine} {Learning} {Algorithms}.
\newblock In \emph{Advances in {Neural} {Information} {Processing} {Systems}}, volume~25. Curran Associates, Inc., 2012.

\bibitem[Sosnovik and Oseledets(2019)]{sosnovik_neural_2019}
Ivan Sosnovik and Ivan Oseledets.
\newblock Neural networks for topology optimization.
\newblock \emph{Russian Journal of Numerical Analysis and Mathematical Modelling}, 34\penalty0 (4):\penalty0 215--223, August 2019.
\newblock ISSN 1569-3988.
\newblock \doi{10.1515/rnam-2019-0018}.
\newblock URL \url{https://www.degruyter.com/document/doi/10.1515/rnam-2019-0018/html}.
\newblock Publisher: De Gruyter.

\bibitem[Takamoto et~al.(2022)Takamoto, Praditia, Leiteritz, MacKinlay, Alesiani, Pfl{\"u}ger, and Niepert]{takamoto2022pdebench}
Makoto Takamoto, Timothy Praditia, Raphael Leiteritz, Daniel MacKinlay, Francesco Alesiani, Dirk Pfl{\"u}ger, and Mathias Niepert.
\newblock Pdebench: An extensive benchmark for scientific machine learning.
\newblock \emph{Advances in Neural Information Processing Systems}, 35:\penalty0 1596--1611, 2022.

\bibitem[Tang et~al.(2019)Tang, Gao, Song, Meng, Zhang, and Zhang]{tang2019topology}
Lei Tang, Tong Gao, Longlong Song, Liang Meng, Chengqi Zhang, and Weihong Zhang.
\newblock Topology optimization of nonlinear heat conduction problems involving large temperature gradient.
\newblock \emph{Computer Methods in Applied Mechanics and Engineering}, 357:\penalty0 112600, 2019.

\bibitem[Toshev et~al.(2023)Toshev, Galletti, Fritz, Adami, and Adams]{toshev2023lagrangebench}
Artur Toshev, Gianluca Galletti, Fabian Fritz, Stefan Adami, and Nikolaus Adams.
\newblock Lagrangebench: A lagrangian fluid mechanics benchmarking suite.
\newblock \emph{Advances in Neural Information Processing Systems}, 36:\penalty0 64857--64884, 2023.

\bibitem[Towers et~al.(2024)Towers, Kwiatkowski, Terry, Balis, Cola, Deleu, Goulão, Kallinteris, Krimmel, KG, Perez-Vicente, Pierré, Schulhoff, Tai, Tan, and Younis]{towers_gymnasium_2024}
Mark Towers, Ariel Kwiatkowski, Jordan Terry, John~U. Balis, Gianluca~De Cola, Tristan Deleu, Manuel Goulão, Andreas Kallinteris, Markus Krimmel, Arjun KG, Rodrigo Perez-Vicente, Andrea Pierré, Sander Schulhoff, Jun~Jet Tai, Hannah Tan, and Omar~G. Younis.
\newblock Gymnasium: {A} {Standard} {Interface} for {Reinforcement} {Learning} {Environments}, November 2024.
\newblock URL \url{http://arxiv.org/abs/2407.17032}.
\newblock arXiv:2407.17032 [cs].

\bibitem[von Platen et~al.()von Platen, Patil, Lozhkov, Cuenca, Lambert, Rasul, Davaadorj, Nair, Paul, Liu, Berman, Xu, and Wolf]{von_Platen_Diffusers_State-of-the-art_diffusion}
Patrick von Platen, Suraj Patil, Anton Lozhkov, Pedro Cuenca, Nathan Lambert, Kashif Rasul, Mishig Davaadorj, Dhruv Nair, Sayak Paul, Steven Liu, William Berman, Yiyi Xu, and Thomas Wolf.
\newblock {Diffusers: State-of-the-art diffusion models}.
\newblock URL \url{https://github.com/huggingface/diffusers}.

\bibitem[W{\"a}chter and Biegler(2006)]{wachter2006implementation}
Andreas W{\"a}chter and Lorenz~T Biegler.
\newblock On the implementation of an interior-point filter line-search algorithm for large-scale nonlinear programming.
\newblock \emph{Mathematical programming}, 106:\penalty0 25--57, 2006.

\bibitem[Wu et~al.(2020{\natexlab{a}})Wu, Kenway, Mader, Jasa, and Martins]{Wu_pyoptsparse_2020}
Ella Wu, Gaetan Kenway, Charles~A. Mader, John Jasa, and Joaquim R. R.~A. Martins.
\newblock pyoptsparse: A python framework for large-scale constrained nonlinear optimization of sparse systems.
\newblock \emph{Journal of Open Source Software}, 5\penalty0 (54):\penalty0 2564, 2020{\natexlab{a}}.
\newblock \doi{10.21105/joss.02564}.

\bibitem[Wu et~al.(2020{\natexlab{b}})Wu, Kenway, Mader, Jasa, and Martins]{PyOptSparse}
Neil Wu, Gaetan Kenway, Charles~A. Mader, John Jasa, and Joaquim R. R.~A. Martins.
\newblock pyoptsparse: A python framework for large-scale constrained nonlinear optimization of sparse systems.
\newblock \emph{Journal of Open Source Software}, 5\penalty0 (54):\penalty0 2564, 2020{\natexlab{b}}.
\newblock \doi{10.21105/joss.02564}.

\bibitem[Xue et~al.(2021)Xue, Wang, Kong, Wang, Liu, Fan, Yuan, Niu, and Li]{xue_deep_2021}
Jie Xue, Zhuo Wang, Deting Kong, Yuan Wang, Xiyu Liu, Wen Fan, Songtao Yuan, Sijie Niu, and Dengwang Li.
\newblock Deep ensemble neural-like {P} systems for segmentation of central serous chorioretinopathy lesion.
\newblock \emph{Information Fusion}, 65:\penalty0 84--94, January 2021.
\newblock ISSN 1566-2535.
\newblock \doi{10.1016/j.inffus.2020.08.016}.

\bibitem[Yildirim et~al.(2019)Yildirim, Kenway, Mader, and Martins]{ANK}
Anil Yildirim, Gaetan~KW Kenway, Charles~A Mader, and Joaquim~RRA Martins.
\newblock A jacobian-free approximate newton--krylov startup strategy for rans simulations.
\newblock \emph{Journal of Computational Physics}, 397:\penalty0 108741, 2019.

\bibitem[Yoo et~al.(2003)Yoo, Jette, and Grondona]{yoo_slurm_2003}
Andy~B. Yoo, Morris~A. Jette, and Mark Grondona.
\newblock {SLURM}: {Simple} {Linux} {Utility} for {Resource} {Management}.
\newblock In Dror Feitelson, Larry Rudolph, and Uwe Schwiegelshohn, editors, \emph{Job {Scheduling} {Strategies} for {Parallel} {Processing}}, Lecture {Notes} in {Computer} {Science}, pages 44--60, Berlin, Heidelberg, 2003. Springer.
\newblock ISBN 978-3-540-39727-4.
\newblock \doi{10.1007/10968987_3}.

\end{thebibliography}
}

\newpage

\newpage
\appendix

\section{Useful Information}
\label{app:useful_info}
\label{app:reproducibility}

\subsection{Links}
The project's parts can be found at the following links:
\begin{itemize}
    \item Documentation website: \url{https://engibench.ethz.ch}. 
    \item \engibench{} code: \url{https://github.com/IDEALLab/EngiBench}. The code used for this paper was tagged with \texttt{v0.0.1}.
    \item \engiopt{} code: \url{https://github.com/IDEALLab/EngiOpt}. The code used for this paper was tagged with \texttt{v0.0.1}.
    \item Datasets: \url{https://huggingface.co/IDEALLab}.
    \item All hyperparameters, learning curves, Python version, OS version, hardware specifications, and run commands for all our experiments are available at \url{https://wandb.ai/engibench/engiopt}.
\end{itemize}

\subsection{Licenses}
Both our codebases are released under the GPL-3.0 license. All the datasets are released under the CC-BY-NC-SA license. 

\subsection{Maintenance And External Contributions}
The \href{https://ideal.ethz.ch/}{IDEAL Lab} and \href{https://sis.id.ethz.ch/}{ETHZ's Scientific IT Services (SIS)} are committed to the long-term maintenance of the project. 

We also hope the open-source community will contribute to its ongoing development and improvement. \revision{To support this, we provide detailed instructions for making contributions to \engibench{} on our website, \eg, \url{https://engibench.ethz.ch/tutorials/new_problem/}.}

\subsection{Experimental setup}
Our experiments were conducted on the Euler cluster from ETH Zurich's high-performance computer. The compute nodes equipped with GPUs contain NVidia GeForce RTX4090 (24GB) and AMD EPYC 9554 CPUs. We used Python 3.11.6, CUDA 12.5, and PyTorch 2.6. Each training job has been allocated one GPU and 4 cores using Slurm~\citep{yoo_slurm_2003}. All datasets have been generated on CPU nodes of the UMD's high-performance computing cluster Zaratan equipped with 128 AMD EPYC 7763 CPUs, except for photonics, which was generated on Euler.

Training all algorithms across all problems with multiple random seeds for our experiments required approximately 80 GPU-hours. Evaluations of the generated or optimized designs for our experiments took 55 hours.
Generating the datasets, however, was significantly more time-consuming and took place over the past few years, as we reused existing simulation data. We estimate that generating the full set of datasets required several thousand CPU-hours in total. A per-problem estimation of the dataset generation time is given below.

\begin{itemize}
    \item Airfoil: $\approx 5000$ hours.
    \item HeatConduction2D: $\approx 1$ hour.
    \item HeatConduction3D: $\approx300$ hours.
    \item ThermoElasticBeams2D: $\approx10$ hours.
    \item Beams2D: $\approx 12$ hours (4 hours per dataset).
    \item Photonics2D: $\approx 200$ hours. 
    \item PowerElectronics: $\approx 20$ hours.
\end{itemize}

\section{Features} 
In this section, we go through some of the features and design choices we have made for the library.

\subsection{Example usage}

\begin{code}
\begin{mintedbox}{python}
from engibench.problems.beams2d.v0 import Beams2D

# Create a problem
problem = Beams2D()
problem.reset(seed=42)

# Inspect problem
problem.design_space  # Box(0.0, 1.0, (50, 100), float64)
problem.objectives  # (("compliance", "MINIMIZE"),)
problem.conditions  # (("volfrac", 0.35), ("forcedist", 0.0),...)
problem.dataset # A HuggingFace Dataset object

# Train your inverse design model or surrogate model
conditions = problem.dataset["train"].select_columns(problem.conditions_keys)
designs = problem.dataset["train"].select_columns("optimal_design")
cond_designs_keys = problem.conditions_keys + ["optimal_design"]
cond_designs = problem.dataset["train"].select_columns(cond_designs_keys)
objs = problem.dataset["train"].select_columns(problem.objectives_keys)

# Train your models
inverse_model = train_inverse(inputs=conditions, outputs=designs)
surr_model = train_surrogate(inputs=cond_designs, outputs=objs)

# Use the model predictions, inverse design here
desired_conds = {"volfrac": 0.7, "forcedist": 0.3}
generated_design = inverse_model.predict(desired_conds)

violated_constraints = problem.check_constraints(generated_design, desired_conds)
if not violated_constraints:
    # Only simulate to get objective values
    objs = problem.simulate(design=generated_design, config=desired_conds)
    problem.reset(seed=41)
    # Or run a gradient-based optimizer to polish the generated design
    opt_design, history = problem.optimize(generated_design, desired_conds)
\end{mintedbox}
\captionof{listing}{API usage with automated column extraction for training.\label{listing:api_longer}}
\end{code}

\cref{listing:api_longer} presents a longer version of our API usage, showing how to automatically extract relevant columns from the datasets (lines 13\textemdash 18) to perform ID or surrogate-based optimization (lines 21\textemdash 26).

\subsection{\revision{Error handling}}
\label{app:error_handling}

\revision{\engibench{} handles errors at multiple levels while preserving flexibility for advanced users. The framework includes several mechanisms:}

\begin{itemize}
  \item \textbf{Constraint checking:} Each benchmark problem includes standardized validation for configuration and constraints (e.g., \cref{listing:api}, line~14). For instance, in the Beams2D task, setting \texttt{volfrac = 2.0} and checking for constraints returns:
  \begin{itemize}
    \item \texttt{Config.volfrac: 2.0 $\notin$ [0.0, 1.0] (Theory, error)} 
    \item \texttt{Config.volfrac: 2.0 $\notin$ [0.1, 0.9] (Implementation, warning)} 
  \end{itemize}
  These pre-simulation checks catch common issues early.
  
  \item \textbf{User control:} While errors are flagged, users can still choose to simulate invalid configurations\textemdash for instance, to explore failure regions or generate negative data.
  
  \item \textbf{Failure tracking:} If a solver crashes (\eg, due to meshing errors or NaNs), the exception is caught and surfaced, even in containerized runs.

  \item \textbf{Failure rates and reporting:} Simulation failure rates vary across tasks and methods. We often track these to compare robustness\textemdash for example, the failure ratio for generative airfoil models is reported in \cref{sec:zoomed_exp}.
\end{itemize}



\section{Extensive Description of Problems}
\label{app:extended_description}
This section describes the implemented problems in more details.

\subsection{Airfoil}

\begin{figure}[H]
    \centering
    \includegraphics[width=0.4\linewidth]{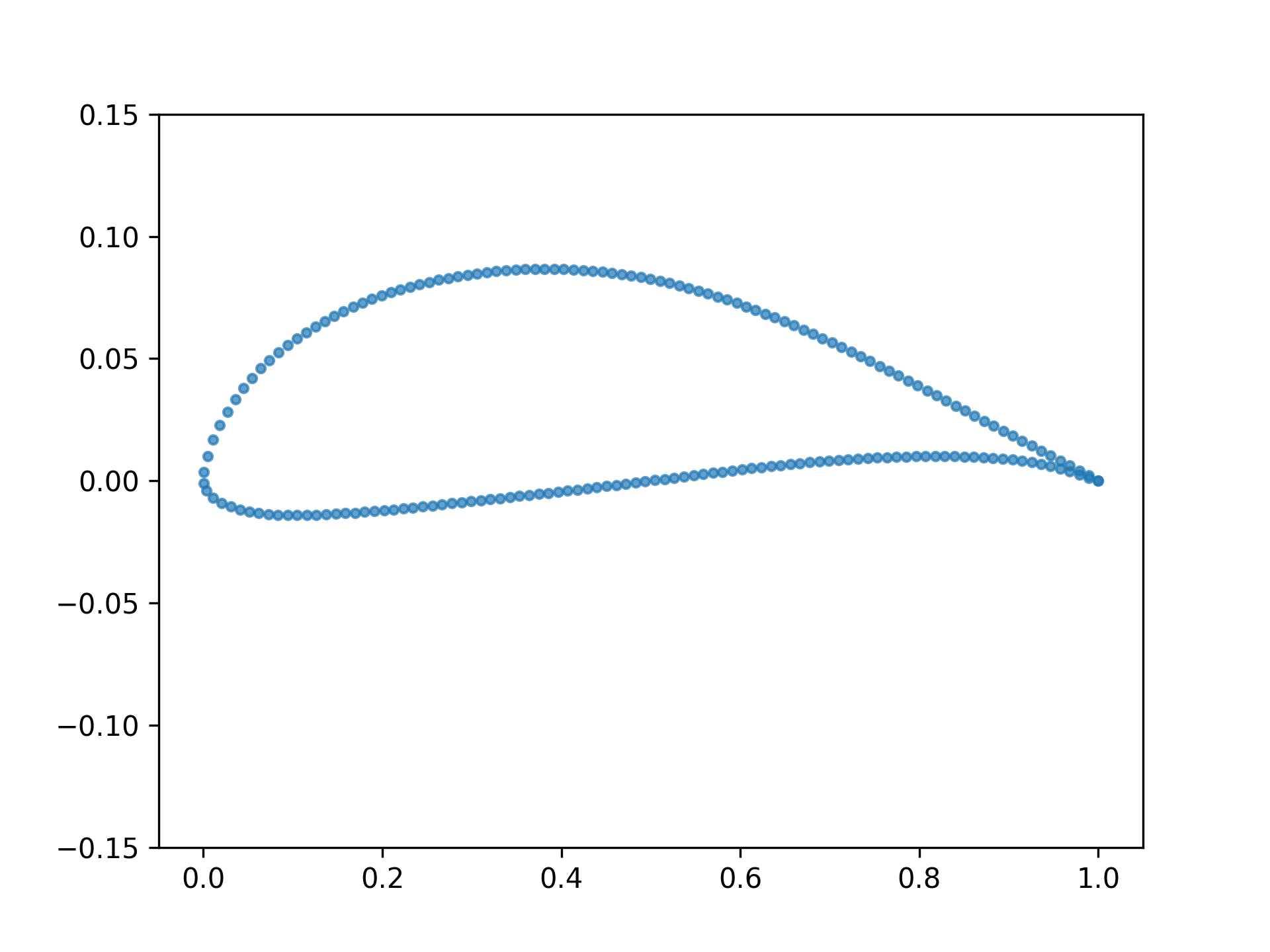}
    \caption{An Airfoil design, plotted via \texttt{problem.render}.}
    \label{fig:viz_airfoil}
\end{figure}

\revision{
\subsubsection*{Motivation}
The field of aerodynamics has always been a challenging testbed for engineering problems. In fact, many optimization and design methodologies were originally developed or perfected specifically for aerodynamic design applications \citep{mdobook}. Part of the reason for this is that even relatively simple aerodynamics problems can be complex, with slight changes in design parameters typically resulting in large changes in performance. In addition, the potential applications derived from solving these problems are quite practical, ranging from fixed-wing aircraft to hydrofoils used in ships, and wind turbine blades \citep{martinsreview}. Here, we present Airfoil, a simple yet sufficiently realistic 2-dimensional aerodynamics benchmark problem.
}

The airfoil problem presents a simple aerodynamic shape optimization routine based on Reynolds' averaged Navier-Stokes equations (RANS). In this problem, the solver attempts to indirectly morph the initial geometry to achieve a certain prescribed lift coefficient (which could correspond to a hypothetical loading requirement) while minimizing the amount of drag generated by the design.
\subsubsection*{Design Space}
The design space is represented as a tuple containing 192 2D points describing the airfoil coordinates and a scalar describing the rotation of the coordinates relative to the chord line needed to achieve a certain incoming direction of flow (the angle of attack, $\alpha$): $\DesignSpace = \left\{\left(x, y\right)^{192}, \alpha \right\} $. \revision{This specific coordinate parameterization (192) and rotational scalar design parameterization have been previously used in \citep{chen_aerodynamic_GAN_2019}. Another benchmark airfoil optimization problem, mentioned in \citep{He2019c},\footnote{Transonic RAE2822} was also used to internally validate the meshing and design parameterization.} \revision{All training data, as well as the original and complete formulation of this problem can be found in \citep{dinizdiffusion}.}

\subsubsection*{Objectives}
The objective is to minimize the coefficient of drag, $c_{d}$, and the optimization problem is defined as follows:

\begin{equation*}
\begin{aligned}
\min_{\Delta y_{i},\alpha} \quad & c_{d}\\
\textrm{s.t.} \quad & c_{l} = c_{l}^{\mathit{con}}  \\
    & -0.025 \leq \Delta y_{i} \leq 0.025 \\
    & 0.0 \leq \alpha \leq 10.0 \\
    &\left( \frac{A}{A_{\mathit{init}}} \right)_{\mathit{min}} \leq \frac{A}{A_{\mathit{init}}} \leq 1.2 \\
    \label{eq:OptProblem}
\end{aligned}
\end{equation*}

The terms used in the definition of the problem are described in \cref{tab:OptProblemAero}. Note that some variables are defined relative to a required initial design input.

\begin{figure}[H]
    \centering
    \begin{subfigure}[b]{0.45\textwidth}
        \centering
        \includegraphics[scale=0.18,trim={8cm 8cm 5cm 8cm},clip]{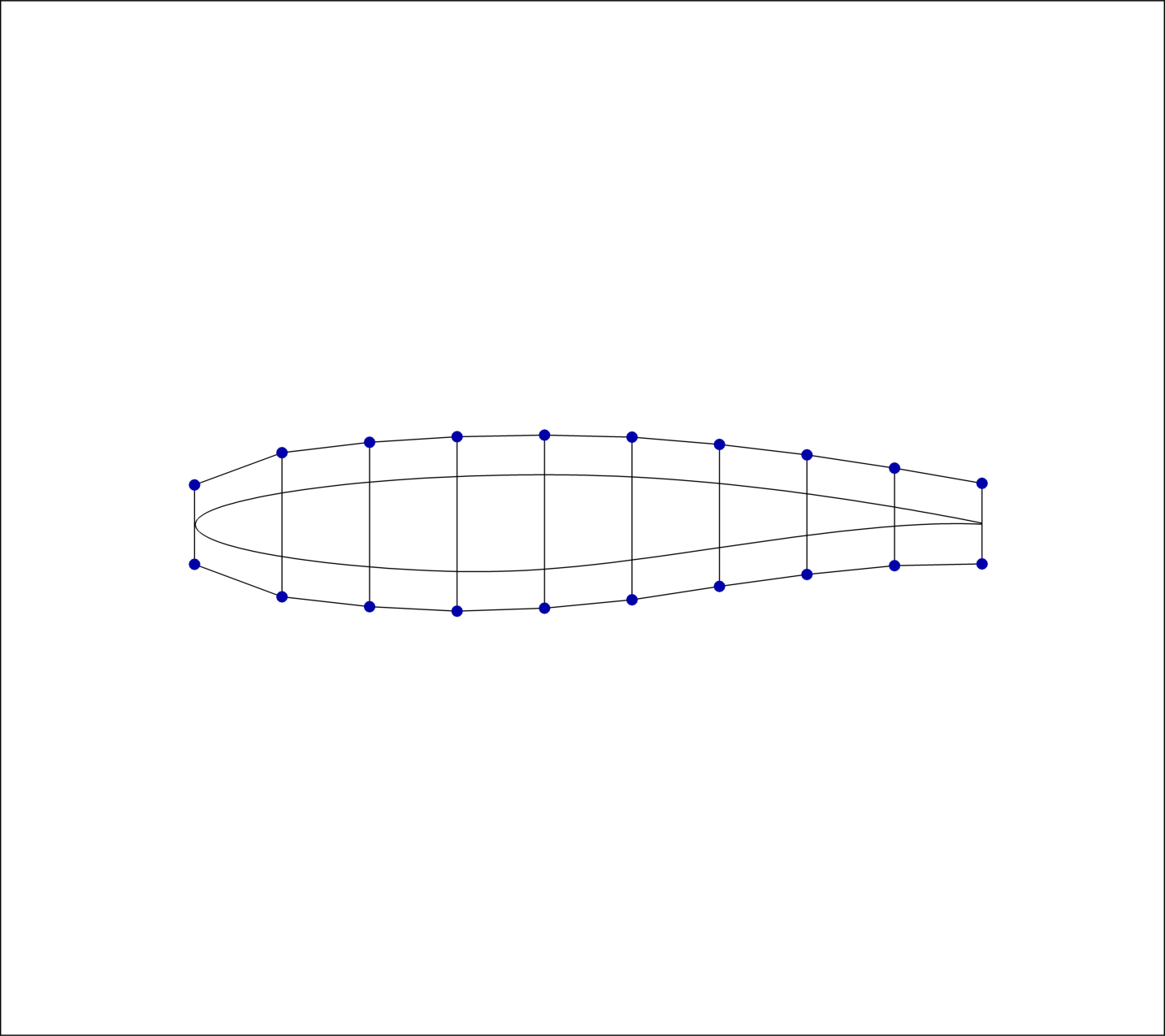}
        \caption{Sample airfoil FFD cage}
        \label{fig:rae_ffd}
    \end{subfigure}
    \hfill
    \begin{subfigure}[b]{0.45\textwidth}
        \centering
        \includegraphics[scale=0.30,trim={4cm 12cm 2cm 0cm},clip]{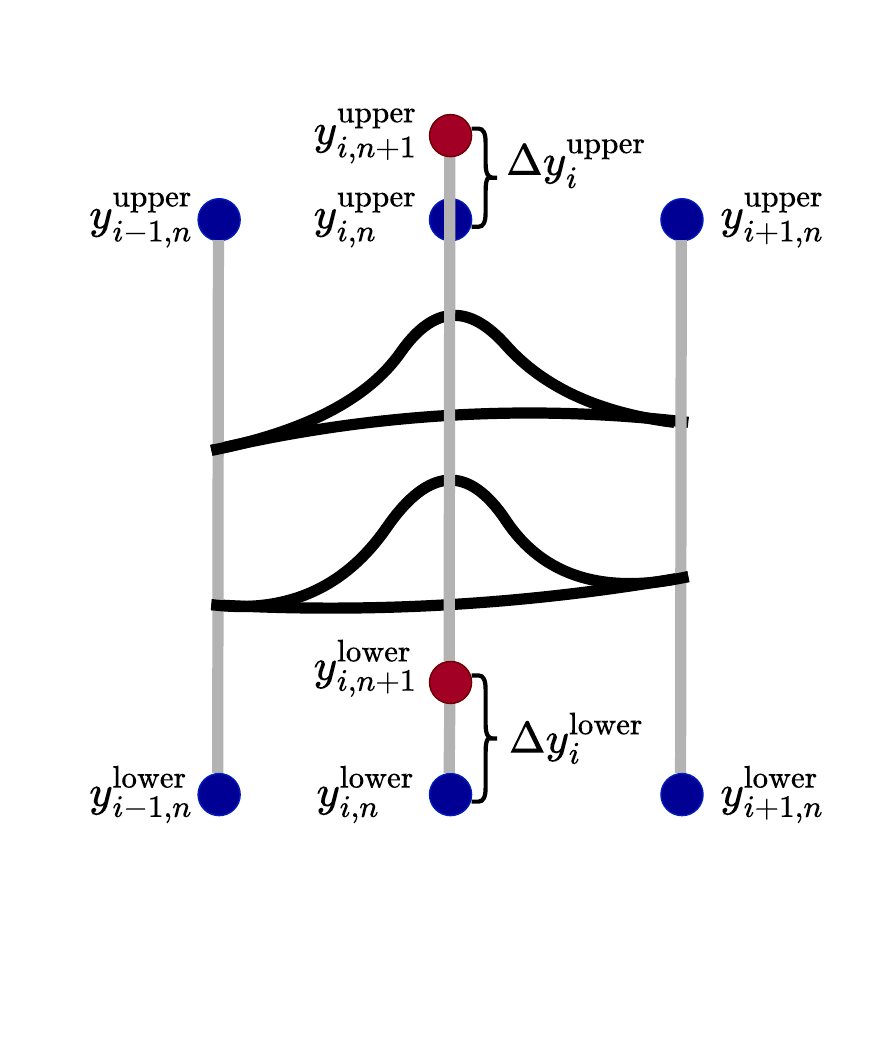}
        \caption{FFD cage morphing process}
        \label{fig:ffd_morph}
    \end{subfigure}
        \caption{Airfoil FFD design variables}
        \label{fig:ffd_vars}
\end{figure}

\begin{table}[H]
\renewcommand{\arraystretch}{1.3}
\centering
\caption{Optimization Problem Parameters}
\label{tab:OptProblemAero}
\resizebox{\columnwidth}{!}{%
\begin{tabular}{lllllll}
\toprule
\textbf{Category}   & \textbf{Parameter}                 & \textbf{Quantity} & \textbf{Lower}                            & \textbf{Upper} & \textbf{Units}  & \textbf{Description}                                                                                              \\ \midrule
\textbf{Objective}  & $c_{d}$                            & 1                 & -                                         & -              & Non-Dim./Counts & Coefficient of drag                                                                                               \\ \hline
\textbf{Variable}   & $\Delta y_{i}$                     & 20                & -0.025                                    & 0.025          & m               & \begin{tabular}[c]{@{}l@{}}Change from initial FFD cage y value:\\ $\Delta y_{i} = y_{i} - y_{init}$\end{tabular} \\
\textbf{}           & $\alpha$                           & 1                 & 0.0                                       & 10.0           & Degrees         & Angle of Attack                                                                                                   \\ \hline
\textbf{Constraint} & $c_{l} = c_{l}^{con}$              & 1                 & 0.0                                       & 0.0            & Non-Dim.        & Coefficient of lift                                                                                               \\
                    & $\frac{A}{A_{\mathit{init}}}$               & 1                 & $\left( \frac{A}{A_{\mathit{init}}} \right)_{\mathit{min}}$ & 1.20           & Non-Dim.        & Area fraction; relative to initial                                                                                \\      \bottomrule               
\end{tabular}%
}
\end{table}
Instead of directly parameterizing all (192) coordinates of the airfoil, a smaller set of 20 control points is used. The collection of these control points forms what is known as a free-form deformation cage (FFD). Changes in the values of the FFD cage smoothly deform the underlying coordinates. In the airfoil problem, modifications to the geometry are parameterized by changes in the FFD control point y coordinates, $\Delta y_{i}$. Figure \ref{fig:ffd_morph} shows how these changes in $i$\textsuperscript{th} FFD y coordinate in the $n$\textsuperscript{th} optimization iteration results in a smooth morphing of the underlying coordinates in the next $n+1$\textsuperscript{th} iteration (in red). Figure \ref{fig:rae_ffd} shows a sample FFD cage for an airfoil from the dataset, with the 20 different FFD $\Delta y_{i}$ design variables (in blue). 

Note that, for simplicity, in this definition we have omitted certain constraints pertaining to thickness as well as those concerned with the shearing of the leading (front) and trailing (tail end) edges.

\subsubsection*{Conditions}
The conditions for the airfoil problem are described in \cref{tab:airfoil_conditons}. 

\begin{table}[H]
\caption{Airfoil Conditions}
\centering
\label{tab:airfoil_conditons}
\renewcommand{\arraystretch}{1.3}
\begin{tabular}{lcl}
\toprule
\textbf{Category}       & \textbf{Condition}                        & \textbf{Description}  \\ \midrule
\textbf{Flow Condition} & M                                         & Mach number           \\
                        & Re                                        & Reynold's number      \\ \hline
\textbf{Constraint}     & $c_{l}^{con}$                             & Coefficient of lift   \\
                        & $\left( \frac{A}{A_{init}} \right)_{min}$ & Minimum area fraction \\ \bottomrule
\end{tabular}
\end{table}

Note that while it may be possible to constrain the design's area ahead of time, this is not typically possible for the prescribed coefficient of lift. In addition it may not be possible to achieve certain combinations of prescribed area and coefficient of lift constraints. 

\subsubsection*{Constraints}

\paragraph{Theoretical constraints (error)}

\begin{align*}
\DesignSpace &\in{\left\{\left(x, y\right)^N, \alpha \right\}}\\
\alpha &\in{[0, 10]}\\
\left( \frac{A}{A_{\mathit{init}}} \right)_{\mathit{min}} &\in{[0, 1.2)}\\
\end{align*}

\paragraph{Theoretical constraints (warning)}
\begin{align*}
\left( \frac{A}{A_{\mathit{init}}} \right) &\in{\left[\left( \frac{A}{A_{\mathit{init}}} \right)_{\mathit{min}} , 1.2\right)}\\
\end{align*}

\paragraph{Implementation constraints (error)}
\begin{align*}
M &\in{(0, \infty)}\\
Re &\in{(0, \infty)}\\
\end{align*}

\paragraph{Implementation constraints (warning)}
\begin{align*}
M &\in{[0.1, 1.0]}\\
Re &\in{[10^{5}, 10^{9}]}\\
\end{align*}

\subsubsection*{Simulator}
All simulations used the open source and differentiable ADflow solver~\cite{Mader_adflow_2020a, ADFLOWADJOINT} as part of the MACH-Aero framework.\footnote{\url{https://github.com/mdolab/MACH-Aero}.} ADflow was configured to run RANS simulations with the Spalart-Allmaras model for turbulence effects. Furthermore, within ADflow, the approximate Newton-Krylov method was used to improve convergence and robustness~\cite{ANK}. pyHyp, a hyperbolic mesh generator~\cite{PyHyp} was used to generate volume meshes automatically. For the optimization problem itself, we use the sequential least squares programming algorithm as implemented in the sparse optimization framework, pyOptSparse~\cite{PyOptSparse}. For geometry parameterization and deformation, module, we used the pyGeo and IDWarp frameworks~\cite{PyGeo, PyHyp}. 

\subsubsection*{Dataset}
A dataset, originally described in \cite{dinizdiffusion}, is integrated within our framework. The limits for the parameters in the data set are listed in \cref{tab:samplingAirfoil}. 1400 parameter combinations were chosen using Latin hypercube sampling (LHS) in the 4-dimensional parameter space. The training, testing and validation sets were randomly split into 748, 140, and 47 airfoil samples, respectively. Finally, \cref{tab:engivarsairfoil} describes each variable in the data set available through the \engibench{} API. 

\begin{table}[H]
\caption{Sampled Parameter Bounds}
\centering
\label{tab:samplingAirfoil}
\renewcommand{\arraystretch}{1.3}
\begin{tabular}{lllll}
\toprule
\textbf{Category}       & \textbf{Parameter}                        & \textbf{Lower} & \textbf{Upper} & \textbf{Description}  \\ \midrule
\textbf{Flow Condition} & M                                         & 0.4            & 0.9            & Mach number           \\
                        & Re                                        & 1E6            & 10E6           & Reynold's number      \\ \hline
\textbf{Constraint}     & $C_{l}^{con}$                             & 0.5            & 1.2            & Coefficient of lift   \\
                        & $\left( \frac{A}{A_{init}} \right)_{min}$ & 0.75           & 1.0            & Minimum area ratio \\ \bottomrule
\end{tabular}
\end{table}

\begin{table}[H]
\caption{Airfoil dataset variables}
\centering
\label{tab:engivarsairfoil}
\begin{tabular}{lll}
\toprule
\textbf{Category}       & \textbf{Variable Name} & \textbf{Description}                        \\ \midrule
\textbf{Flow Condition} & mach                   & Mach number                                 \\
                        & reynolds               & Reynold's number                            \\ \midrule
\textbf{Constraint} & cl\_target & \begin{tabular}[c]{@{}l@{}}Coefficient of lift \\ (targeted constraint during optimization)\end{tabular} \\
                        & area\_ratio\_min       & Minimum area ratio                          \\ \midrule
\textbf{Design Value}   & area\_initial          & Area of the initial design                  \\
                        & cd                     & Coefficient of drag for the current design  \\
                        & cl                     & Coefficient of lift for the selected design \\
                        & area\_ratio            & Area ratio of the selected design           \\ \bottomrule
\end{tabular}
\end{table}

\subsection{HeatConduction2D/HeatConduction3D}

\begin{figure}[H]
    \centering
    \includegraphics[width=0.4\linewidth]{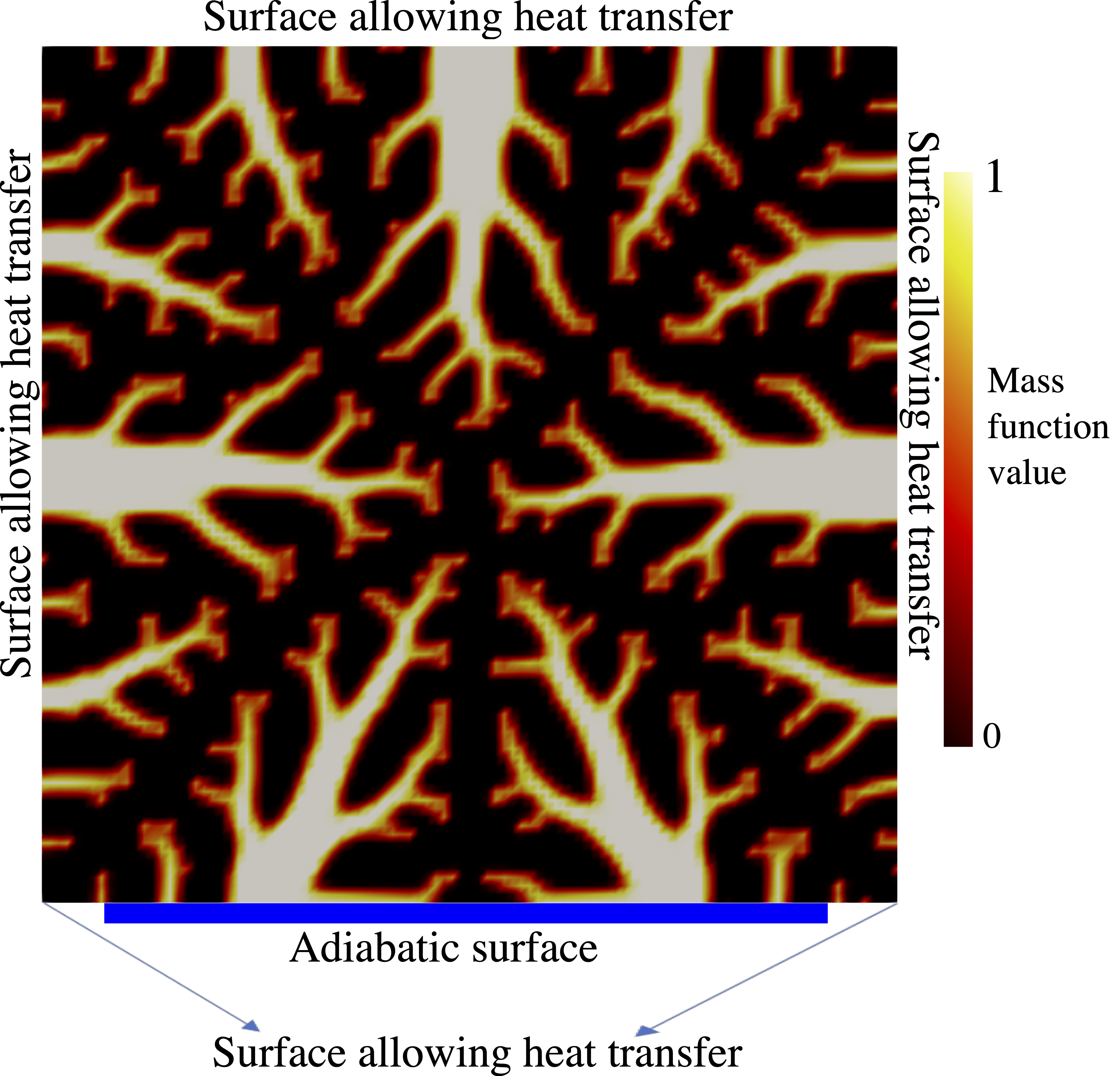}
    \caption{The dendrite-like topology emerging from optimizing heat conduction.}
    \label{fig:viz_heatconduction}
\end{figure}

\revision{
\subsubsection*{Motivation}
Heat conduction problems serve as fundamental benchmarks for the development and evaluation of design optimization methods, with applications ranging from thermal management in electronic devices to insulation systems and heat exchangers in industrial applications~\cite{mo2021topology,fawaz2022topology}. As thermal management has become critical in fields such as aerospace, automotive, and consumer electronics,  both industry and academia have shown growing interest in advanced thermal design systems~\cite{tang2019topology}. In response to this demand, topology optimization has become popular as a powerful approach for improving heat dissipation while minimizing material usage.  In addition, the development of additive manufacturing technologies has made the complex geometries produced by topology optimization more feasible to fabricate in real-world applications~\cite{meng2020topology}.
}

\subsubsection*{Design Space}
These problems are a specific subset of topology optimization problems aimed at minimizing thermal compliance within a unit square (2D) or unit cube (3D), subject to: a constraint on the \textit{volume} of highly conductive material used, and given boundary conditions, particularly the location of the adiabatic region. The adiabatic region refers to a symmetric \textit{length} on the bottom side of the 2D problem space or a prescribed symmetric \textit{area} on the bottom surface of the 3D problem space. The design space for the 2D problem consists of a 2D array representing solid densities, which is parametrized by resolution, that is, $\DesignSpace = [0,1]^{\text{resolution}\times \text{resolution}}$. By default, a $101 \times 101$ space is used for the 2D problem. The 3D design space is similarly represented as a 3D tensor of densities. By default, a $51 \times 51 \times 51$ space is used for the 3D problem.

\subsubsection*{Objectives}
In this problem, we aim to minimize the thermal compliance (\text{minimize} ${C_{T}}$) 
\begin{equation*}
C_{T}=\int_{\Omega}^{} hT + \alpha \int_{\Omega}^{} \nabla x \cdot \nabla x,
\end{equation*}
subject to the Poisson equation with mixed Dirichlet–Neumann conditions:
\begin{align*}
\nabla \cdot (k(a) \nabla T) + h &= 0 \quad \text{in } \Omega,  \\
T &= 0 \quad \text{on } \Gamma_D, \\
(k(a) \nabla T) \cdot n &= 0 \quad \text{on } \Gamma_N. 
\end{align*}
The design variable $x$ $\in$ [0,1] represents the spatial distribution of thermally conductive material, constrained by a total volume budget:
\begin{equation*}
\int_{\Omega}^{}  x \leq V
\label{vol_bound}
\end{equation*}
wherein $\Omega$ denotes the design domain (unit square in 2D, unit cube in 3D), $h$ is the heat source term, $T$ is the temperature field, $\alpha$  a regularization parameter, $k(a)$ is the material conductivity interpolated from $x$. The boundaries  $\Gamma_D$ and $\Gamma_N$ correspond to fixed-temperature (Dirichlet) and insulated (Neumann) regions, respectively.

\subsubsection*{Conditions}
The conditions ($\CondSpace$) include two key factors: the specified volume fraction of highly conductive material (\textit{volume}), and the location of the adiabatic region. The adiabatic region refers to a specified length on the bottom side of the 2D problem space or a prescribed symmetric area on the bottom surface of the 3D problem space.

\subsubsection*{Constraints}
This section outlines the constraints relevant to the 2D and 3D heat conduction problems.

\paragraph{Theoretical constraints (error)}
\begin{align*}
\DesignSpace^{2D} &\in{[0, 1]}^{\text{resolution} \times \text{resolution}}\\
\DesignSpace^{3D} &\in{[0, 1]}^{\text{resolution} \times \text{resolution} \times \text{resolution}}\\
\mathit{resolution} &\in{[1, \infty)}\\
\mathit{volume} &\in{[0, 1]}\\
\mathit{length} &\in{[0, 1]}\\
\mathit{area} &\in{[0, 1]}\\
\end{align*}


\paragraph{Implementation constraints (warning)}
\begin{align*}  
\mathit{resolution} &\in{[10, 1000]} \\
\mathit{volume} &\in{[0.3, 0.6]} \\
\end{align*}

\subsubsection*{Simulator}
We employ a modified version of the open-source Dolfin-adjoint example to solve the 2D and 3D heat conduction topology optimization problems~\cite{mitusch2019dolfin}. Rather than formulating the problem as an integer optimization task, we use a continuous relaxation approach, which is the standard technique in topology optimization~\cite{bendsoe2013topology}. This allows the material distribution to change smoothly between solid and void, enabling efficient gradient-based optimization. The solver relies on the interior-point method implemented in Ipopt~\cite{wachter2006implementation}, which handles the nonlinear constraints.

\subsubsection*{Dataset}
The datasets for the 2D and 3D heat conduction problems have been originally introduced in~\citet{habibi2023actually, habibi2025mean}. Each dataset entry contains the optimal design solution (\texttt{optimal\_design}) along with its corresponding set of input conditions. To generate our 2D/3D dataset, we sampled two input conditions: volume and the adiabatic region's length (2D) or area (3D). We divided each parameter range into 20 equal intervals, yielding 21 values per parameter (\eg, area: 0.0, 0.05, ..., 1.0). By combining all pairs, we generated 441 optimized topologies. For test-train splitting, we randomly excluded two unique values of the volume limit and two unique values of the adiabatic region from the training set. This exclusion resulted in 361 designs for training, with the remaining 80 designs evenly split into 40 for validation and 40 for testing. The parameter sampling bounds used to generate the datasets are summarized in \cref{tab4:sampling}.
\begin{table}[H]
\caption{Sampled Parameter Bounds}
\centering
\label{tab4:sampling}
\renewcommand{\arraystretch}{1.3}
\begin{tabular}{lllll}
\toprule
\textbf{Problem}       & \textbf{Parameter}                        & \textbf{Lower} & \textbf{Upper} \\ \midrule
\textbf{HeatConduction2D} & volume                                         & 0.3            & 0.6            \\
                        & length                                        & 0            & 1           \\ \hline
\textbf{HeatConduction3D}     & volume                             & 0.3            & 0.6           \\
                        & area & 0           & 1.0\\ \bottomrule
\end{tabular}
\end{table}

\subsection{ThermoElasticBeams2D}

\begin{figure}[H]
    \centering
    \includegraphics[width=0.35\linewidth]{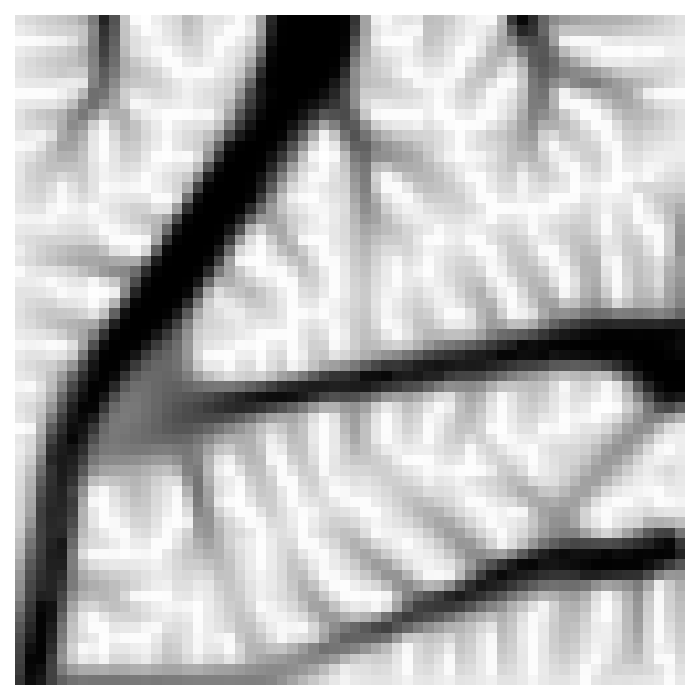}
    \caption{Dendrite-like topology emerging when optimizing a thermo-elastic beam for both structural stiffness and thermal diffusion.}
    \label{fig:viz_thermoelasticbeams}
\end{figure}

\revision{
\subsubsection*{Motivation}
As articulated in their respective sections, both the Beams2D and HeatConduction2D problems are fundamental engineering design problems that have historically served as benchmarks for the development and testing of optimization methods. While their relevance is supported by needs in real engineering design scenarios (aerospace, automotive, consumer electronics, \etc), their mono-domain nature ignores the reality that coupling between domains exists, and should be accounted for in scenarios where performance in one domain significantly impacts performance in another \cite{giraldo2020multi}. To address this distinction, a multi-physics topology optimization problem is developed that captures the coupling between structural and thermal domains.
}

\subsubsection*{Design Space}

This multi-physics topology optimization problem is governed by linear elasticity and steady-state heat conduction with a one-way coupling from the thermal domain to the elastic domain. The problem is defined over a square 2D domain, where load elements and support elements are placed along the boundary to define a unique elastic condition. Similarly, heatsink elements are placed along the boundary to define a unique thermal condition. The design space is then defined by a 2D array representing density values (parameterized by $\DesignSpace = [0,1]^{\text{nelx}\times \text{nely}}$, where \textit{nelx} and \textit{nely} denote the x and y dimensions).

\subsubsection*{Objectives}

The objective of this problem is to minimize total compliance $C$ under a volume fraction constraint $V$ by placing a thermally conductive material. Total compliance is defined as the sum of thermal compliance (${C_{T}}$) and structural compliance (${C_{S}}$):

\begin{equation*}
C = C_T + C_S
\end{equation*}

Above, thermal compliance is defined by:

\begin{equation*}
C_{T}=\int_{\Omega}^{} hT + \alpha \int_{\Omega}^{} \nabla x \cdot \nabla x,
\end{equation*}

where the domain assumes uniform heat generation. The corresponding thermal boundary conditions are defined by mixed Dirichlet-Neumann conditions, namely:

\begin{align*}
\nabla \cdot (k(a) \nabla T) + h &= 0 \quad \text{in } \Omega,  \\
T &= 0 \quad \text{on } \Gamma_D, \\
(k(a) \nabla T) \cdot n &= 0 \quad \text{on } \Gamma_N. 
\end{align*}

where the terms are defined as follows: $\Omega$ represents the 2D domain, $h$ denotes uniform heat generation, $T$ is the temperature field, $\alpha$ is a regularization parameter, and $k(a)$ is material conductivity interpolated from design variable $x$. Finally, $\Gamma_D$ and $\Gamma_N$ represent the Dirichlet and Neumann boundary conditions respectively.

As the problem has a one-way coupling from the thermal domain to the elastic domain, structural compliance is defined by two components: compliance due to the external loads imposed on the system $C_{S1}$, and compliance due to the forces induced by element-wise thermal expansion $C_{S2}$. Both components are captured in the following calculation as long as element displacement vector $\mathbf{u}_e$ captures displacements due to both external loads and thermal expansion forces:

\begin{equation*}
    \underset{\mathbf{x} \in [0, 1]^{\text{nelx} \times \text{nely}}}{\text{minimize}} \, 
    C_{S}(\mathbf{x}) = \sum_{e=1}^{N} E_e(x_e) \, \mathbf{u}_e^{\mathrm{T}} \mathbf{k}_0 \mathbf{u}_e 
    \quad \text{where} \quad 
    \begin{aligned}
        & \mathbf{K} \mathbf{U} - \mathbf{F} = 0.
    \end{aligned}
\end{equation*}

where $\mathit{N}$ is the number of total elements ($\mathit{nelx} * \mathit{nely}$), $E_e(x_e)$ is the penalized Young's Modulus based on the Solid Isotropic Material with Penalization method (SIMP) formulation~\cite{sigmund2007morphology}, $\mathbf{u}_e$ is the element displacement vector, and $\mathbf{k}_0$ is the reference stiffness matrix. 

For the equation above, the displacement vector is calculated using the relation $\mathbf{K} \mathbf{U} - \mathbf{F} = 0$, where $F$ captures the forces due to both external loads and the thermal expansion of elements. The forces due to thermal expansion ($f^{\mathit{th}}$) are calculated as follows for each element:

\begin{equation*}
    f^{\textit{th}}_e = x^{e}_p C_{e} (\mathbf{T_e} - T_{\textit{ref}})
\end{equation*}

where $T_e$ represents the nodal temperatures of the element, $T_{\textit{ref}}$ represents the stress-free reference temperature, $x^{e}_p$ represents the SIMP density‑penalisation factor, and $C_e$ denotes the coupling matrix mapping the temperature delta to a normal force representing thermal expansion.

Finally, a volume fraction constraint is imposed to restrict the amount of material that can be used for a design. This constraint is defined by:

\begin{equation*}
\int_{\Omega}^{}  x \leq V
\end{equation*}

where $\Omega$ denotes the design domain, and $V$ is the volume fraction constraint value.

\subsubsection*{Conditions}

The set of conditions ($\CondSpace$) defining a unique problem formulation includes: the volume fraction constraint value ($V$), the filter radius for the design variables $\mathit{rmin}$, and the corresponding locations of the fixed elements, loaded elements, and heatsink elements.

\subsubsection*{Constraints}

The constraints for this problem are defined as follows:

\paragraph{Theoretical constraints (error)}
\begin{align*}
\DesignSpace &\in{[0, 1]}^{\text{nelx} \times \text{nely}}\\
\mathit{nelx} &\in{[1, \infty)}\\
\mathit{nely} &\in{[1, \infty)}\\
V &\in{[0, 1]}\\
\mathit{rmin} &\in{(0, \infty)}
\end{align*}

\paragraph{Theoretical constraints (warning)}
\begin{align*}
    \frac{\sum_{i=0}^{\text{nelx}}\sum_{j=0}^{\text{nely}} x_{i,j}}{\text{nelx} \times \text{nely}} \leq V
\end{align*}

\paragraph{Implementation constraints (error)}
\begin{align*}
\mathit{rmin} &\in{(0, min(nelx, nely))}
\end{align*}

\paragraph{Implementation constraints (warning)}
\begin{align*}  
\mathit{nelx} &\in{[10, 1000]} \\
\mathit{nely} &\in{[10, 1000]} \\
V &\in{[0.1, 0.9]} \\
\mathit{rmin} &\in{[1.0, 10.0]} 
\end{align*}

\subsubsection*{Simulator}

The simulation code is based on a Python adaptation\footnote{\href{https://github.com/arjendeetman/TopOpt-MMA-Python}{GitHub: TopOpt-MMA-Python}} of the popular 88-line topology optimization code~\cite{andreassen_88lines_2011}, modified to handle the thermal domain in addition to thermal-elastic coupling. Optimization is conducted by reformulating the integer optimization problem as a continuous one (leveraging a SIMP approach), where a density filtering approach is used to prevent checkerboard-like artifacts. The optimization process itself operates by calculating the sensitivities of the design variables with respect to total compliance (done efficiently using the Adjoint method), calculating the sensitivities of the design variables with respect to the constraint value, and then updating the design variables by solving a convex-linear subproblem and taking a small step (using the method of moving asymptotes). The optimization loop terminates when either an upper bound of the number of iterations has been reached or if the magnitude of the gradient update is below some threshold. 

\subsubsection*{Dataset}

The dataset for this problem contains a set of 1000 optimized thermoelastic designs in a 64x64 domain, where each design is optimized for a unique set of conditions. Each datapoint contains: the conditions under which the design was optimized (fixed elements, loaded elements, heatsink elements, volume fraction constraint, ...), the optimized design, and the objective values for the optimized design. Each datapoint's conditions are randomly generated by arbitrarily placing: a single loaded element along the bottom boundary, two fixed elements (fixed in both the x and y direction) along the left and top boundary, and heatsink elements along the right boundary. Furthermore, values for the volume fraction constraint are randomly selected in the range $[0.2, 0.5]$. The full dataset is available at: \url{https://huggingface.co/datasets/IDEALLab/thermoelastic_2d_v0}.

\subsection{Beams2D}
\begin{figure}[H]
    \centering
    \includegraphics[width=0.5\linewidth]{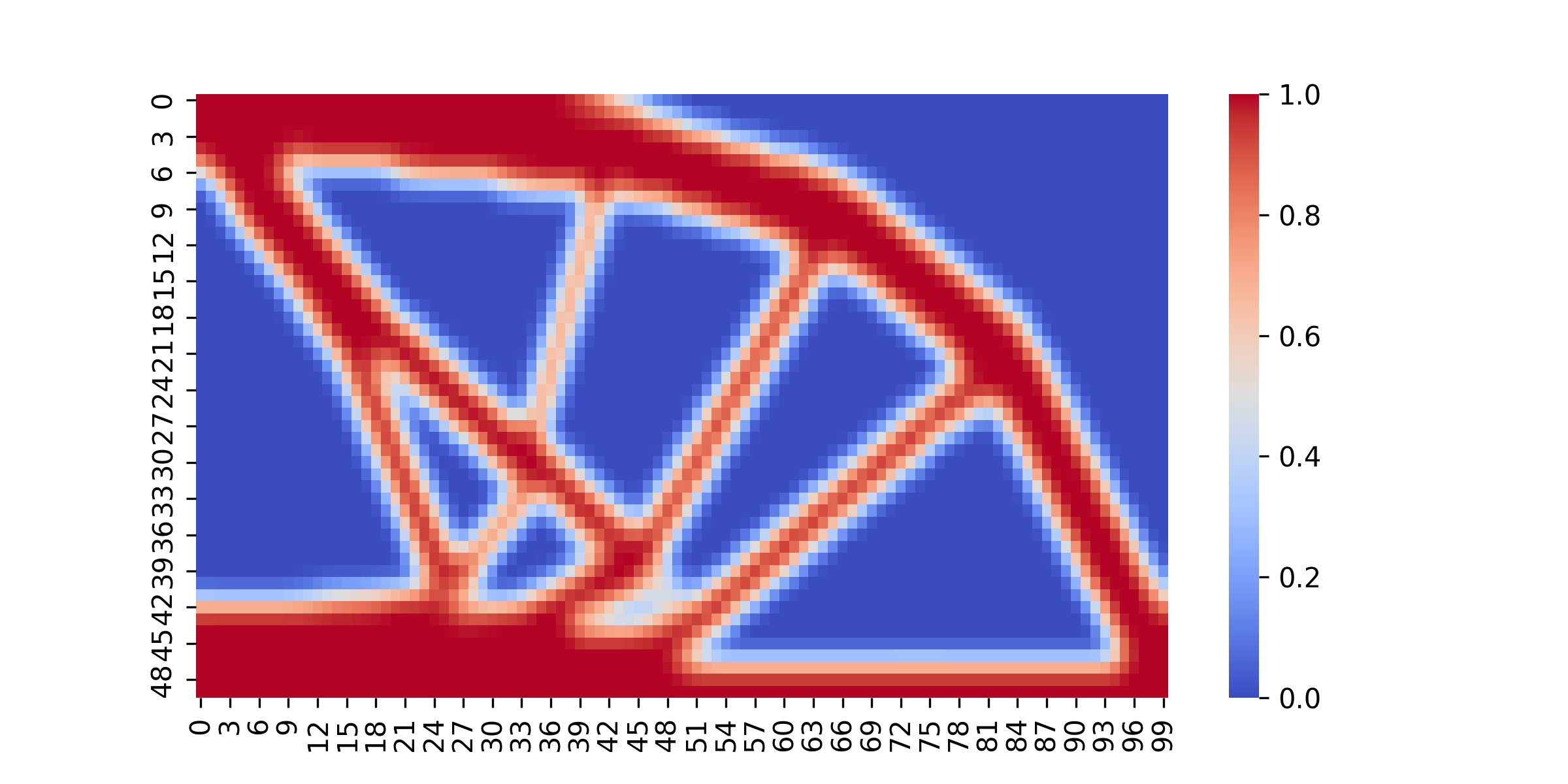}
    \caption{An optimized beam, representative of Beams2D.}
    \label{fig:viz_beams}
\end{figure}

\revision{
\subsubsection*{Motivation}
The optimization of beam cross-sections is one of a fundamental problem in engineering, aiming to maximize the structural stiffness under some applied force. This objective is usually formulated as minimizing the compliance, which is the inverse of stiffness. In particular, TO frames the problem as one of optimal material distribution, defining a grid of elements for which the material densities must be determined on a scale from 0 to 1, where 1 represents the presence of material. After applying the beam loads and other boundary conditions, designs are typically optimized using a gradient-based approach with the help of the finite element method (FEM)~\cite{bendsoe1989optimal,andreassen_88lines_2011}. While this is one of the simplest TO applications, it is still a computationally expensive process requiring many iterations, opening the door for faster approximation methods such as generative inverse design.} 

\revision{One of the most common beam types in TO is the Messerschmitt-B\"olkow-Blohm (MBB) beam, which is supported at the bottom-right and bottom-left corners, with a downward force applied on the top-center. Given this symmetric configuration, one half of the design may be optimized while representing the entire structure. We implement the MBB beam in \engibench{} for the most accessible comparison to previous works in this domain.}

\subsubsection*{Design Space}

This problem simulates the right half-section of a MBB beam under bending. This half-beam is subjected to a force at its top-left corner (corresponding to the top-center of the entire design) which may also be shifted to the right to simulate different loading conditions. A roller support at the bottom-right corner prevents vertical movement, and a symmetric boundary condition is enforced on the left edge. The design space consists of a 2D array representing solid densities, parameterized by \textit{nelx} and \textit{nely}, i.e., $\DesignSpace = [0,1]^{\text{nelx}\times \text{nely}}$. By default, a $100 \times 50$ space is used.

\subsubsection*{Objectives}


We aim to minimize compliance $c(\mathbf{x})$, which may be thought of as the inverse of structural stiffness. This objective is calculated as the sum of strain energy over the structure:

\begin{equation*}
    \underset{\mathbf{x} \in [0, 1]^{\text{nelx} \times \text{nely}}}{\text{minimize}} \, 
    c(\mathbf{x}) = \sum_{e=1}^{N} E_e(x_e) \, \mathbf{u}_e^{\mathrm{T}} \mathbf{k}_0 \mathbf{u}_e 
    \quad \text{subject to} \quad 
    \begin{aligned}
        & \frac{V(\mathbf{x})}{V_0} - volfrac = 0, \\
        & \mathbf{K} \mathbf{U} - \mathbf{F} = 0.
    \end{aligned}
\end{equation*}

Where $N = \text{nelx} \times \text{nely}$ is the total number of array elements, $E_e(x_e)$ is the penalized Young's Modulus based on the Solid Isotropic Material with Penalization method (SIMP) formulation~\cite{sigmund2007morphology}, $\mathbf{u}_e$ is the element displacement vector, $\mathbf{k}_0$ is the reference stiffness matrix for a solid element $x_e = 1$, $V(\mathbf{x})$ and $V_0$ are respectively the material volume and design domain volume, $volfrac$ is the desired volume fraction, $\mathbf{K}$ is the global stiffness matrix, $\mathbf{U}$ is the global displacement vector, and $\mathbf{F}$ is the global force vector~\cite{andreassen_88lines_2011}.

\subsubsection*{Conditions}

The conditions ($\CondSpace$) consist of the desired volume fraction of solid material in the design (\textit{volfrac}), the minimum feature length of beam members (\textit{rmin}), the fractional distance of the downward force from the default top-left corner to the top-right corner (\textit{forcedist}), and a boolean overhang constraint: when true, this removes unsupported structures with an overhang angle of greater than 45 degrees from the vertical axis, preserving manufacturability. In practice, this constraint often leads to convergence issues when combined with an insufficient \textit{volfrac}, so we have excluded it from the dataset for the moment.

\subsubsection*{Constraints}  

\paragraph{Theoretical constraints (error)}
\begin{align*}
\DesignSpace &\in{[0, 1]}^{\text{nelx} \times \text{nely}}\\
\mathit{nelx} &\in{[1, \infty)}\\
\mathit{nely} &\in{[1, \infty)}\\
\mathit{volfrac} &\in{[0, 1]}\\
\mathit{rmin} &\in{(0, \infty)}\\
\mathit{forcedist} &\in{[0, 1]}
\end{align*}

\paragraph{Theoretical constraints (warning)}
\begin{align*}
    \frac{\sum_{i=0}^{\text{nelx}}\sum_{j=0}^{\text{nely}} x_{i,j}}{\text{nelx} \times \text{nely}} \leq \mathit{volfrac}
\end{align*}

\paragraph{Implementation constraints (error)}
\begin{align*}
\mathit{rmin} &\in{(0, 0.5 \times \max \{nelx, nely\}}
\end{align*}

\paragraph{Implementation constraints (warning)}
\begin{align*}  
\mathit{nelx} &\in{[10, 1000]} \\
\mathit{nely} &\in{[10, 1000]} \\
\mathit{volfrac} &\in{[0.1, 0.9]} \\
\mathit{rmin} &\in{[1.0, 10.0]} 
\end{align*}

\subsubsection*{Simulator}


Our simulation code is based on a Python adaptation\footnote{\href{https://github.com/arjendeetman/TopOpt-MMA-Python}{GitHub: TopOpt-MMA-Python}} of the popular 88-line topology optimization code~\cite{andreassen_88lines_2011}. It uses the more versatile density filtering approach in combination with a standard Optimality Criteria (OC) optimization method. Two primary sensitivity matrices, one with respect to compliance ($dc$) and the other with respect to volume fraction ($dv$), are continuously updated and used to calculate a given design's compliance value. We have also ensured that during the required Lagrange multiplier search within OC, the inner optimization loop terminates if the absolute difference upper and lower bounds diminishes to a value smaller than machine precision. This prevents the code from becoming stuck at this point, which we observed in some warm-starting instances with noisy initial designs.

\subsubsection*{Dataset}

This problem offers multiple datasets for various sizes of \textit{nelx} and \textit{nely}. Each dataset includes columns for \texttt{optimal\_design}, all conditions listed above, and the corresponding objective values. For advanced usage, we also provide a column containing the optimization history. The datasets have been generated by sampling conditions over a structured grid for various problem sizes. The full set of parameter combinations and dataset generation script are available at \url{https://github.com/IDEALLab/EngiBench/blob/main/engibench/problems/beams2d/README.md}.

\subsection{Photonics2D}

\begin{figure}[H]
    \centering
    \includegraphics[width=0.8\linewidth]{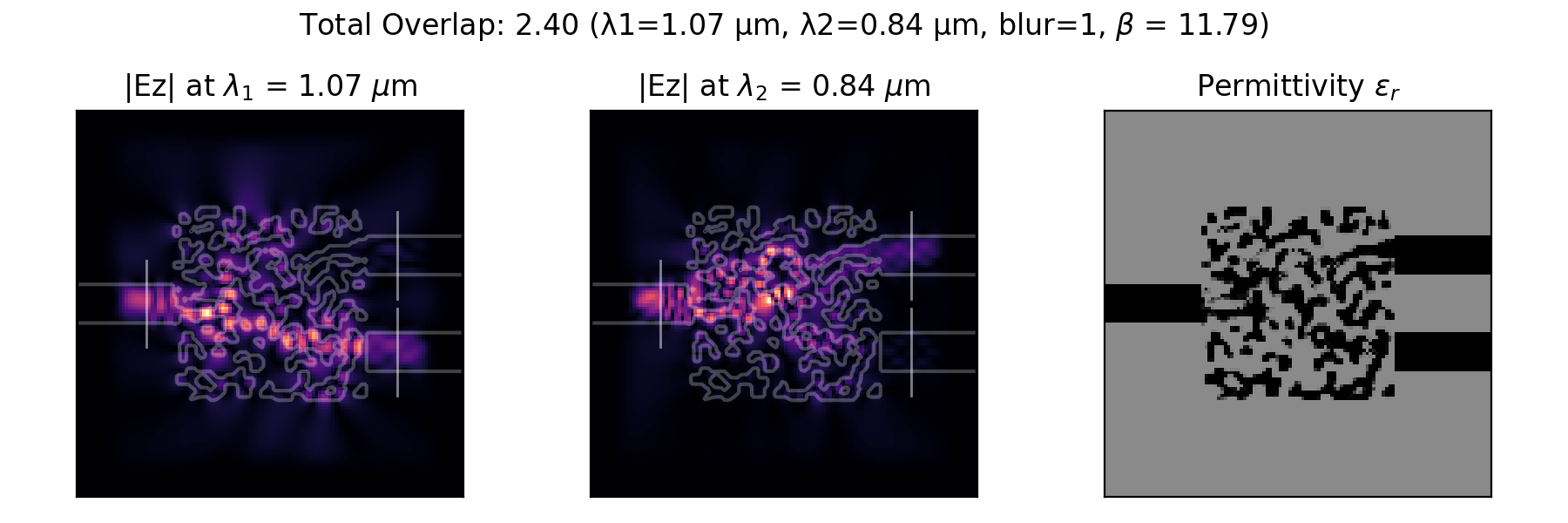}
    \caption{Example visualization of our Photonics2D problem. The two left plots show where different wavelengths are routed (source is on the left). The rightmost plot shows the material distribution.}
    \label{fig:viz_photonics}
\end{figure}

\revision{
\subsubsection*{Motivation}
The optimization of photonic circuits in general, and multiplexers in particular, was one of the initial and most widely studied problems in the inverse design of electromagnetic/optical devices~\cite{piggott2015inverse}. In part, this is because multiplexer devices have several interesting properties that make them more difficult to create generative models of, compared to other problems in EngiBench. This includes the fact that, due to the wave properties of the physical phenomena there are usually multiple solutions with equivalent or similar performance, which results from shifting or inverting the phase profile of the electromagnetic wave. This adds complexity to the generative model in that the solution may not have a single unique global minimum. Another motivating factor for including this problem is the complexity of the structures/designs themselves: unlike in structural or thermal compliance problems, which lead to connected structures, the photonics solutions often involve several disconnected elements whose relative position and spacing is governed by the specific wavelengths it needs to demultiplex. This is a difficult prediction and generation task, compared to, \eg, generating a connected beam structure. Thus, it acts as a good counterpoint to add to the library and provides a mechanism to benchmark generative algorithms that can perform well on both connected and disconnected design topologies.}

\subsubsection*{Design Space}

This problem simulates a wavelength demultiplexer where the optimized device will direct an electromagnetic wave to one of two possible output ports depending on the wavelength/frequency of the incoming wave~\citep{hughes2018adjoint,piggott2015inverse}. Specifically, the demultiplexer targets two specific wavelengths (referred to $\lambda_1$ and $\lambda_2$ in the library), and the performance of the device is how well it can bend or direct the energy toward two specific locations in the device, as measured by how much of the electric field of each wavelength overlaps with the desired output port locations. The design space consists of a 2D array representing the presence of either a high or low permittivity, parameterized by \textit{nelx} and \textit{nely}, i.e., $\DesignSpace = [0,1]^{\text{nelx}\times \text{nely}}$. By default, the library uses a $120 \times 120$ space, however this can be modified to non-square design spaces by the user.

\subsubsection*{Objectives}


The main objective is to maximize the overlap of the electric field of the simulated wavelength at the target output location, with an optional penalty for the amount of material used (this penalty weight is set to a small default value ($1e^{-2}$) for consistency, but can be altered for advanced usage):

\begin{equation*}
    \underset{\mathbf{x} \in [0, 1]^{\text{nelx} \times \text{nely}}}{\text{maximize}} \, 
    c(\mathbf{x}) = \left[\int_{p_1} |m_1^Te_{\omega_1}|^2 \times \int_{p_2}|m_2^Te_{\omega_2}|^2\right] - w||\mathbf{x}||^2 
\end{equation*}


Where $N = \text{nelx} \times \text{nely}$ is the total number of array elements, $w$ is the penalty weight for usage of material (set by default to $1e^{-2}$), $p1$ and $p2$ are the output port locations, $m_1$ and $m_2$ are the output mode profiles, and $e_{\omega_1}$ and $e_{\omega_2}$ are the computed fields under an input wavelength $\lambda_1$ and $\lambda_2$, respectively.

In practice, the problem uses a SIMP formulation (similarly as described in Beams2D) to gradually interpolate between two material permittivities by using a Heaviside projection operator (where the projection strength is governed by a $\beta$ strength parameter) to project the designed density field onto the two target permittivities. As with other SIMP-style optimization approaches, to ensure smooth optimization toward binary designs, this projection strength undergoes a continuation scheme where it is initially set to a low value ($\beta=1.0$) and then polynomially (quadratically) increased throughout optimization until a final maximum value ($\beta=300$), where the projection operator is essentially outputting binary designs.

\subsubsection*{Conditions}

The conditions ($\CondSpace$) consist of the two input wavelengths to be demultiplexed\textemdash $\lambda_1$ and $\lambda_2$, as well as a desired \texttt{blur\_radius} ($r_{blur}$) parameter, which blurs (using a circular convolution) the pixelized design field for a chosen number of integer pixels\textemdash this blurring essentially creates a penalty on the minimum feature size of the design. The size of the device\textemdash expressed as \textit{nelx} and \textit{nely}\textemdash is also adjustable, and could be viewed as a possible condition for multi-resolution problems, but in practice, as with Beams2D above, this is built into the problem definition since it produces a different dataset.

\subsubsection*{Constraints}
We now list the constraints for the Photonics2D problem.

\paragraph{Theoretical constraints (error)}
\begin{align*}
\DesignSpace &\in{[0, 1]}^{\text{nelx} \times \text{nely}}\\
\mathit{nelx} &\in{[1, \infty)}\\
\mathit{nely} &\in{[1, \infty)}\\
\lambda_i &> 0\\
r_{blur} &\geq 0
\end{align*}

\paragraph{Implementation constraints (error)}
\begin{align*}
\lambda_i \in [0.5,\infty]\\
r_{blur} \geq 0\\
\mathit{nelx} >60\\
\mathit{nely} \geq 105
\end{align*}

\paragraph{Implementation constraints (warning)}
\begin{align*}  
\lambda_i \in [0.5,1.5]\\
r_{blur} \in [0,5]\\
\mathit{nelx} \in [90,200]\\
\mathit{nely} \in [110,300]
\end{align*}

\subsubsection*{Simulator}


The simulation code uses the \texttt{ceviche} library~\citep{hughes2019forward} and specifically, the wave demultiplexer demonstration case provided by the library authors\footnote{\url{https://github.com/fancompute/workshop-invdesign/blob/master/04_Invdes_wdm_scheduling.ipynb}} based on their related publication~\citep{hughes2018adjoint}, which uses a similar formalism to an earlier demultiplexer paper by~\citet{piggott2015inverse}. The optimization method is first-order and uses the Adam optimizer. Beyond the baseline implementation already available via \texttt{ceviche}, we implemented a polynomial $\beta$ continuation scheme that performed more reliably that the step-wise continuation scheme used in the original implementation, and \engibench{} also possesses the ability to change the starting and ending continuation values, for future research cases where one wishes to estimate or optimize the continuation schedule themselves. Other than these changes, the implementation of this problem is as consistent as possible with that of the original \texttt{ceviche} library.

\subsubsection*{Dataset}

This problem offers a single datasets of \textit{nelx}=120 and \textit{nely}=120, although various sizes of \textit{nelx} and \textit{nely} could be generated from the library if desired. The dataset includes columns for the \texttt{optimal\_design}, all conditions listed above, and the corresponding objective value history as the optimizer optimized toward the optimal design provided in the dataset. The dataset was generated by sampling by sampling at random $\lambda_1$, $\lambda_2$ and $r_{blur}$ over the conditions mentioned above. The full set of parameter combinations used for the dataset are contained in the docstring and problem documentation available at \url{https://github.com/IDEALLab/EngiBench/blob/main/engibench/problems/photonics2d/v0.py}.

\subsection{Power Electronics}
\revision{
\subsubsection*{Motivation}
Optimizing circuit parameters is a critical aspect of circuit design but remains challenging, particularly for power converter circuits that contain diodes and switches, which introduce significant nonlinearity and discontinuity. These characteristics make circuit simulation results and key objectives such as \textit{DcGain} and \textit{Voltage Ripple} highly sensitive to even small parameter variations.
}

\revision{
Because the circuit simulator NgSpice operates as a black box and is non-differentiable, Bayesian optimization is commonly employed for parameter tuning. Surrogate models also offer a promising alternative for this task.
In this work, we present experiments using a fixed circuit topology. Even under this constraint, optimizing circuit parameters to minimize the objectives remains a challenging problem for surrogate models, as we show in \cref{sec:experiments}.
}

\begin{figure}[H]
    \centering
    \includegraphics[width=0.5\linewidth]{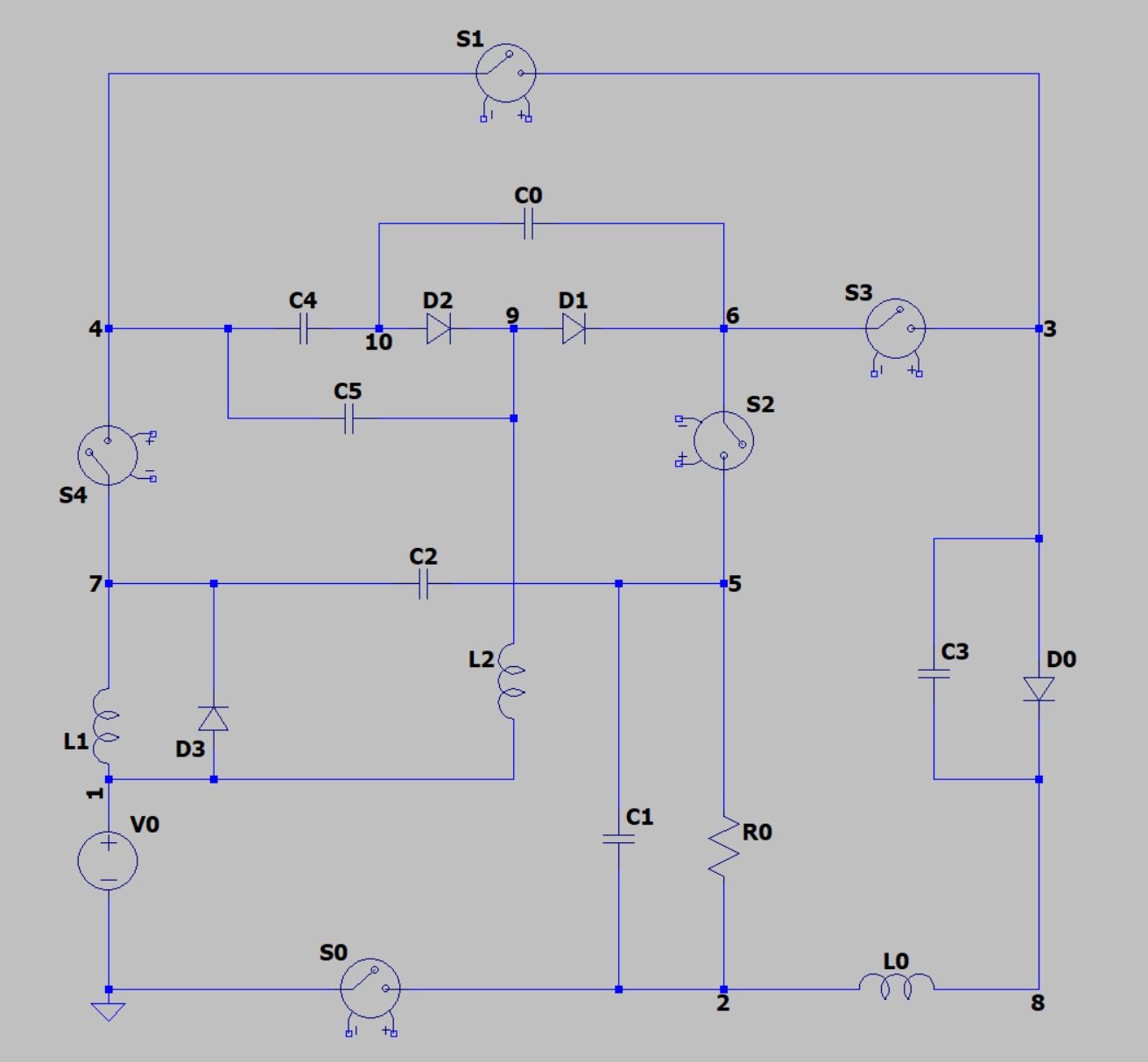}
    \caption{The fixed electric circuit  with 5 switches, 6 capacitors, 3 inductors and 4 diodes that we optimize in PowerElectronics.}
    \label{fig:viz_powerelec}
\end{figure}

This problem models a DC-DC power converter circuit with a fixed topology. The circuit comprises 5 switches, 4 diodes, 3 inductors, and 6 capacitors. To simulate a circuit, we first encapsulate the circuit topology and parameter information into a text-based file known as a ``netlist''. This file is then processed by NgSpice, an open-source, cross-platform simulator widely used for analyzing analog, digital, and mixed-signal circuits~\citep{ngspice}. For resistors ``R'', inductors ``L'', capacitors ``C'' (RLC) circuits, we apply transient analysis, where NgSpice formulates the system as a set of differential equations based on Kirchhoff’s laws. These equations are discretized using numerical integration methods such as the Backward Euler or trapezoidal rule and are solved iteratively at each time step to compute performance metrics, as detailed in the Objectives section below.

Since NgSpice functions as a black-box simulator and does not offer differentiability with respect to input parameters, gradient-based optimization methods are not applicable. Therefore, we formulate the problem as one with a fixed topology, ensuring that optimization is restricted to parameter tuning rather than structural modifications. Furthermore, we choose a specific on-off pattern for the 5 switches, with \textit{Switch 1, Switch 2, Switch 3, Switch 4, Switch 5} set to \textit{on, off, on, on, off} at the beginning and switching their states simultaneously. This setting avoids NgSpice simulation failure.
Despite the fixed topology and simplification of the switch parameters, determining the optimal values for the remaining circuit parameters to minimize the objectives remains a challenging task for surrogate models, see \cref{sec:zoomed_exp,app:additional_info_expes}.

\subsubsection*{Design Space}
The design space for this problem is represented as a 10-dimensional bounded box, where each dimension corresponds to a specific circuit parameter. These parameters include values for capacitors, inductors, and a shared duty cycle for all switches. Each design can be expressed as a vector $x$ of the form:
\[
x = \begin{bmatrix} C_1,\dots,C_6,L_1,L_2,L_3,T_1 \end{bmatrix}^{\!\top} \in \mathcal{X}
\quad \text{where} \quad
\mathcal{X} = [1\text{e}{-6}, 2\text{e}{-5}]^6 \times [1\text{e}{-6}, 1\text{e}{-3}]^3 \times [0.1, 0.9]
\]
Here, $C_1,\dots,C_6$ are the capacitance values (in Farads), $L_1,L_2,L_3$ are the inductance values (in Henries), and $T_1$ is the duty cycle shared across all 5 switches. The duty cycle $T_1$ denotes the fraction of time during which the switches are in the ``on'' state and governs a periodic on-off pattern repeated at high frequency throughout the simulation.

\subsubsection*{Objectives}
\label{app:PEobjectives}

The simulation outputs two scalar values: \textit{DcGain} and \textit{Voltage Ripple}. The former represents the ratio of load to input voltage and should ideally approximate a predefined constant, such as $0.25$, as closely as possible. Meanwhile, the latter quantifies the voltage fluctuation at the load. Let $t_1, \dots t_N$ denote the time steps during the simulation, then the objectives are characterized by the following parameters:

\begin{equation*}
\begin{aligned}    
    \min_{\mathbf{x} \in \mathcal{X}} |\text{DcGain(x)} - 0.25| &= |\frac{\overline{V_{load}(t)}}{V_{Source}} - 0.25|\\ 
    &= |\frac{1}{V_{Source}} \cdot \frac{1}{T} \sum_{i=1}^{N-1} \frac{V_{load}(t_{i+1}) + V_{load}(t_{i})}{2} \cdot (t_{i+1} - t_i) - 0.25| 
\end{aligned}
\end{equation*}
where $\overline{V_{load}(t)}$ is the average voltage of the load calculated by the simulator during the transient analysis and $V_{source}$ is the voltage source of $1000$ volt. $T = t_N - t_1$ is the duration of the simulation.

\begin{align*}
    \min_{\mathbf{x} \in \mathcal{X}}  \text{Voltage Ripple} &= \frac{V_{pp}(t)}{\overline{V_{load}(t)}}\\ &= \frac{\max_{i\in[1, N]}V_{load}(t_i) - \min_{i\in[1, N]}V_{load}(t_i)}{\overline{V_{load}(t)}}
\end{align*}
where ${V_{pp}(load)}$ is the peak-to-peak value on the load calculated during the transient analysis.

\subsubsection*{Conditions}

This problem does not include environmental or operational conditions as part of its input specification. Unlike other domains where the simulation setup may vary based on conditions (\eg, load configurations or external temperatures), the PowerElectronics circuit is simulated under fixed source voltage and switching behavior. As a result, the design optimization task focuses solely on tuning internal circuit parameters, with no external conditions to vary. More complex variants of this problem\textemdash involving multiple topologies or variable source voltages\textemdash may be considered in future releases.

\subsubsection*{Constraints}

We list the theoretical constraints, which represent physically infeasible conditions that always result in an error.

\paragraph{Theoretical constraints (error)}
\begin{align*}
\mathit{C_i} &\in{[1e-6, 2e-5]}, &i=0,1,2,...,5\\
\mathit{L_i} &\in{[1e-6, 1e-3]}, &i=0,1,2\\
\mathit{T_1} &\in{[0.1, 0.9]} &
\end{align*}

\subsubsection*{Simulator}

We use the well-established NgSpice Circuit Simulator in this problem.

\subsubsection*{Dataset}

All simulation starts from 1 millisecond to 1.06 milliseconds. The dataset consists of three parts as follows:
\begin{itemize}
  \item \textbf{Part 1:} The 6 capacitors and 3 inductors only take their minimum and maximum values to map the edge of the design space. $T_1$ ranges over $\{0.1, 0.2, 0.3, 0.4, 0.5, 0.6, 0.7, 0.8, 0.9\}$. This results in:
  \begin{equation*}
  2^6 \times 2^3 \times 9 = 4608 \text{ samples.}
  \end{equation*}
  \item \textbf{Part 2:} Randomly sample 4608 points in the 10-dimensional space (6 capacitors, 3 inductors, and $T_1$). The minimum and maximum values for each dimension are excluded from the sampling.
  \item \textbf{Part 3:} To enhance the sampling coverage in the design space, we use Latin hypercube sampling to generate 4608 points. Each dimension is divided into 10 intervals, and the minimum and maximum values in each dimension are excluded from the sampling.
\end{itemize}

These three parts are randomly shuffled and split into training, validation, and test datasets with the ratio of $7/2/1$.

\section{Additional Information on Experiments}
\label{app:additional_info_expes}

\subsection{Cross-domain study}

\begin{table}[ht]
  \centering
  \caption{Hyperparameters used for GAN2D and CGAN2D in the cross-domain study.}
  \label{tab:hparams_gans2d}
  \begin{tabular}{cc}
    \toprule
    \textbf{Hyperparameter} & \textbf{Value} \\
    \midrule
    \texttt{lr\_disc}        & $0.0004$  \\
    \texttt{lr\_gen}        & $0.0001$  \\
    \texttt{n\_epochs} & $100$               \\
    \texttt{hidden\_layers}                         & $4$ \\
    \texttt{hidden\_size}                       & $128$ \\
    \texttt{batch\_size}                            & $32$ \\
    \texttt{b1}            & $0.5$                \\
    \texttt{b2}            & $0.999$                \\
    \texttt{latent\_dim}            & $32$                \\
    \bottomrule
  \end{tabular}
\end{table}

\begin{table}[ht]
  \centering
  \caption{Hyperparameters used for CDiffusion2D in the cross-domain study.}
  \label{tab:hparams_cdiffusion2d}
  \begin{tabular}{cc}
    \toprule
    \textbf{Hyperparameter} & \textbf{Value} \\
    \midrule
    \texttt{lr}        & $0.0004$  \\
    \texttt{n\_epochs} & $100$               \\
    \texttt{batch\_size}                            & $32$ \\
    \texttt{b1}            & $0.5$                \\
    \texttt{b2}            & $0.999$                \\
    \texttt{latent\_dim}            & $100$                \\
    \texttt{layers\_per\_block}            & $2$                \\
    \texttt{noise\_schedule}            & linear                \\
    \bottomrule
  \end{tabular}
\end{table}

Hyperparameters used for our experiments in \cref{sec:3by3_exp} can be found in \cref{tab:hparams_gans2d,tab:hparams_cdiffusion2d}. They were chosen based on our prior experience and without an extensive search, \textit{i.e.}, reducing the number of channels and removing the transformer blocks of the diffusion model.\footnote{Our implementations draw from 
 \url{https://huggingface.co/learn/diffusion-course/unit1/2} and \url{https://github.com/togheppi/cDCGAN}.} Conducting a systematic hyperparameter sweep proved impractical, since each problem and evaluation metric (beyond just MSE) would require its own tuning regime. We view the challenge of hyperparameter optimization across heterogeneous engineering problems and metrics as an open and valuable direction for future research. We ran the experiments with seeds in $\{1,\dots,10\}$.\footnote{Code for GAN2D: \url{https://github.com/IDEALLab/EngiOpt/tree/main/engiopt/gan_cnn_2d}.\\
 Code for CGAN2D: \url{https://github.com/IDEALLab/EngiOpt/tree/main/engiopt/cgan_cnn_2d}.\\
 Code for CDiffusion2D: \url{https://github.com/IDEALLab/EngiOpt/tree/main/engiopt/diffusion_2d_cond}.
 } 

Here is an example command line used to train our CDiffusion model on the heat conduction problem. Our framework makes it possible to train this model on a different problem by only changing the \texttt{problem\_id}:
\begin{lstlisting}[escapeinside={(*@}{@*)}, basicstyle=\ttfamily,breaklines=true]
python engiopt/diffusion_2d_cond/diffusion_2d_cond.py 
    (*@\aftergroup\speciallstcolor@*)--problem_id heatconduction2d(*@\aftergroup\endspeciallstcolor@*) 
    --seed 10 
    --save_model 
    --n_epochs=100
\end{lstlisting}
For conditioning in the model evaluations, 50 conditions were sampled uniformly at random with replacement from the configurations in the test datasets. Then, these configurations were given to the conditional model generators as conditions. For the unconditional models, we used the same sampling procedure\textemdash both to ensure broader coverage of test conditions and to supply conditions for simulation and optimization experiments\textemdash but the generators themselves received only noise as input. Metrics were then computed based on the outputs relative to their associated conditions.

\begin{figure*}[h]
  \centering
  \def\figwidth{0.48\textwidth}

  \begin{subfigure}{\figwidth}
    \centering
    \includegraphics[width=\linewidth]{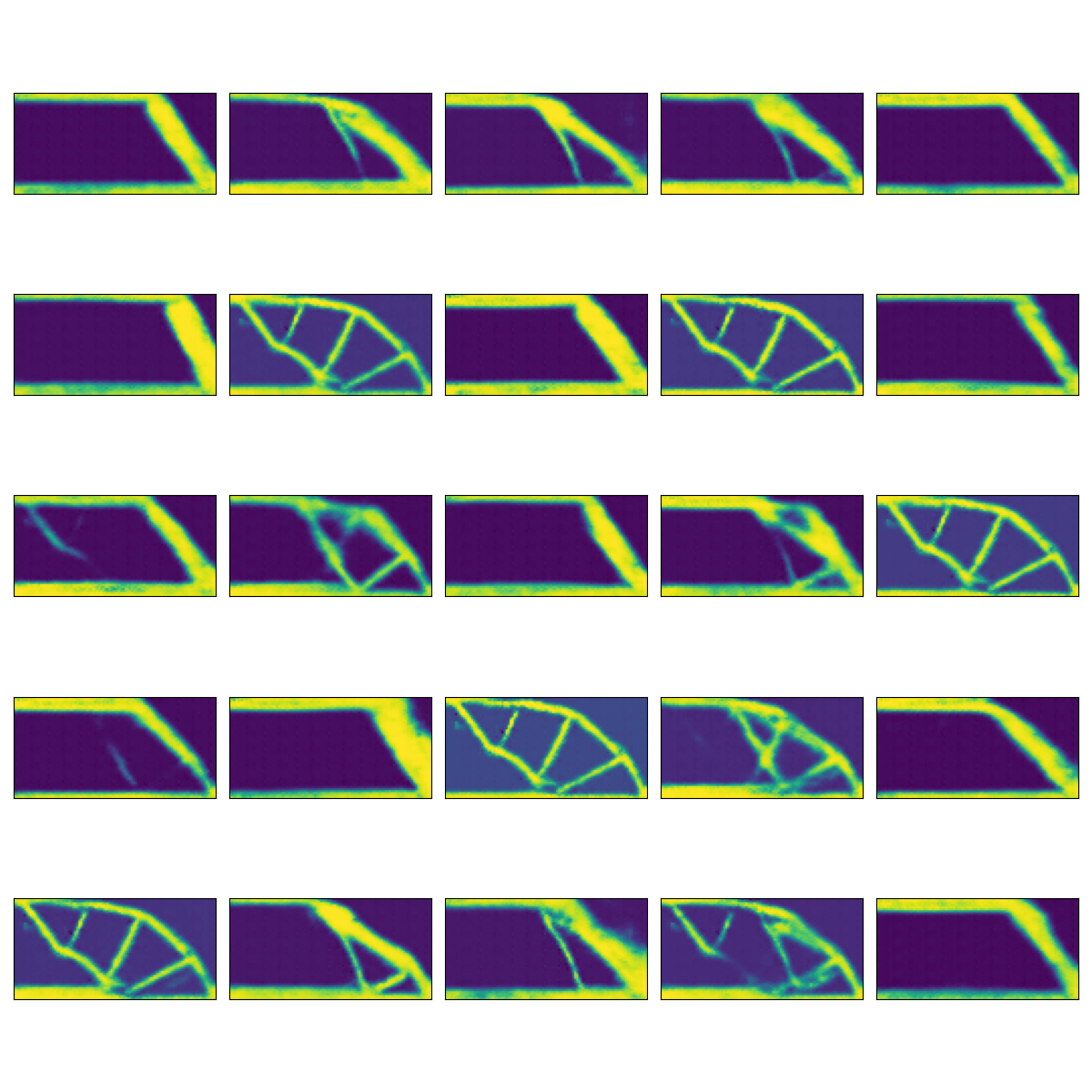}
    \caption*{\footnotesize GAN2D}
  \end{subfigure}\hfill
  \begin{subfigure}{\figwidth}
    \centering
    \includegraphics[width=\linewidth]{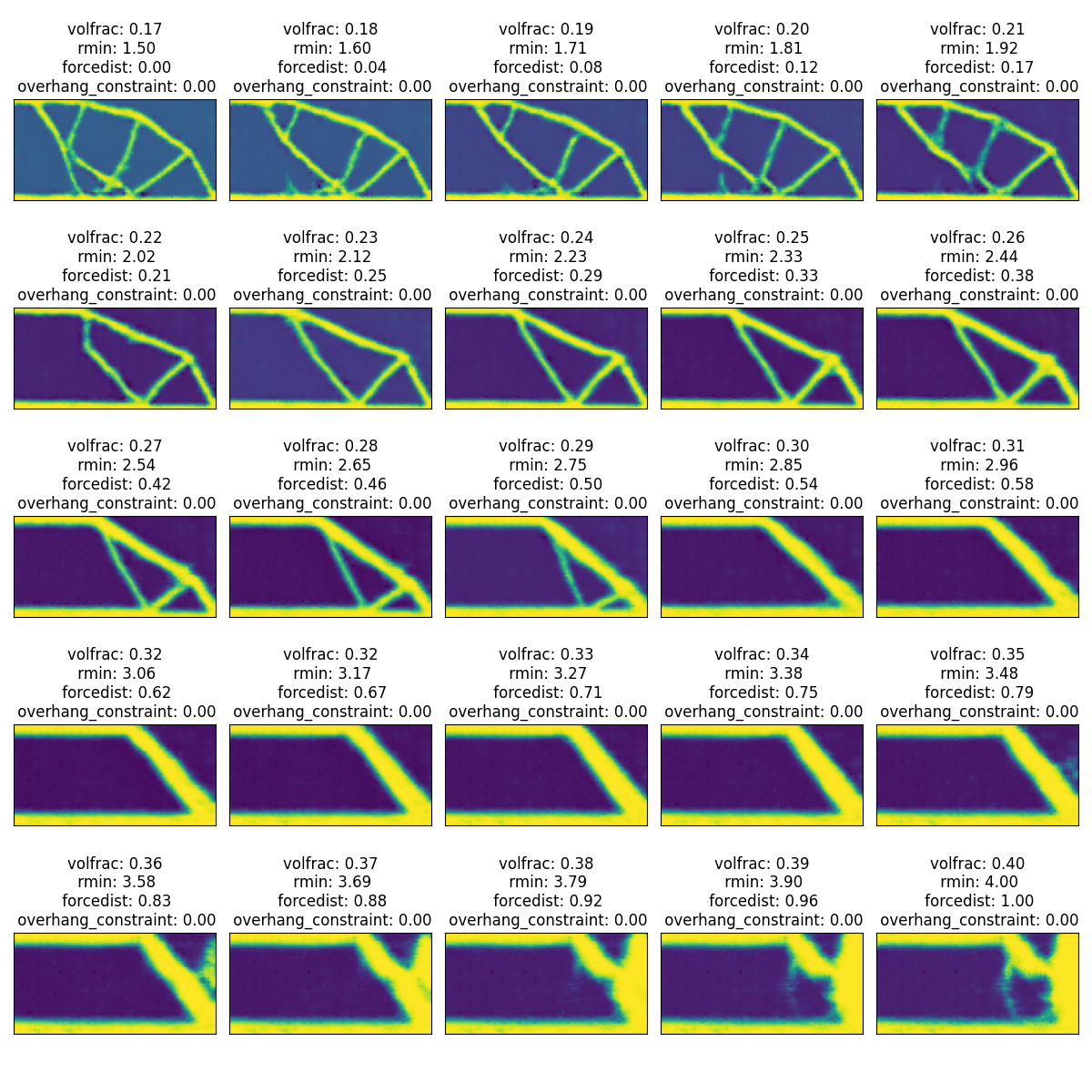}
    \caption*{\footnotesize CGAN2D}
  \end{subfigure}\hfill
  \begin{subfigure}{\figwidth}
    \centering
    \includegraphics[width=\linewidth]{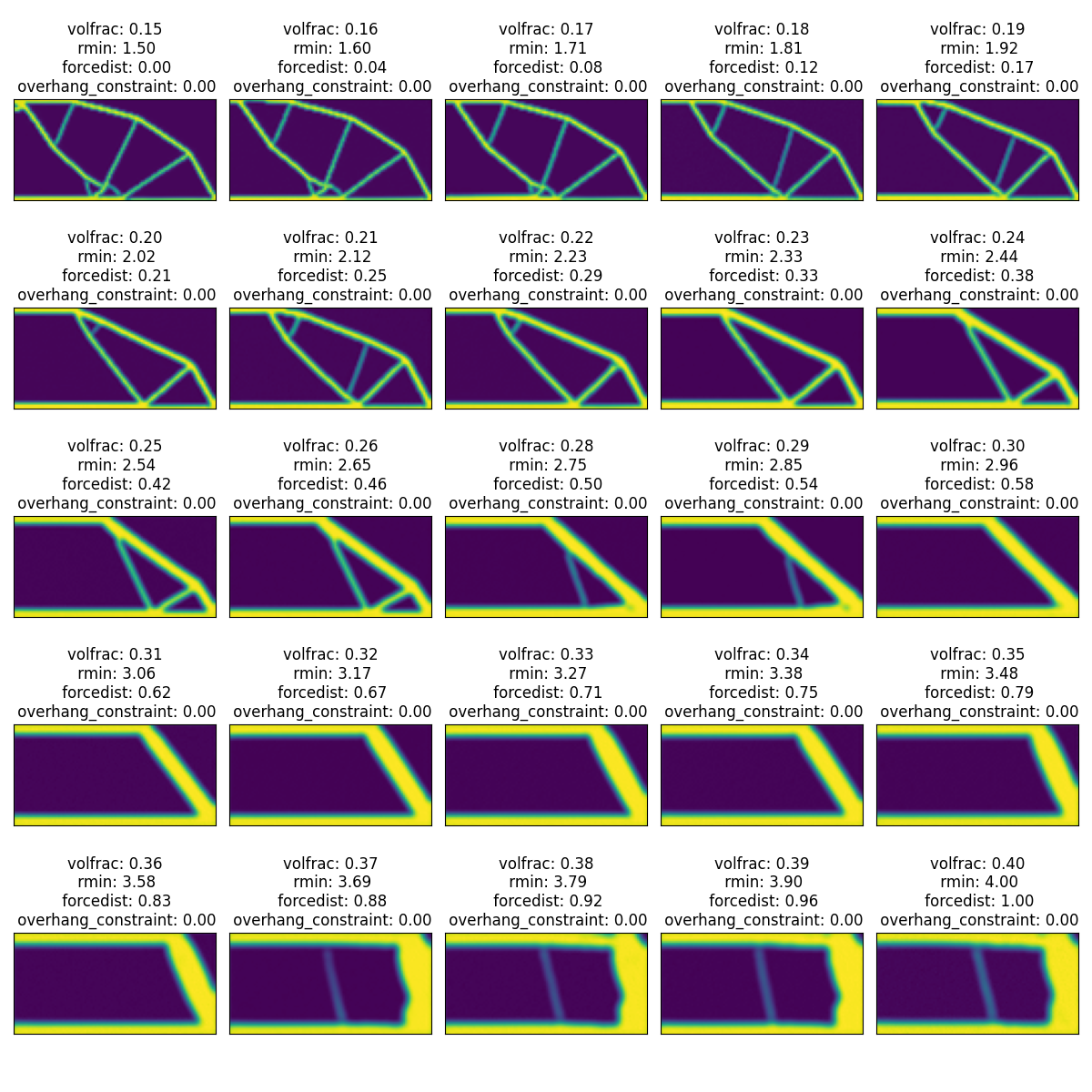}
    \caption*{\footnotesize CDiffusion2D}
  \end{subfigure}

  \caption{Illustrative outputs of our generative models on the Beams2D task.}
  \label{fig:beams}
\end{figure*}

\begin{figure*}
  \centering
  \def\figwidth{0.48\textwidth}

  \begin{subfigure}{\figwidth}
    \centering
    \includegraphics[width=\linewidth]{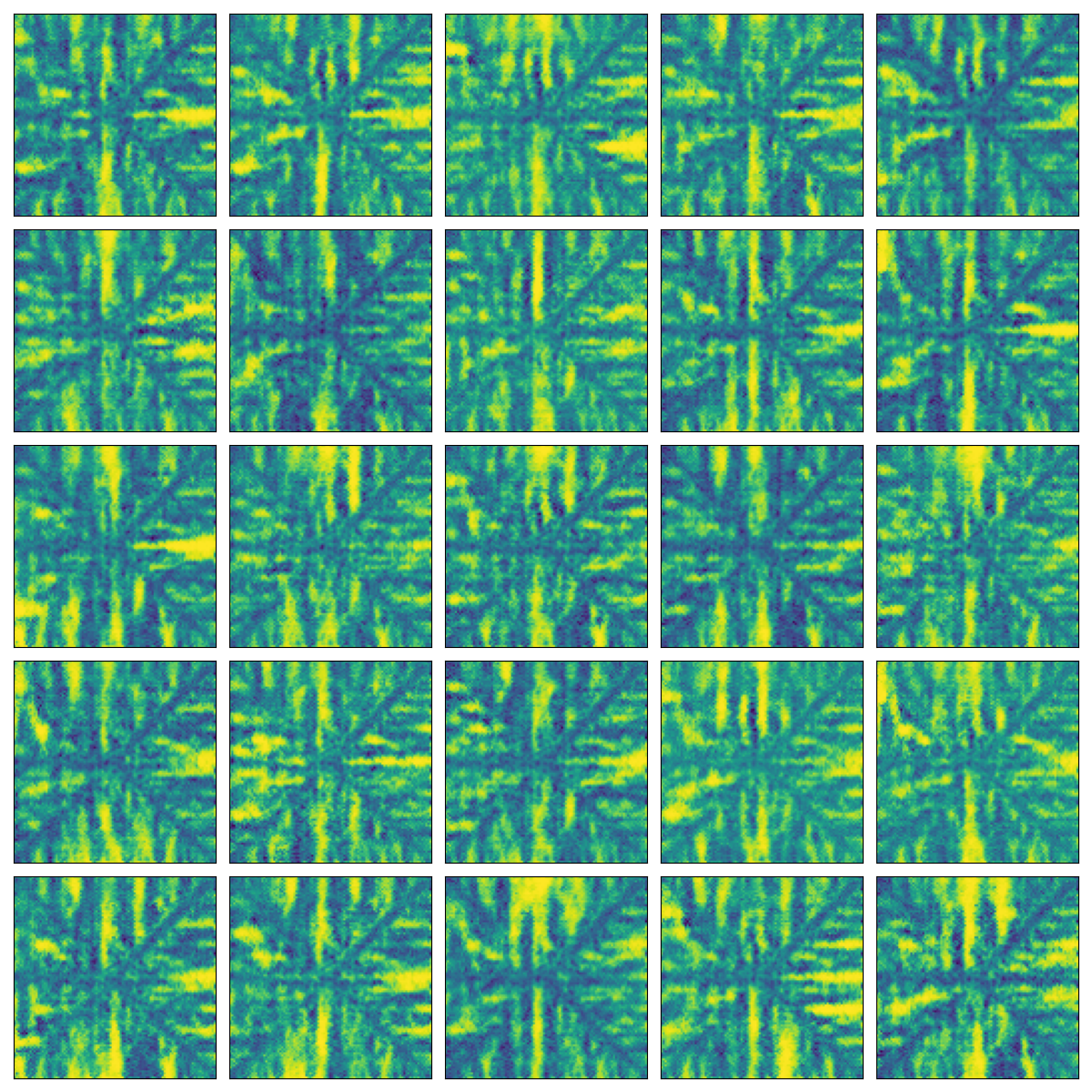}
    \caption*{\footnotesize GAN2D}
  \end{subfigure}\hfill
  \begin{subfigure}{\figwidth}
    \centering
    \includegraphics[width=\linewidth]{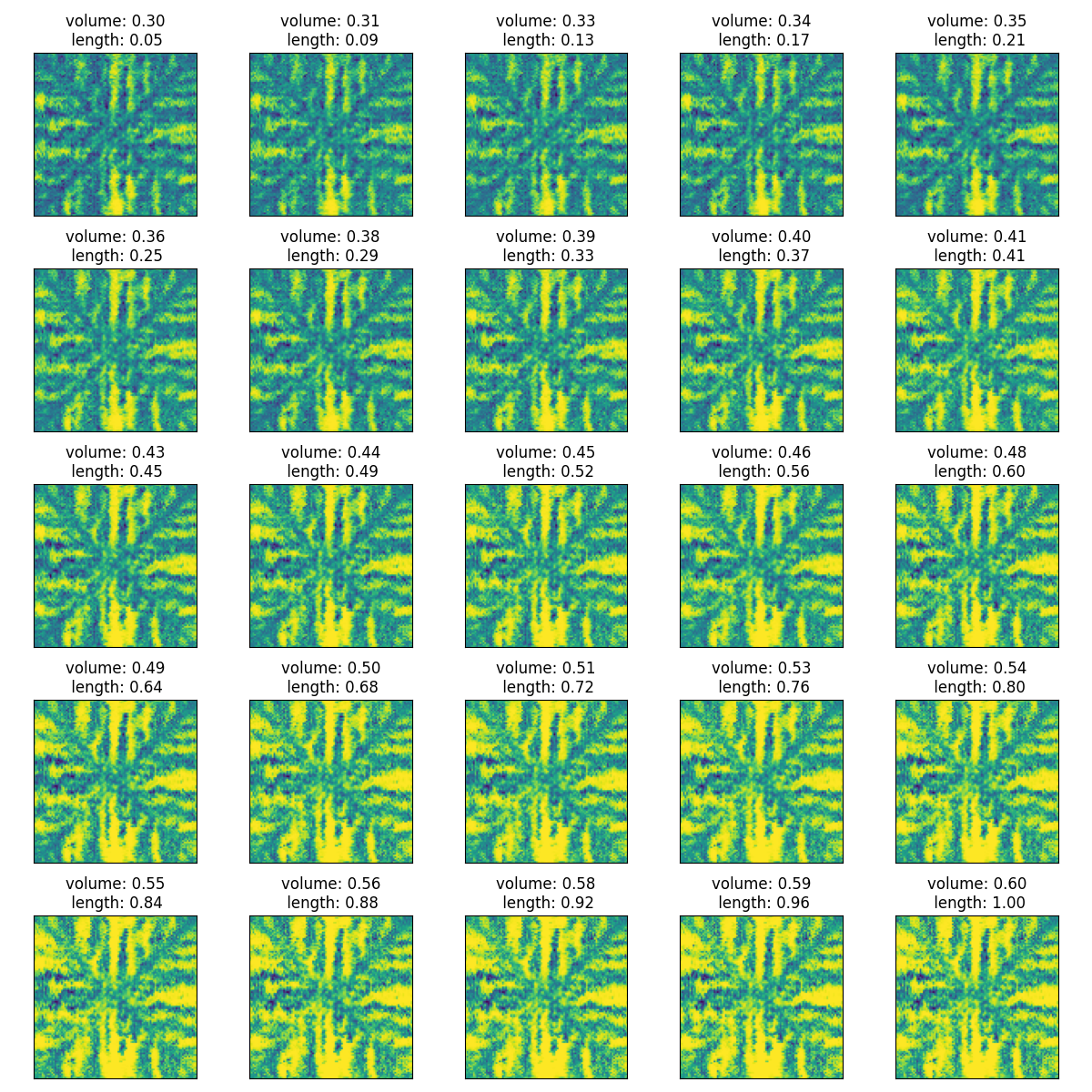}
    \caption*{\footnotesize CGAN2D}
  \end{subfigure}\hfill
  \begin{subfigure}{\figwidth}
    \centering
    \includegraphics[width=\linewidth]{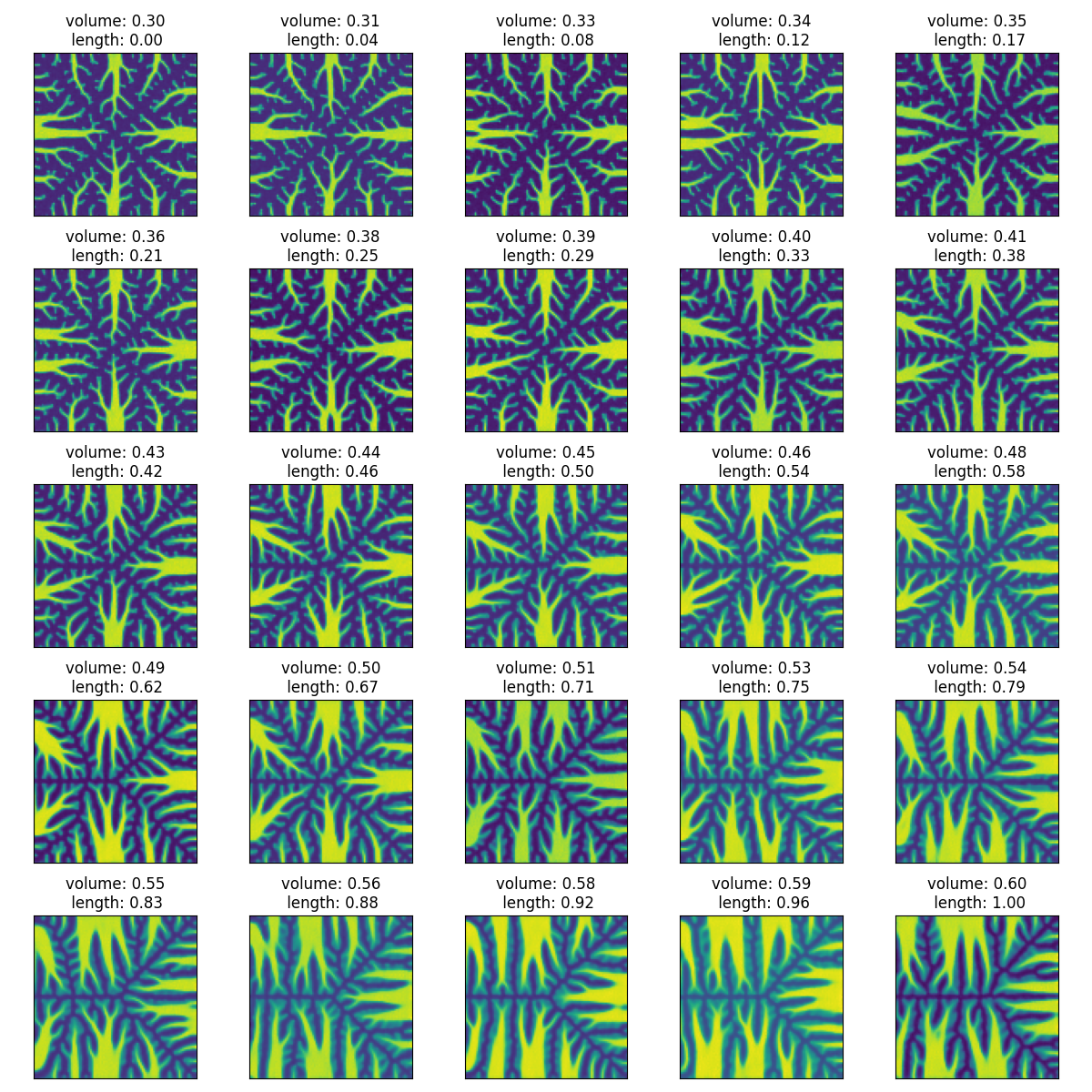}
    \caption*{\footnotesize CDiffusion2D}
  \end{subfigure}

  \caption{Illustrative outputs of our generative models on the HeatConduction2D task.}
  \label{fig:heatconduction}
\end{figure*}

\begin{figure*}[h]
  \centering
  \def\figwidth{0.48\textwidth}

  \begin{subfigure}{\figwidth}
    \centering
    \includegraphics[width=\linewidth]{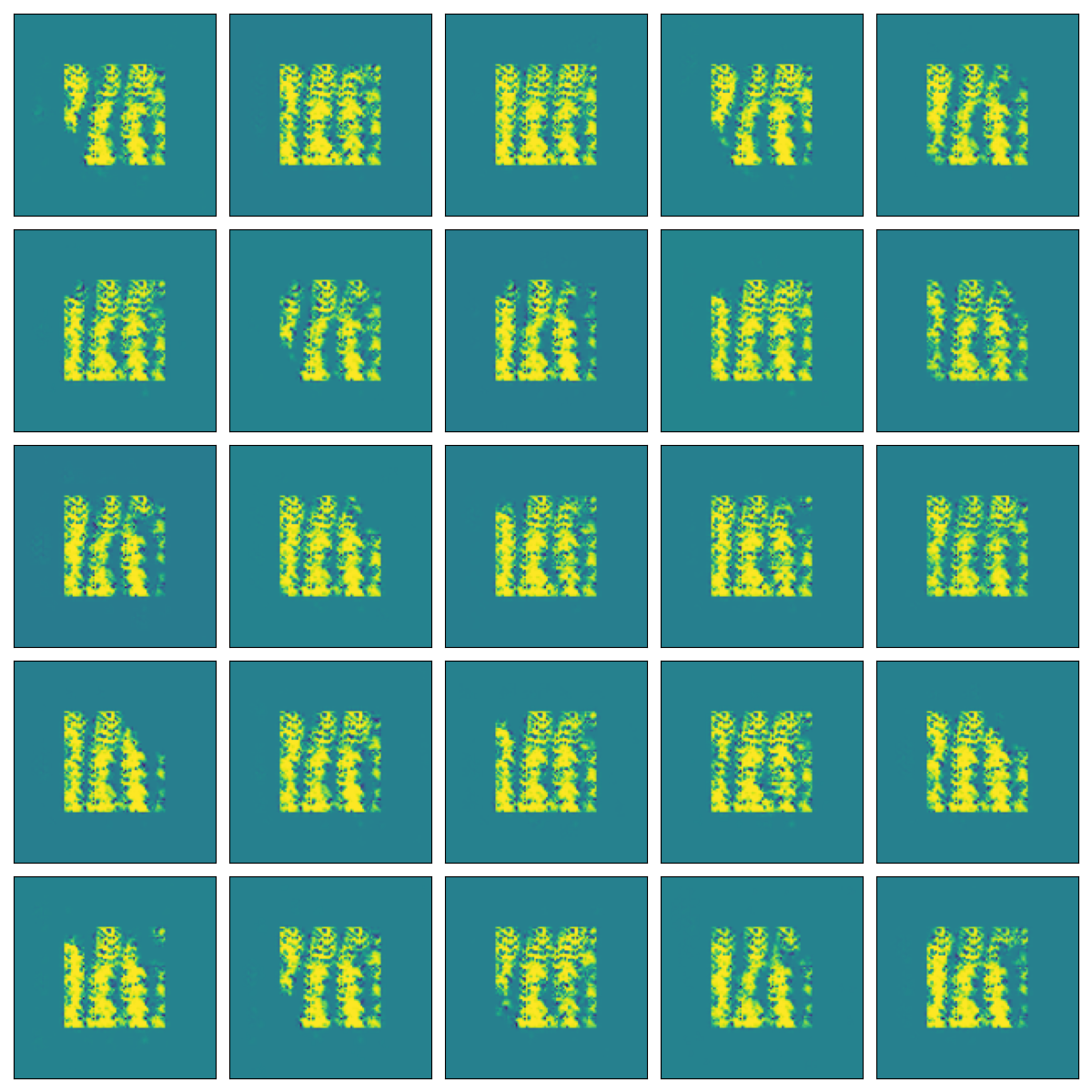}
    \caption*{\footnotesize GAN2D}
  \end{subfigure}\hfill
  \begin{subfigure}{\figwidth}
    \centering
    \includegraphics[width=\linewidth]{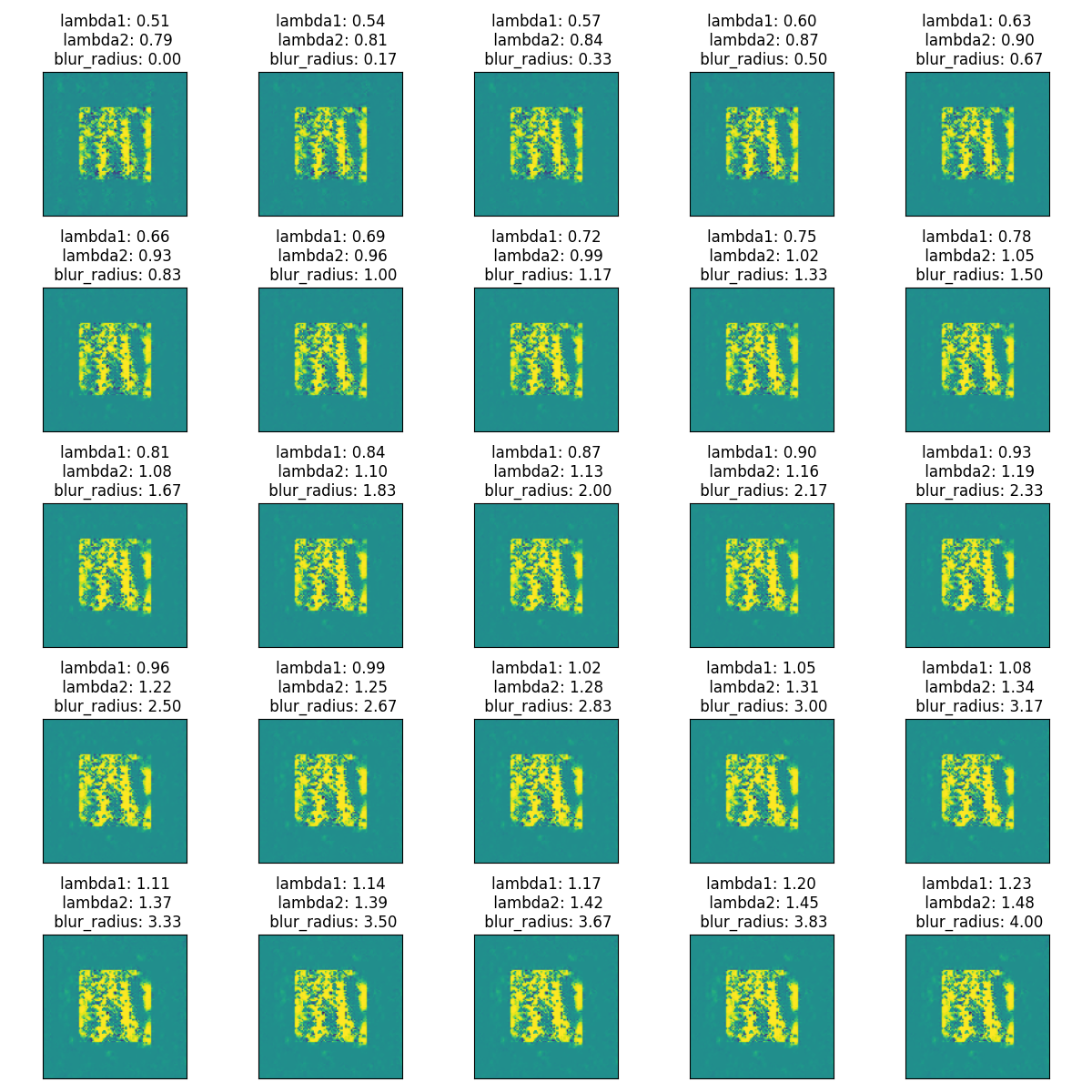}
    \caption*{\footnotesize CGAN2D}
  \end{subfigure}\hfill
  \begin{subfigure}{\figwidth}
    \centering
    \includegraphics[width=\linewidth]{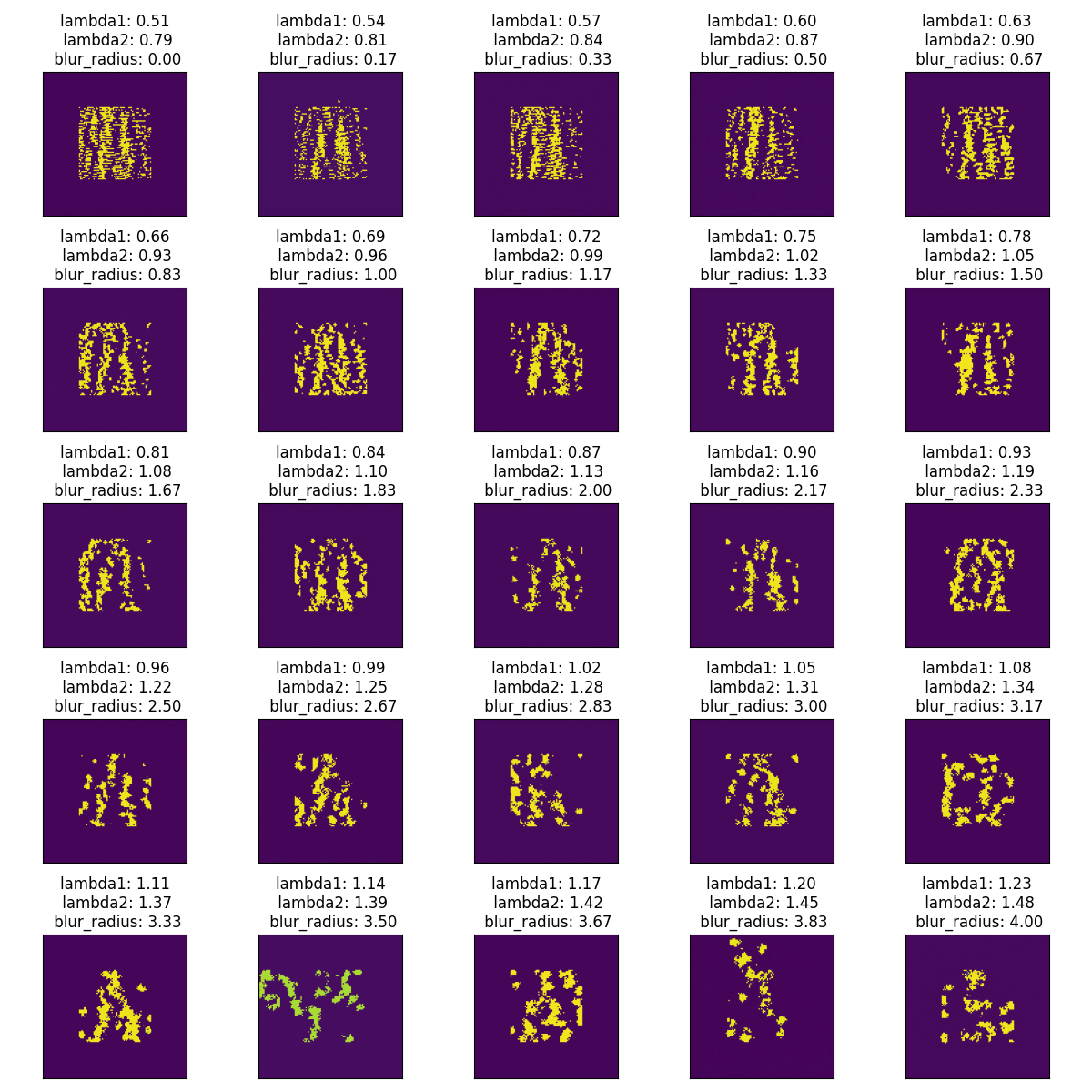}
    \caption*{\footnotesize CDiffusion2D}
  \end{subfigure}

  \caption{Illustrative outputs of our generative models on the Photonics task.}
  \label{fig:photonics}
\end{figure*}

Some example outputs generated from the trained algorithms are illustrated in \cref{fig:heatconduction,fig:beams,fig:photonics}. Diffusion models tend to produce more detailed and structured outputs, whereas GANs yield blurrier designs. Interestingly, although GAN outputs may appear less refined visually, they are often easier to optimize, as reflected by lower COG values in HeatConduction2D and Beams2D (\cref{sec:3by3_exp}). The figures also highlight the contrast between unconditional and conditional models: while conditional models generate designs tailored to specific conditions, unconditional models sample freely across the entire design space. This explains the higher DPP scores observed for unconditional models, as they naturally exhibit greater diversity. Conversely, the conditioned models should better respect the constraints, as is the case for the CGAN. It is unclear why the Diffusion models violate constraints almost $100\%$ of the time in the results. Similarly, we expected conditional models to achieve lower MMD scores, as they are evaluated against the conditional distributions aligned with the sampled test conditions. These results are surprising and should not be overinterpreted: we did not perform extensive hyperparameter tuning, and the number of seeded runs is insufficient for strong statistical claims. Rather, these findings serve as a proof of concept and highlight important directions for future investigation.

\vfill
\newpage

\vfill
\newpage

\clearpage
\subsection{Airfoils}
\begin{table}[ht]
  \centering
  \caption{Hyperparameters used for GAN and BézierGAN in the Airfoil experiment.}
  \label{tab:hparams_gans1d}
  \begin{tabular}{cc}
    \toprule
    \textbf{Hyperparameter} & \textbf{Value} \\
    \midrule
    \texttt{lr\_disc}        & $0.0002$  \\
    \texttt{lr\_gen}        & $0.00005$  \\
    \texttt{n\_epochs} & $5000$               \\
    \texttt{batch\_size}                            & $32$ \\
    \texttt{b1}            & $0.5$                \\
    \texttt{b2}            & $0.999$                \\
    \texttt{latent\_dim}            & $4$                \\
    \texttt{noise\_dim}            & $10$                \\ \midrule
    \texttt{bezier\_control\_pts}            & $32$                \\
    \bottomrule
  \end{tabular}
\end{table}

\begin{table}[ht]
  \centering
  \caption{Hyperparameters used for the Diffusion model in the Airfoil experiment.}
  \label{tab:hparams_diffusion1d}
  \begin{tabular}{cc}
    \toprule
    \textbf{Hyperparameter} & \textbf{Value} \\
    \midrule
    \texttt{lr}        & $0.0003$  \\
    \texttt{n\_epochs} & $100$               \\
    \texttt{batch\_size}                            & $64$ \\
    \texttt{b1}            & $0.5$                \\
    \texttt{b2}            & $0.999$                \\
    \texttt{unet\_dim}            & $32$                \\
    \bottomrule
  \end{tabular}
\end{table}

Hyperparameters used for our experiments in \cref{sec:zoomed_exp} can be found in \cref{tab:hparams_gans1d,tab:hparams_diffusion1d}. For the GANs, we used the same hyperparameter values as in \citet{chen_airfoil_2020}. For the diffusion model, we reused hyperparameters from \url{https://github.com/lucidrains/denoising-diffusion-pytorch}. We ran the experiments with seeds in $\{1,\dots,10\}$. \footnote{Code for GAN: \url{https://github.com/IDEALLab/EngiOpt/tree/main/engiopt/gan_1d}.\\
Code for B\'ezierGAN: \url{https://github.com/IDEALLab/EngiOpt/tree/main/engiopt/gan_bezier}.\\
Code for Diffusion: \url{https://github.com/IDEALLab/EngiOpt/tree/main/engiopt/diffusion_1d}.}

\begin{figure*}[h]
  \centering
  \def\figwidth{0.48\textwidth}

  \begin{subfigure}{\figwidth}
    \centering
    \includegraphics[width=\linewidth]{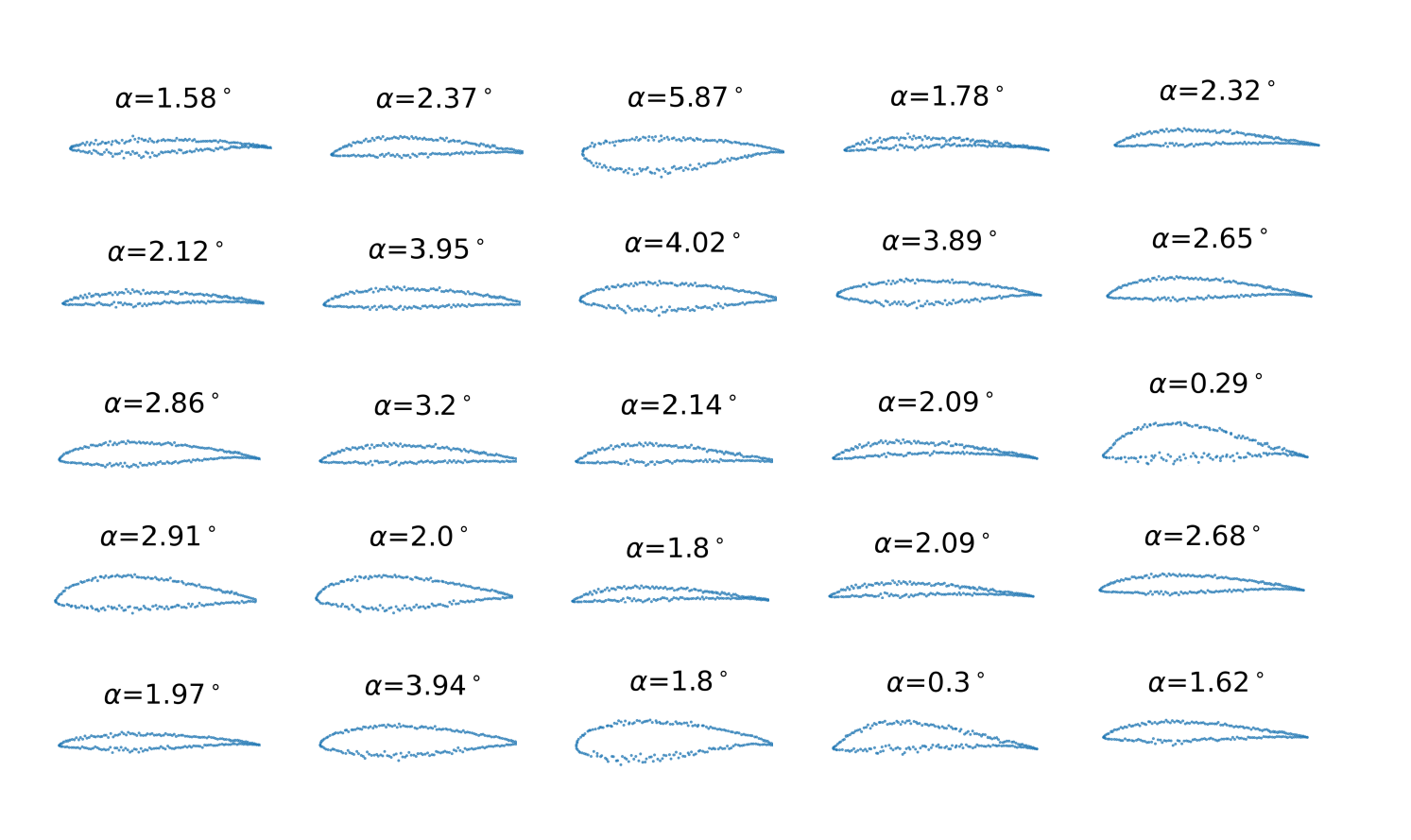}
    \caption*{\footnotesize GAN}
  \end{subfigure}\hfill
  \begin{subfigure}{\figwidth}
    \centering
    \includegraphics[width=\linewidth]{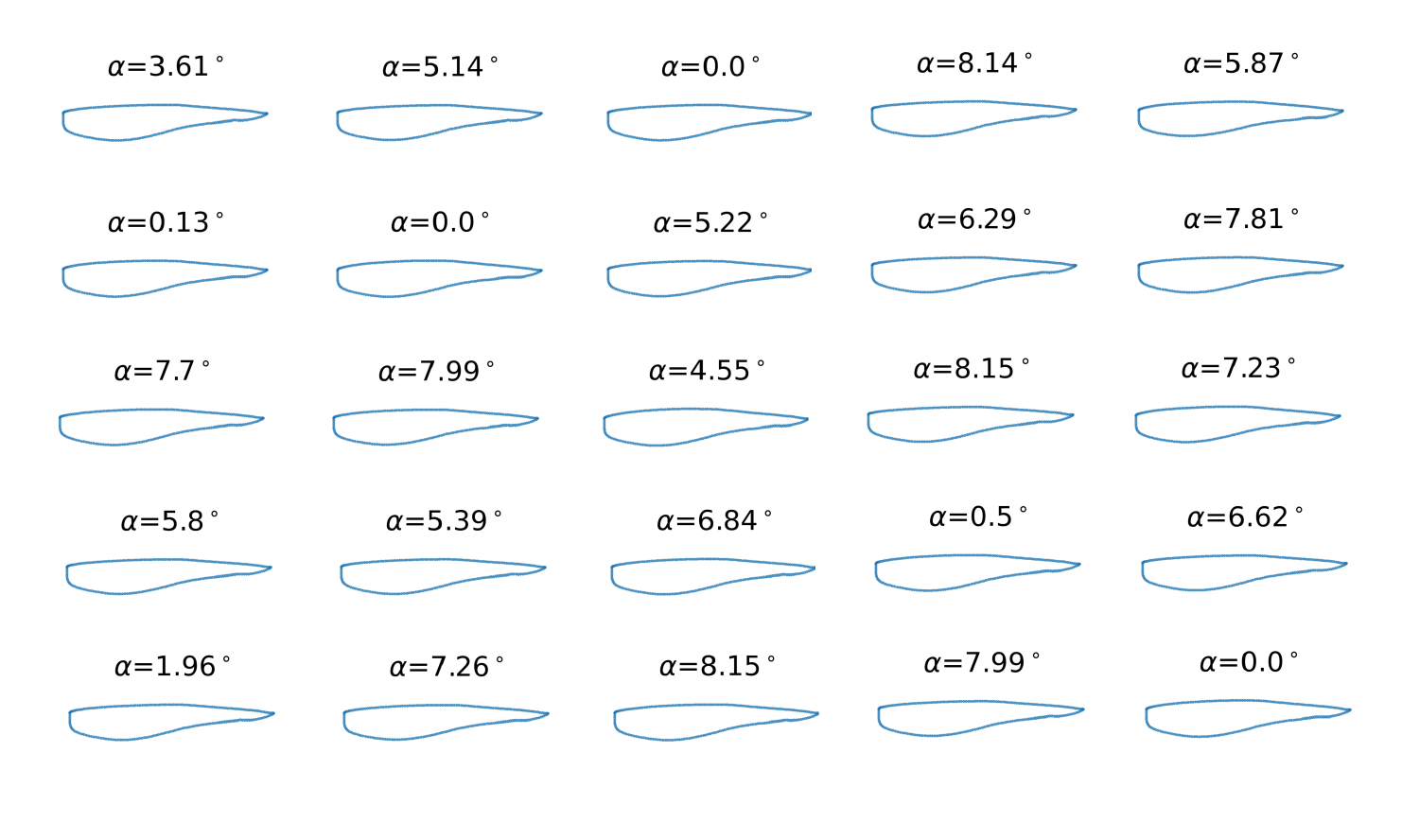}
    \caption*{\footnotesize Diffusion}
  \end{subfigure}\hfill
  \begin{subfigure}{\figwidth}
    \centering
    \includegraphics[width=\linewidth]{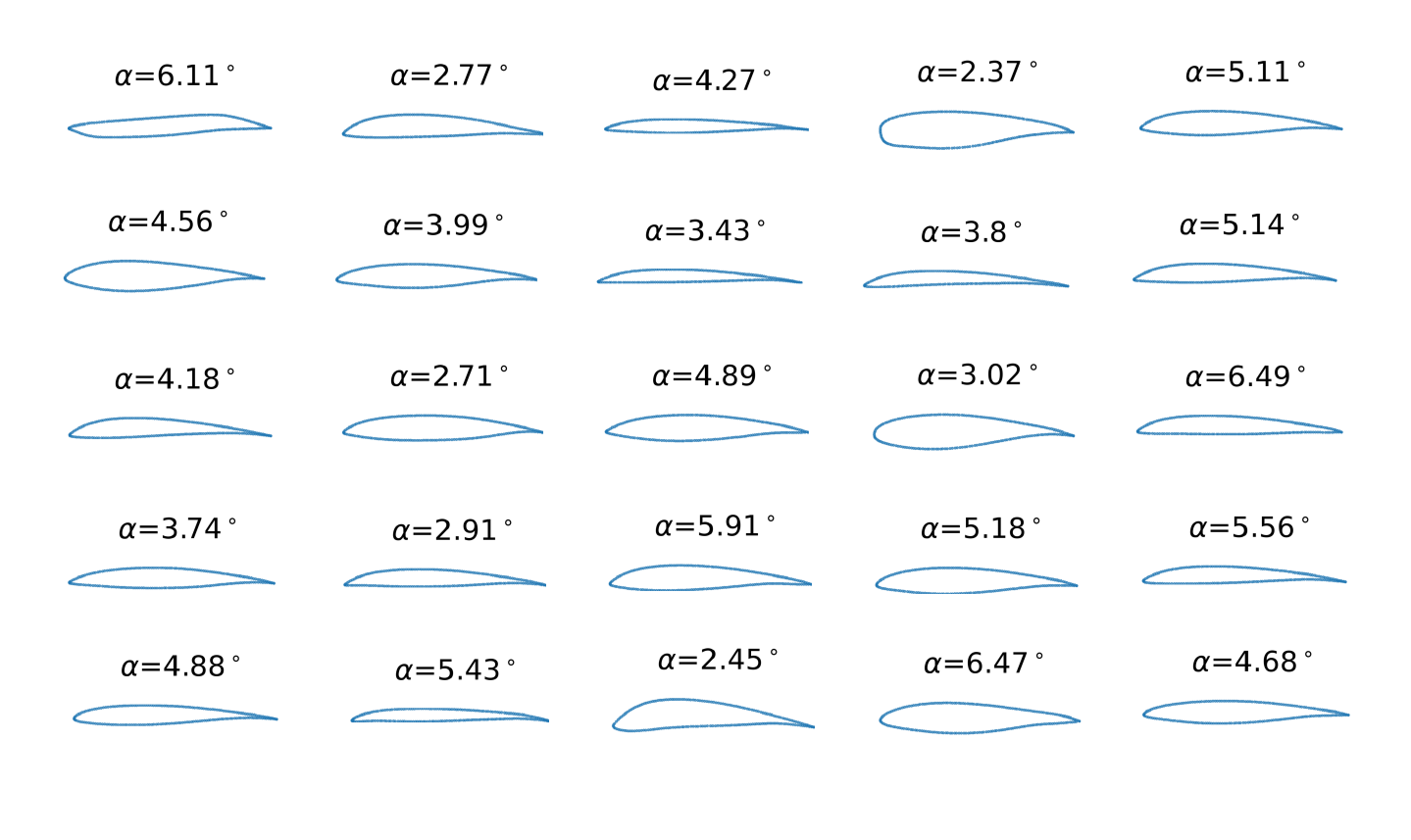}
    \caption*{\footnotesize BézierGAN}
  \end{subfigure}

  \caption{Illustrative outputs of our generative models on the Airfoil task.}
  \label{fig:airfoils_outputs}
\end{figure*}

Example outputs from the trained models are shown in \cref{fig:airfoils_outputs}. As illustrated, the standard GAN frequently produces invalid designs characterized by discontinuities or noisy coordinates along the airfoil spline, resulting in a $30\%$ simulation failure rate. In contrast, the B\'ezierGAN and diffusion models consistently generate designs that are smooth and valid for simulation. However, the diffusion model suffers from mode collapse, producing highly similar outputs across different samples.

\clearpage
\subsection{Surrogate-assisted multi-objective optimization}
\label{app:surrogate}

\subsubsection*{Problem setting}
For a fixed power-converter topology, the continuous design vector
\[
x = \begin{bmatrix} C_1,\dots,C_6,L_1,L_2,L_3,T_1 \end{bmatrix}^{\!\top} \in \mathcal{X}
\quad \text{where} \quad
\mathcal{X} = [1\text{e}{-6}, 2\text{e}{-5}]^6 \times [1\text{e}{-6}, 1\text{e}{-3}]^3 \times [0.1, 0.9]
\]
is mapped by the NGSpice simulator to two responses:
\textit{DcGain} \(y_1\) and \textit{Voltage Ripple} \(y_2\).
We aim to \emph{simultaneously} minimize
\[
  g_1(x) = \lvert y_1(x) - 0.25\rvert,
  \qquad
  g_2(x) = y_2(x).
\]



\subsubsection*{Methods}

We performed several steps for training our surrogate models and optimizing designs using these. \footnote{Code for surrogate-based optimization: \url{https://github.com/IDEALLab/EngiOpt/tree/main/engiopt/surrogate_model}.}

\begin{figure}[t]
  \centering
  \includegraphics[width=0.8\linewidth]{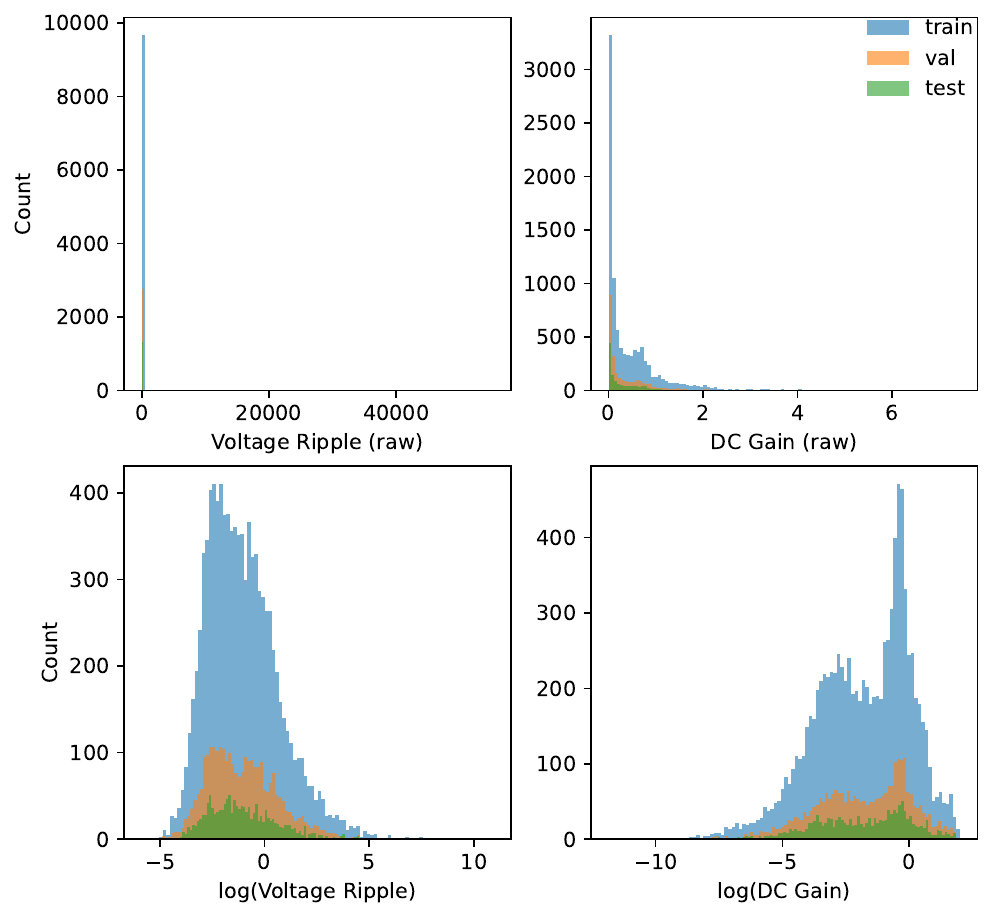}
  \caption{Train/val/test distributions of \textit{Voltage Ripple} and
           \textit{DcGain} in raw (top) and log (bottom) domains.
           Heavy skew and extreme outliers motivate the log transform
           and robust scaling.}
  \label{fig:raw-log-dists}
\end{figure}

\textbf{Pre-processing:}
Every raw feature \(x_j\) has been converted to a robust-scaled value \(\tilde{x}_j = (x_j - Q_{1,j}) / (Q_{3,j}-Q_{1,j})\), where \(Q_{1,j}\) and \(Q_{3,j}\) are the 25th / 75th percentiles of feature \(j\) in the training set (inter-quartile range scaling). Each target \(y_k\) has been first stabilized by \(\log(y_k)\) and then scaled with the same formula, mitigating the heavy-tailed distribution (see \cref{fig:raw-log-dists}).

\textbf{Surrogate model:}
For each target \(k\in\{1{:}2\}\) (\,1: \textit{DcGain}, 2: \textit{Voltage Ripple}\,) we fit an MLP \(f_{\boldsymbol{\theta}_k} : \mathbb{R}^{10}\!\to\!\mathbb{R}\) with parameters \(\boldsymbol{\theta}_k\).  Given the training set \(\mathcal{D}_{\text{tr}}=\{(\tilde{{x}}^{(i)},y_k^{(i)})\}_{i=1}^{N_{\text{tr}}}\), where \(\tilde{{x}}\) is the robust-scaled design vector and \(N_{\text{tr}}=0.7N\), the parameters are obtained by

\[
  \boldsymbol{\theta}_k^{\star}
  \;=\;
  \underset{\boldsymbol{\theta}}{\arg\min}\;
  \frac{1}{N_{\text{tr}}}
  \sum_{i=1}^{N_{\text{tr}}}
  \mathcal{L}_{\text{sL1}}\!
  \bigl(f_{\boldsymbol{\theta}}(\tilde{{x}}^{(i)}),\,y_k^{(i)}\bigr),
\]

where \(\mathcal{L}_{\text{sL1}}\) denotes the Smooth-\(L_1\) loss. Optimization uses Adam (PyTorch 2.6) with early stopping (patience 20) to prevent over-fitting.

\textbf{Bayesian hyperparameter search:}
We used BoTorch+Ax~\cite{balandat2020_botorch, snoek_practical_2012} to minimize the \emph{ensemble‑averaged}
validation loss over the search space in \cref{tab:hparam-space-surrogate-model}.
Each trial trains an \emph{implicit ensemble} of three MLPs obtained by
different initialization seeds \((s,s\!+\!1,s\!+\!2)\).
Similar to other hyperparameter optimization works~\citep{eimer_hyperparameters_2023,parker-holder_automated_2022,felten_hyperparameter_2023}, we tuned hyperparameters on different seeds \(s\in\{100,101,102\}\) for each trial, giving \(3\times50\) trials, or \(3\times3\times50=450\) (seed, ensemble size, trials) MLPs trained per target output; every model has been trained for 50 epochs with early stopping patience set at 15.
The optimized hyperparameters are \texttt{learning\_rate},
\texttt{hidden\_layers}, \texttt{hidden\_size}, \texttt{batch\_size} penalty \texttt{l2\_lambda}, and \texttt{activation}. Hyperparameters have intentionally been biased to limit the size of the networks and avoid over-fitting.

\begin{table}[t]
  \centering
  \caption{Hyperparameter search space ($\mathcal{Z}$) and optimization budget.}
  \label{tab:hparam-space-surrogate-model}
  \begin{tabular}{lc}
    \toprule
    \textbf{Hyperparameter} & \textbf{Value range} \\
    \midrule
    \texttt{learning\_rate}          & $[10^{-5},\,10^{-3}]$ \\
    \texttt{hidden\_layers}           & $\{2,\,3,\,4\}$ \\
    \texttt{hidden\_size}            & $\{16,\,32,\,64,\,128\}$ \\
    \texttt{batch\_size}             & $\{8,\,16,\,32,\,64\}$ \\
    \texttt{l2\_lambda}  & $[10^{-6},\,10^{-3}]$ \\
    \texttt{activation}              & $\{\text{ReLU},\,\text{Tanh}\}$ \\
    \midrule
    \textit{Budget} (trials) & 50 \\
    Seeds \(s\) & $\{100, 101, 102\}$
 \\    \bottomrule
  \end{tabular}
\end{table}

\textbf{Implicit ensembles:}
Using the best hyperparameters (best validation loss across seeds), we trained \(E=10\) independent ensembles.
Each ensemble contains \(M=7\) replicas initialized with contiguous seeds
\((s,\dots,s\!+\!6)\) for \(s\in\{0,10,\dots,90\}\). Note that these seeds are different from the seeds used for the hyperparameter search to avoid seed overfitting~\citep{eimer_hyperparameters_2023}.
Every replica is trained for at most 300 epochs with early stopping
(patience 50). Following the implicit ensemble paradigm \cite{xue_deep_2021,ganaie_ensemble_2022}, predictions are averaged \(\hat{y}_k(x)=\frac1M\sum_{m=1}^{M}f_{\theta_k^{(m)}}(x)\), which lowers epistemic error. Variance can be used alongside mean prediction, but it is out of the scope of this work.

\textbf{Multi‑objective search:}
For each surrogate ensemble, we launched a seed‑matched NSGA‑II run (population 500, generations 100) via \emph{pymoo}
\cite{deb_fast_2002,blank_pymoo_2020}. Seeds \(s\in\{0,10,\dots,90\}\) generate ten independent approximations \(\hat{\mathcal{P}}^{(s)}\subset\mathcal{X}\) of the Pareto set of optimal designs.


\textbf{Validation:}
Designs in \(\hat{\mathcal{P}}\) have then been re-evaluated with the simulator, giving surrogate- and simulator-based Pareto fronts that we treat as distributions
\(\hat{\mathbf{g}},\mathbf{g}\).
The unbiased squared Maximum Mean Discrepancy \cite{gretton_kernel_2012}, \(\text{MMD}^2(\hat{\mathbf{g}},\mathbf{g})\), with a Gaussian kernel rejects the null of identical distributions (\emph{p}\!<\!0.05), exposing extrapolation bias.

\subsubsection*{Results}
Bayesian optimization converged to different optima for the two targets (\cref{tab:hparam-best-surrogate-model}): the \textit{DcGain} surface favored a small learning rate and tanh activations, whereas the harder \textit{Voltage Ripple} surface required an order of magnitude larger learning rate and ReLU activations. The learning rate decays were empirically selected. Both networks architectures ended up with the largest number of \texttt{hidden\_layers} and \texttt{hidden\_size}, signaling a possible overfitting trend.

\begin{figure*}[ht]
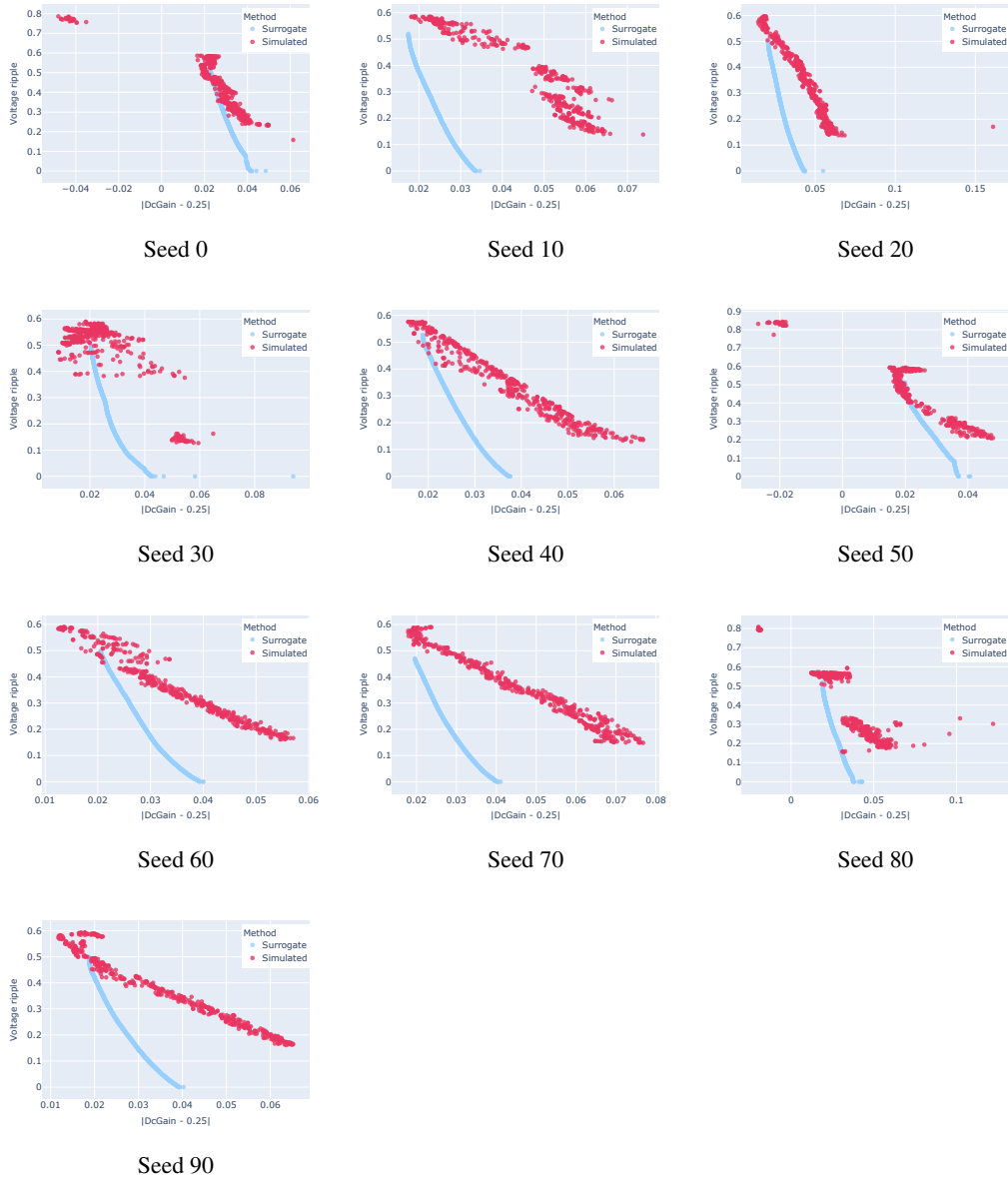

  \centering
  \def\figwidth{0.33\textwidth}

  \foreach \s in {0,10,20}{%
    \begin{subfigure}{\figwidth}
      \centering
      \includegraphics[width=\linewidth]{images/pareto_front_\s.pdf}
      \caption*{\footnotesize Seed \s}
    \end{subfigure}\hfill
  }
  \par\vspace{0.5em}

  \foreach \s in {30,40,50}{%
    \begin{subfigure}{\figwidth}
      \centering
      \includegraphics[width=\linewidth]{images/pareto_front_\s.pdf}
      \caption*{\footnotesize Seed \s}
    \end{subfigure}\hfill
  }
  \par\vspace{0.5em}

  \foreach \s in {60,70,80}{%
    \begin{subfigure}{\figwidth}
      \centering
      \includegraphics[width=\linewidth]{images/pareto_front_\s.pdf}
      \caption*{\footnotesize Seed \s}
    \end{subfigure}\hfill
  }
  \par\vspace{0.5em}

  \foreach \s in {90}{%
    \begin{subfigure}{\figwidth}
      \centering
      \includegraphics[width=\linewidth]{images/pareto_front_\s.pdf}
      \caption*{\footnotesize Seed \s}
    \end{subfigure}\hfill
  }

  \caption{NSGA-II Pareto fronts for the ten seed-matched surrogate ensembles.  
           Blue scatter: surrogate prediction; red scatter: NGSpice evaluation.}
  \label{fig:pareto-ten-seeds}
\end{figure*}

With these settings, each surrogate ensemble was able to learn a coherent Pareto front (blue curves, \cref{fig:pareto-ten-seeds}); reevaluation in NgSpice (red scatter) revealed systematic offsets for every seed. A two-sample MMD test (1000 permutations, Gaussian kernel) confirmed that the surrogate and NgSpice-generated fronts stem from different distributions: across ten runs the squared statistic ranged from \( 9.97 \times 10 ^{-3}\) to \( 6.18 \times 10 ^{-2}\), while every p-value hit the permutation floor of \( 0.001 \). Hence, the null hypothesis of equal clouds is rejected for all seeds, demonstrating that the surrogate fails to generalize to the NSGA-II–proposed regions despite extensive hyperparameter tuning and ensembling.

\begin{table}[t]
  \centering
  \caption{Best hyperparameter configuration selected by Bayesian optimization for each output. The learning-rate decay is the per-epoch multiplicative factor.}
  \label{tab:hparam-best-surrogate-model}
  \begin{tabular}{@{}lcc@{}}
    \toprule
    \textbf{Hyperparameter} & \textbf{DC\,Gain} & \textbf{Voltage\,Ripple} \\
    \midrule
    \texttt{learning\_rate}        & $2.67\times10^{-4}$ & $1.00\times10^{-3}$ \\
    \texttt{learning\_rate\_decay} & $0.95$              & $1.00$ \\
    \texttt{hidden\_layers}        & $4$                 & $4$ \\
    \texttt{hidden\_size}          & $128$               & $128$ \\
    \texttt{batch\_size}           & $8$                 & $8$ \\
    \texttt{l2\_lambda}            & $1.0\times10^{-6}$  & $3.07\times10^{-4}$ \\
    \texttt{activation}            & Tanh                & ReLU \\
    \bottomrule
  \end{tabular}
\end{table}

\subsubsection*{Why is this problem difficult for surrogate models?}

Despite a fixed topology and well-defined design bounds, this problem remains challenging for surrogate-assisted optimization. Several intertwined factors contribute:

\textbf{Stiff dynamics from mixed time constants:} The circuit includes both microfarad-level capacitors and millihenry-level inductors, creating vastly different time scales in transient responses. This stiffness can lead to sharp transitions that may be difficult for approximators like MLPs to capture.

\textbf{Discontinuities and threshold effects:} Diodes and switches introduce strong nonlinearity and discontinuous behavior. Small parameter changes near diode conduction thresholds can result in large output shifts (e.g., clipping, ringing, or switching of current paths), making the response surface non-smooth or even piecewise.

\textbf{High variance and sparse coverage:} The target \textit{Voltage Ripple} exhibits heavy-tailed behavior, with a small subset of designs leading to large, high-frequency oscillations. These regions are sparsely represented in the dataset, making them hard to learn accurately even with ensembling.

\textbf{Simulation-induced numerical noise:} NgSpice's numerical integration (\eg, trapezoidal or backward Euler) and tolerances can introduce small but structured numerical artifacts. These manifest as ``noise'' from the surrogate's perspective, even though they arise from deterministic simulations.

\textbf{Hidden constraints and failure modes:} Although obvious simulation failures are avoided by fixing the switch pattern, near-failure designs can behave erratically or saturate, leading to sharp distortions in the response space that are hard to interpolate.

In contrast to smooth benchmark functions, this domain exhibits the highly sensitive and constrained design manifold characteristic of real-world physical systems. It demonstrates the modeling challenges that traditional surrogate models can encounter, calling for physics-informed features or structure-aware priors to perform well.

\subsubsection*{Take-aways}
\begin{enumerate}
    \item \textit{Voltage Ripple} is intrinsically harder to approximate than \textit{DcGain} due to stiffness (high frequency transients arise from stiff systems dynamics, a regime where standard MLPs struggle to generalize accurately), discontinuities, and skewed distributions driven by switching and nonlinear components;
    \item Ensembling reduces variance but does not resolve extrapolation bias in undersampled or near-discontinuous regions;
    \item Sparse representation of unstable behaviors (\eg, large ripple) and overfitting in smooth regions can mislead optimization, resulting in systematic surrogate error near the predicted Pareto front;
    \item Domain-aware enhancements\textemdash such as physics-informed architectures, adaptive sampling near switching thresholds, or hybrid surrogate-simulation loops\textemdash may be needed to reliably optimize high-fidelity circuit models.
\end{enumerate}

Future work could explore hybrid strategies that combine adaptive space-filling design, physics-informed normalization, outlier-aware modeling (\eg, classification of unstable regimes), and autoregressive architectures to couple dependent objectives.

\end{document}